\documentclass[aps,prl,twocolumn,superscriptaddress,raggedbottom,showpacs,floatfix]{revtex4-1}
\usepackage{color}
\usepackage{graphicx,latexsym}
\usepackage{epstopdf}
\usepackage{comment,bm}
\usepackage{amsmath}
\usepackage{amssymb}
\usepackage{mathtools}
\usepackage{graphics}
\usepackage{comment}
\usepackage{tikz}

\newcommand{\beq}{\begin{equation}}
\newcommand{\eeq}{\end{equation}}
\newcommand{\bei}{\begin{itemize}}
\newcommand{\eei}{\end{itemize}}
\newcommand{\ben}{\begin{enumerate}}
\newcommand{\een}{\end{enumerate}}

\newcommand{\be}{{\mathbf e}}

\usepackage{hyperref}
\hypersetup{
   pdfpagemode=None, 
   pdfstartpage=1,
   pdfmenubar=true,
   pdftoolbar=true,
   colorlinks = true,
   linkcolor=blue,
   citecolor=blue,
   urlcolor=blue,
   bookmarksopen=false
 }

\definecolor{darkblue}{rgb}{0.,0.24,0.51}
\definecolor{britishracinggreen}{rgb}{0.0, 0.26, 0.15}
\definecolor{darkgreen}{rgb}{0,0.60,.2}

\def\be{\begin{equation}}
\def\ee{\end{equation}}

\begin{document}
\renewcommand{\vec}{\mathbf}
\renewcommand{\Re}{\mathop{\mathrm{Re}}\nolimits}
\renewcommand{\Im}{\mathop{\mathrm{Im}}\nolimits}
\title{Topological fluctuating electron-hole Cooper pairs in graphene-GaAs heterostructures}

%\author{Dmitry K. Efimkin$^{1,2}$ and Mehdi Kargarian$^{3}$}
%\affiliation{$^1$School of Physics and Astronomy, Monash University, Victoria 3800, Australia}
%\affiliation{$^2$ARC Centre of Excellence in Future Low-Energy Electronics Technologies, Monash University, Victoria 3800, Australia}
%\affiliation{$^3$Department of Physics, Sharif University of Technology, Tehran 14588-89694, Iran}
\author{Dmitry K. Efimkin}
\email{dmitry.efimkin@monash.edu}
\affiliation{School of Physics and Astronomy, Monash University, Victoria 3800, Australia}
\affiliation{ARC Centre of Excellence in Future Low-Energy Electronics Technologies, Monash University, Victoria 3800, Australia}

\begin{abstract}
Fluctuating Cooper pairs formed by spatially separated electrons and holes are precursors of their equilibrium condensation. Their presence strongly impacts transport phenomena and interlayer tunneling in double-layer systems above the transition temperature. Here, we consider a hybrid graphene/quantum well double-layer system and focus on the dynamics of fluctuating Cooper pairs formed by conventional electrons and Dirac holes. We demonstrate that the chiral nature of Dirac holes is manifested in the presence of two (almost) degenerate competing pairing channels, which are intertwined by effective pseudospin-orbit interactions. We argue that the spectrum of the Ginzburg-Landau Hamiltonian describing the energetics of fluctuating Cooper pairs is geometrically nontrivial and can be characterized by the half-integer topological Chern number. We derive a kinetic equation for fluctuating Cooper pairs and demonstrate that their nontrivial geometries generate two anomalous velocities of distinct geometrical origins. These velocities are intricately connected with the Berry curvature and the quantum metric for the Ginzburg-Landau Hamiltonian, respectively. The resulting anomalous contributions to conductivity are singular at the transition temperature, and we discuss possible setups for their experimental observation.  
\end{abstract}
%\date{\today}
\maketitle

\section{I. Introduction}
Over the past decades, there has been a growing interest in the geometrical Berry phase for electrons in solids. The presence of the phase has a profound effect on material properties and is responsible for a variety of phenomena, such as polarization, orbital magnetism, and various (quantum, anomalous, or spin) Hall effects (see reviews Ref.~\cite{BFReview1,BFReview2,AHEReview} and references therein). Additionally, the corresponding Berry curvature, which functions as an analog of a magnetic field in reciprocal space, is a cornerstone concept for the topological classification of solids~\cite{TopologicalClassification}. The spatial distribution of the Berry curvature in reciprocal space distinguishes topological insulators, semimetals, superconductors, and electron-hole superfluids from their topologically trivial counterparts~\cite{TopologicalReview1,TopologicalReview2,TopologicalReview3}.

Superfluids formed by closely spaced electrons and holes in bilayer systems (e.g., semiconductor GaAs/AlGaAs quantum well (QW) systems~\cite{LozovikYudson1,LozovikYudson2,Shevchenko,GaAsRecent1,GaAsRecent2,EHreview}, heterostructures formed by monolayer materials~\cite{EH1,EH2,EH3,EH4,EH5,EH6,EHBG1,EHBG2,EHBG3,EHBG4,EHBG5,EHBG6,EHBG4twisted}, topological insulator thin films~\cite{EHTI1,EfimkinEHTI}) are close relatives of superconductors. These superfluids can maintain phase coherence via a Cooper pair (CP) condensate and support dipolar superfluidity, which is potentially useful for various applications. Thus far, the nontrivial geometry of electron-hole superfluids has been considered only as a feature of Bogoliubov
quasiparticles. The spectrum of these quasiparticles is shaped by the interplay between
the band geometry of electrons and holes and the momentum distribution of the static order parameter describing the equilibrium CP condensate. Here, we argue that there is another geometrical aspect of electron-hole superfluids (as well as superconductors) that has been previously overlooked. 

Above the transition temperature for electron-hole CP condensation, Gaussian fluctuations of the order parameter occur, which are usually interpreted as partly coherent fluctuating CPs~\footnote{It should be noted, that these CPs are formed not by electrons, but by spatially separated electrons and holes}. The presence of these fluctuations results in the strong enhancement and critical behavior of interlayer tunneling~\cite{TunnelingEfimkin1,TunnelingEfimkin2} and the Coulomb drag resistance~\cite{DragHu,DragMink1,DragMink2,DragEfimkin} reminiscent of the Aslamazov-Larkin effect and related effects in superconductors~\cite{Varlamov,SkocpolTinkham,LarkinVarlamov,AslamazovlarkinDiamagnetsim,AslamazovlarkinConductivity}. The behavior consistent with the presence of these fluctuations has been reported in double-bilayer graphene~\cite{TunnelingExp}, $\hbox{MoSe}_2$-$\hbox{WSe}_2$ heterostructures~\cite{LumExp}, conventional semiconductor QWs~\cite{DragExp1,DragExp1,DragExp3,DragExp4}, and hybrid graphene-GaAs bilayers~\cite{HybridDrag}. Here, we demonstrate  that nontrivial geometries of electrons and holes pass on to the fluctuating CPs and that their spectrum can also be classified as topological. 
%\footnote{It should be noted that remarkable progress in observing spontaneous coherence of exciton-polaritons in semiconductor microcavities has been recently achieved (see Refs. [83–85] for a review and references therein).} 

In the present paper, we consider the hybrid graphene-GaAs bilayer reported in Ref.~\cite{HybridDrag} and sketched in Fig~\ref{Fig1}. We focus on the dynamics of fluctuating CPs formed by conventional electrons and Dirac holes. We demonstrate that the chiral nature of Dirac holes is manifested in the presence of two (almost) degenerate competing pairing channels, which are intertwined by effective pseudospin-orbit interactions. We argue that the spectrum of the Ginzburg-Landau (GL) Hamiltonian for fluctuating CPs is geometrically nontrivial and can be characterized by the half-integer Chern number. We derive a kinetic equation for fluctuating CPs and demonstrate that their nontrivial geometries generate two anomalous velocities, which are intricately related to the Berry curvature and quantum metric for the GL Hamiltonian, respectively. The resulting anomalous contributions to conductivity are singular at the transition temperature, and we discuss possible setups for their experimental observation.

The remainder of the paper is organized as follows. In Section II, we present a low-energy model describing the hybrid graphene-GaAs double-layer system. In Section III, we discuss the Cooper instability and classify the pairing channels. In Section IV, we introduce the GL Hamiltonian and discuss the nontrivial geometries of its spectrum. In Section V, we derive a kinetic equation for fluctuating CPs. The temperature dependence of the anomalous paraconductivity is presented in Section VI, and different experimental setups for its detection are overviewed in Section VII. Section VIII is devoted to discussions and conclusions.

\section{II. Model}

\begin{figure}[t]
	\begin{center}
		\includegraphics[trim=2cm 12.5cm 16.5cm 4cm, clip, width=1.0\columnwidth]{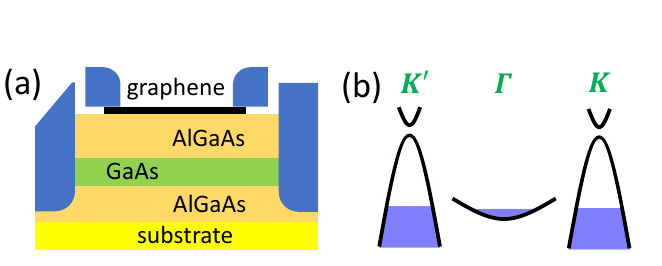}
		\caption{Schematics of the hybrid graphene/GaAs bilayer. A graphene sheet is deposited on the surface of a semiconductor AlGaAs, underneath which a GaAs quantum well hosts two-dimensional electron gas. Carriers in the two layers are induced by gating or by a doping layer~\cite{HybridDrag} (not shown in the sketch) and have independent contacts (dark blue) required for transport measurements). 	(b) The low-energy band structure of the hybrid graphene-GaAs bilayer includes a conduction band with conventional electrons and a pair of valence bands (K and K' valleys) with Dirac holes.}
		\label{Fig1}
	\end{center}
\end{figure}

The hybrid graphene-GaAs bilayer and its low-energy band structure are sketched in Fig.~\ref{Fig1} (a). Following the experimental setup~\cite{HybridDrag}, we assume that graphene sheet with excess of holes is deposited on the surface of a semiconductor AlGaAs, underneath which a GaAs quantum well hosts two-dimensional electron gas. The physics of electron-hole bilayers formed by monolayer materials is very rich and has been considered in a number of recent papers~\cite{EH1,EH2,EH3,EH4,EH5,EH6,EHBG1,EHBG2,EHBG3,EHBG4,EHBG5,EHBG6,EHMultiband1,EHMultiband2,EHMultiband3,EHBG4twisted} (see Ref.~\cite{EHreview} for a review) aiming to provide realistic predictions of the density/temperature phase diagram. Here, we follow a different route and consider a minimal phenomenological model that properly accounts for the chiral nature of Dirac holes. 

The nature of correlations in the hybrid electron-hole bilayer depends on three dimensionless parameters. The first two are the Wigner-Seitz interaction strength parameters $r^{\mathrm{e}}_\mathrm{s}=m_\mathrm{e} e^2/\hbar p_\mathrm{F} \kappa_{\mathrm{e}}$ and $r^{\mathrm{h}}_\mathrm{s}=e^2/\hbar v \kappa_{\mathrm{h}}$ that scales the ratio of interactions and kinetic energy in GaAs QW and graphene, respectively. The third parameter $p_\mathrm{F}d/\hbar$ scales the interlayer Coulomb interactions respect to the intralayer one. Here $m_\mathrm{e}$ is mass for electron mass in GaAs QW and $p_\mathrm{F}$ is their Fermi momentum; $v$ is the velocity of Dirac holes in graphene;  $\kappa_\mathrm{e}$ and $\kappa_\mathrm{h}$ are the effective dielectric constants for each layer in the considered multilayer heterostructure and $d$ is separation distance between electrons and holes. If $r_\mathrm{s}^{\mathrm{e}(\mathrm{h})}\ll1$ and $p_\mathrm{F}d/\hbar\gg1$ the system is in the weak coupling regime with pairing correlations only in the vicinity of Fermi level for electrons and holes. This regime can be described by the Bardeen-Cooper-Schrieffer (BCS) theory. Its range of the applicability in the considered hybrid bilayer also extends to the moderate coupling strength regime, $r_\mathrm{s}^{\mathrm{e}(\mathrm{h})}\sim1 $ and $p_\mathrm{F}d/\hbar\sim1$,  and is wider compared to the conventional QW bilayers. The reason is that the 
gapless nature of Dirac spectrum~\footnote{these arguments are valid in the presence of substrate induced gap if is magnitude is small} does not allow the bilayer to host interlayer excitons. As a result, the strong coupling regime, $r_\mathrm{s}^{\mathrm{e}(\mathrm{h})}\gg 1$ and $p_\mathrm{F}d/\hbar\sim1$, is not the Bose-Einstein condensate of indirect excitons, but is anticipated to be multiband BCS-like paired state~\cite{EHMultiband1,EHMultiband2,EHMultiband3} where pairing correlations also span to remote bands (empty conduction band in graphene). In reported hybrid graphene-GaAs bilayers~\cite{HybridDrag} the above-listed controlling parameters can be estimated as $r_\mathrm{s}^\mathrm{e}\sim 1.1$ and $r_\mathrm{s}^\mathrm{h}\sim 0.3$ and $p_\mathrm{F}d/\hbar\sim 2.6$. That is why the system is in the weak-to-moderate coupling regime and can therefore be described by the phenomenological weak coupling BCS theory we develop below.

Electrons in the QW can be described by the field operator $e_\vec{r}$, and their kinetic energy is given by
\begin{equation}
\begin{split}
\hat{H}_\mathrm{e}= \int \vec{dr} \; e^\dagger_{\vec{r}} \left(\frac{\hat{\vec{p}}^2-p_\mathrm{F}^2}{2 m_\mathrm{e}} \right)  e_{\vec{r}}.
\end{split}
\end{equation}
The spin degree of freedom is of little importance here and does not need to be treated explicitly. The low-energy electronic states in graphene are concentrated near two inequivalent valleys  ($\hbox{K}$ and $\hbox{K}'$), which are labeled by the index $\zeta =\pm 1$. These states are described by the spinor field operator $h_{\vec{r}}=(h^{\mathrm{A}}_{\vec{r}},h^{\mathrm{B}}_{\vec{r}})$, and their pseudospin corresponds to the sublattice (A and B) degree of freedom of the honeycomb lattice. The kinetic energy of the Dirac states is given by 
\begin{equation}
 H_\mathrm{h}=\int \vec{dr} \; \hat{h}^\dagger_{\vec{r}}\begin{pmatrix} \delta+\epsilon_\mathrm{F}^\mathrm{h} & v( \zeta \hat{p}_x -i \hat{p}_y) \\  v(\zeta \hat{p}_x + i \hat{p}_y)   & -\delta+\epsilon_\mathrm{F}^\mathrm{h} \end{pmatrix}  \hat{h}_{\vec{r}}.
\end{equation}
Here $\epsilon_\mathrm{F}^\mathrm{h}$ is the Fermi energy. A small energy asymmetry between sublattices $\delta\ll \epsilon_\mathrm{F}^\mathrm{h}$ can be induced by substrate engineering. As shown in Fig.~\ref{Fig1} (b), asymmetry opens the gap $2|\delta|$ between the partly filled valence band and the empty conduction band, which has a massive Dirac dispersion $\pm\epsilon_\vec{p}^\mathrm{h}$ with $\epsilon_\vec{p}^\mathrm{h}=(v^2 \vec{p}^2+\delta^2)^{1/2}$. We will refer to the empty states in the valence band as holes; however, it is instructive to not perform the formal particle-hole transformation~\footnote{It should be noted that after the formal electron-hole transformation the connections to the BCS theory of superconductivity established in early works on electron-hole Cooper pair condensation~\cite{LozovikYudson1,LozovikYudson2,Shevchenko} become apparent.}. In the weak-to-moderate coupling regime, electron-hole correlations occur only within the vicinity of the Fermi level, and the empty conduction band can simply be truncated.

The chiral nature of Dirac holes is encoded in their spinor wave function, which is given by 
\begin{equation}|\vec{p}\rangle_\mathrm{h} =\begin{pmatrix} \;\zeta \sin\left(\frac{\theta_\vec{p}}{2}\right) e^{- i \frac{\zeta \phi_\vec{p}}{2}} \\ - \cos\left(\frac{\theta_\vec{p}}{2}\right) \; e^{i \frac{\zeta \phi_\vec{p}}{2} } \;\;\end{pmatrix}.
\end{equation}
Here, $\phi_\vec{p}$ is the polar angle for vector $\vec{p}$ and $\cos(\theta_\vec{p})=\delta/\epsilon_\vec{p}^\mathrm{h}$. It is instructive to introduce the compact notations $c_{\vec{p}}\equiv\cos(\theta_{\bm p}/2)$ and $s_{\vec{p}}=\sin(\theta_{\bm p}/2)$.  The important feature of Dirac holes is their geometrically nontrivial spectrum~\cite{BFReview1}. The geometry is characterized by the Berry connection $\vec{A}^\mathrm{h}_{\vec{p}}=i\vphantom{0}_{\mathrm{h}}\langle \vec{p}|\nabla_\vec{p}|\vec{p}\rangle_\mathrm{h}$, and the Berry curvature $\Omega^\mathrm{h}_{\vec{p}}= [{\bm \nabla_\vec{p}}\times \vec{A}^\mathrm{h}_{ \vec{p}}]_\mathrm{z}$. The latter is valley-dependent and is given by \begin{equation}
\Omega_{\vec{p}}^{\mathrm{h}}=\frac{\zeta v^2 \delta}{(v^2 \vec{p}^2+\delta^2)^{3/2}}, \quad \quad 
\mathcal C^\mathrm{h}=\int\frac{d\vec{p}}{2\pi} \Omega_{\vec{p}}^{\mathrm{h}}=\frac{\zeta \delta}{2 |\delta|}.    
\end{equation}
The low-energy spectrum around each of two valleys can be characterized by the half-integer Chern number
and is therefore topologically nontrivial. When the total Chern number of the two valleys is zero, the large momentum separation between the valleys provides partial topological protection and results in a number of phenomena (e.g., helical states at edges and domain walls, where the asymmetry energy $\delta$ flips its sign) known for topological insulators. 

We further assume that the Fermi momentum of Dirac holes matches the momentum $p_\mathrm{F}$ for electrons in a QW. This is the most favorable regime for electron-hole Cooper pairing driven by an attractive Coulomb interaction. Being effectively screened by charge carriers in both layers, the attractive interactions can be approximated by the contact pseudopotential with momentum-independent Fourier transform $V$. If we neglect intervalley scattering, the interactions can be described by the following Hamiltonian:
\begin{equation}
H_{\mathrm{int}}=\int d\vec{r} \; V e^\dagger_{\vec{r}} \hat{h}^\dagger_{\vec{r}}\cdot  \hat{h}_{\vec{r}}e_{\vec{r}}. 
\end{equation} 
Due to the presence of the pseudospin degree of freedom for Dirac holes, the contact interactions drives the Cooper instability in multiple pairing channels. 

\section{III. Cooper pairing channels}
To analyze possible pairing channels, we assume that the CP condensate is at rest. The condensate can be described by the order parameter $\Delta^\mathrm{v}_\vec{p}=-V\langle h_{\vec{p}\mathrm{v}}^\dagger e_\vec{p}\rangle$, where $h_{\vec{p}\mathrm{v}}$ is the annihilation operator for the valence band states in graphene. Within the mean field theory, the pairing channels can be addressed with the help of the linearized self-consistent equation for the order parameter $\Delta^\mathrm{v}_\vec{p}$, which can be written as 
\begin{equation}
\label{DeltaSC}
\Delta_\vec{p}^\mathrm{v}=V \sum_{\vec{p}'} \Lambda_{\vec{p}\vec{p}'}    \frac{\mathrm{th}\left(\frac{\xi^\mathrm{e}_{\vec{p}'}}{2T}\right)+\mathrm{th}\left(\frac{\xi^\mathrm{h}_{\vec{p}'}}{2T}\right)}{2 (\xi_{\vec{p}'}^\mathrm{e}+\xi_{\vec{p}'}^\mathrm{h})} \Delta_{\vec{p}'}^\mathrm{v}. 
\end{equation}
Here, $\xi_{\vec{p}}^{\mathrm{e}(\mathrm{h})}=v_{\mathrm{e}(\mathrm{h})}(p-p_\mathrm{F})$ is the dispersion of electrons (holes) linearized in the vicinity of the Fermi momentum $p_\mathrm{F}$. The overlap of the spinor wave functions $\Lambda_{\vec{p}\vec{p}'}=\vphantom{0}_{\mathrm{h}}\langle \vec{p}'|\vec{p}\rangle_{\mathrm{h}}$ reflects the chiral nature of Dirac holes and is given by 
\begin{equation}
\label{Lambda}
\Lambda_{\vec{p}\vec{p}'}=s_\vec{p} s_{\vec{p}'}e^{-i  \frac{\zeta\phi_{\vec{p}\vec{p}'}}{2}}+c_\vec{p} c_{\vec{p}'} e^{i \frac{\zeta\phi_{\vec{p}\vec{p}'}}{2}}.   
\end{equation}
Here, $\phi_{\vec{p}\vec{p}'}$ is the angle between momenta $\vec{p}$ and $\vec{p}'$. The presence of the factor $\Lambda_{\vec{p}\vec{p}'}$ reorganizes the Cooper pairing channels and plays a very important role.  If we approximate $c_\vec{p}$ and $s_\vec{p}$ by their values at the Fermi level $c_\mathrm{F}$ and $s_\mathrm{F}$, the order parameter $\Delta^\mathrm{v}$ is momentum-independent 
 and Eq.~(\ref{DeltaSC}) becomes algebraic. As clearly shown, there are two competing channels with orbital quantum numbers $l=\pm 1/2$, and the corresponding dimensionless coupling constants are given by
\begin{equation}
\begin{split}
\lambda_{-\frac{\zeta}{2}}=s_\mathrm{F}^2 V \nu_0= \frac{V \nu_0}{2}\left(1-\frac{\delta}{\epsilon_\mathrm{F}^\mathrm{h}}\right),\\
\lambda_{\frac{\zeta}{2}}=c_\mathrm{F}^2 V \nu_0= \frac{V \nu_0}{2}\left(1+\frac{\delta}{\epsilon_\mathrm{F}^\mathrm{h}}\right).
\end{split}    
\end{equation}
The coupling constant $\lambda_l$ determines the transition temperature in the corresponding channel as
\begin{equation}
\label{CriticalTemperature}
T^0_{l}=\frac{2 \epsilon_0e^\mathrm{C}}{\pi} e^{-\frac{1}{\lambda_l}}.    
\end{equation}
Here, $C=0.577$ is the Euler constant, $\nu_0=p_\mathrm{F}/\pi\hbar^2 (v_\mathrm{e}+v_\mathrm{h})$ is the effective density of states at the Fermi level, and $\epsilon_0=\sqrt{v_\mathrm{e} v_\mathrm{h}} p_0$ is determined by the momentum cutoff $p_0$. In the absence of sublattice asymmetry, $\delta=0$, the coupling constants match each other $\lambda_{1/2}=\lambda_{-1/2}$, and two channels are therefore degenerate (We demonstrate below that in this regime the system is unstable towards the finite momentum Cooper pairing). Due to the exponential dependence of the transition temperature on the coupling constant, the degeneracy is effectively lifted even in the presence of a very small sublattice asymmetry $\delta\ll\epsilon_\mathrm{F}^\mathrm{h}$. 

The presence of two (almost) degenerate competing Cooper pairing channels is a unique feature of this hybrid bilayer~\footnote{The presence of two (almost) degenerate competing channels also occurs for other hybrid double-layer systems, including bilayer graphene-GaAs and monolayer graphene-bilayer graphene} that clearly distinguishes this system from QW-QW and graphene-graphene bilayers.
In the former case, $\Lambda_{\vec{p} \vec{p}'}\rightarrow 1$, and the contact interactions drive Cooper pairing only in the s-wave channel. For the case of a graphene/graphene bilayer with a gapless spectrum, $\delta=0$, the factor must be modified as $\Lambda_{\vec{p} \vec{p}'}\rightarrow \Lambda_{\vec{p} \vec{p}'}^2$~\footnote{In a graphene/graphene bilayer, the angle factor must be modified as $\Lambda_{\vec{p}\vec{p'}}\rightarrow \Lambda_{\vec{p}\vec{p'}}^2$ for intravalley CPs and as $\Lambda_{\vec{p}\vec{p'}}\rightarrow |\Lambda_{\vec{p}\vec{p'}}|^2$ for intervalley CPs. In the absence of sublattice asymmetry, $\delta=0$, the modified angle factor is the same,  $\Lambda_{\vec{p}\vec{p'}}=\cos^2(\phi_{\vec{p},\vec{p}'}/2)$}. As a result, there are three channels $l=0,\pm1$, and the corresponding coupling constants are $\lambda_0=\nu_0 V/2$ and $\lambda_{\pm1}=\nu_0 V/4$. In a similar manner as for the QW-GaAs bilayer, the s-wave channel is the dominant Cooper pairing channel. 

The origin of two competing channels can be tracked to the presence of the pseudospin degree of freedom for Dirac holes. In the sublattice basis, the order parameter  $\hat{\Delta}_{\vec{p}}=\{\Delta_{\vec{p}}^\mathrm{A}, \Delta_{\vec{p}}^\mathrm{B}\}$ has two components: ($\Delta^\mathrm{A}_\vec{p}=-V\langle h_{\vec{p}}^{\mathrm{A} \dagger} e_\vec{p}\rangle$ and $\Delta^\mathrm{B}_\vec{p}=-V\langle h_{\vec{p}}^{\mathrm{B}\dagger} e_\vec{p}\rangle$). These components describe selective Cooper pairing with correlations only at one of two sublattices. These components are related to the order parameter $\Delta_\vec{p}^\mathrm{v}$ introduced above, as follows:
\begin{equation}
\Delta_\vec{p}^\mathrm{v}=s_\vec{p} \Delta^\mathrm{A}_\vec{p} e^{-i\frac{\zeta \phi_\vec{p}}{2}}+c_\vec{p} \Delta^\mathrm{B}_\vec{p} e^{i \frac{\zeta \phi_\vec{p}}{2}}.
\end{equation}
The order parameter in the sublattice picture has the only s-wave component and is therefore momentum-independent $\hat{\Delta}$. As a result, there is a one-to-one correspondence between channels in the band basis and components of the order parameter in the sublattice basis. The relation between them is valley-dependent and can be presented as 
\begin{equation}
\label{ChannelConnections}
\Delta_{-\zeta/2}^{\mathrm{v}}=s_\mathrm{F}\Delta^{\mathrm{A}}, \quad \quad \Delta_{\zeta/2}^{\mathrm{v}}=c_\mathrm{F}\Delta^{\mathrm{B}}.
\end{equation}
The basis choice is a matter of convenience, and in the remainder of the paper, we will deal only with the order parameter in the sublattice basis. 

Thus far, we have considered only the instability toward Cooper pairing with zero momentum, $\vec{q}=0$. As we demonstrate below, the chiral nature of Dirac electrons results in intertwining between $\Delta^\mathrm{A}$ and $\Delta^\mathrm{B}$ at finite $\vec{q}$. 

\section{IV. Ginzburg-Landau Hamiltonian}
Above the transition temperature, the order parameter vanishes, but Gaussian fluctuations remain, which are usually interpreted as fluctuating CPs. These fluctuations are described by the dynamical bosonic field   $\hat{\Delta}_{t\vec{r}}=\{\Delta_{t\vec{r}}^\mathrm{A}, \Delta_{t\vec{r}}^\mathrm{B} \}$, and their energetics are determined by the GL functional $F_\mathrm{GL}[\Delta]$, which is derived in Appendix A. After a proper rescaling~\footnote{It is instructive to rescale the fields as $\Delta^\mathrm{A}\rightarrow \Delta^\mathrm{A}/\sqrt{\nu_0} s_\mathrm{F}$ and $\Delta^\mathrm{B}\rightarrow \Delta^\mathrm{B}/\sqrt{\nu_0} c_\mathrm{F}$. See Appendix A for detailed derivations.}, $F_\mathrm{GL}[\Delta]$ can be presented as 
\begin{equation}
F_\mathrm{GL}=\int \vec{d r} \; \Delta^\dagger_{t\vec{r}} \hat{H}_\mathrm{GL} \Delta_{t\vec{r}}.    
\end{equation}
The dimensionless Hermitian matrix $\hat{H}_\mathrm{GL}$ is the GL Hamiltonian. This matrix describes two bosonic modes ($\Delta_{t\vec{r}}^\mathrm{A}$ and $\Delta_{t\vec{r}}^\mathrm{B}$) intertwined by effective pseudospin-orbit interactions as 
\begin{equation}  
\hat{H}_\mathrm{GL}=\begin{pmatrix} \varepsilon_\mathrm{A}  + \xi^2\vec{q}^2 & - \xi_\star (\zeta q_x-iq_y) \\ - \xi_\star (\zeta q_x+iq_y)& \varepsilon_\mathrm{B}  + \xi^2\vec{q}^2
\end{pmatrix}.
\end{equation}
Each mode has a temperature-dependent energy gap $\varepsilon_{\mathrm{A}(\mathrm{B})}=\ln [T/T^0_{\mathrm{A}(\mathrm{B})}]$. The gap vanishes at the transition temperature $T^0_{\mathrm{A}(\mathrm{B})}$ given by Eq.~(\ref{CriticalTemperature}) (combined with the channel labeling rule, Eq.~(\ref{ChannelConnections})), which signals Cooper instability in the corresponding channel. The largest of them $T_0=\max[T^0_{\mathrm{A}},T^0_{\mathrm{B}}]$ determines the critical temperature of electron-hole Cooper pairing. The lengths
\begin{equation} \xi= \sqrt{\frac{7\zeta(3)}{2}}\frac{\hbar v_\mathrm{e} v_\mathrm{h}}{2\pi (v_\mathrm{e}+v_\mathrm{h}) T}, \; \quad \xi_{\star}=\frac{\hbar v_\mathrm{e}}{4\lambda (v_\mathrm{e}+v_\mathrm{h}) p_\mathrm{F}}
\end{equation} 
scale the quadratic kinetic energy of the modes and their pseudospin-orbit interactions, respectively. Here, $\lambda=\mathrm{max}[\lambda_\mathrm{A}, \lambda_\mathrm{B}]$ is the coupling constant in the leading channel. The ratio of scales $\xi_\star/\xi\sim  T_0/\lambda v_\mathrm{h} p_\mathrm{F}$ is small; thus, the contribution of the effective pseudospin-orbit interactions to the kinetic energy is small compared with the quadratic term. 

The effective pseudospin-orbit interactions mix and intertwine components of the order parameter $\Delta_{\vec{q}}^\mathrm{A}$ and $\Delta_{\vec{q}}^\mathrm{B}$. The spectrum of the GL Hamiltonian $H_{\mathrm{GL}}$ has two eigenmodes ($\gamma=\pm 1$) and is given by
\begin{equation}
\label{CPEnergy}
\varepsilon_{\gamma\vec{q}}=\varepsilon_\mathrm{s}+\xi^2 \vec{q}^2+\gamma d_\vec{q},\quad \quad  d_\vec{q}= \sqrt{\varepsilon^2_\mathrm{z}+\xi_\star^2 \vec{q}^2}.
\end{equation}
Here, $\varepsilon_{\mathrm{s}} =(\varepsilon_{\mathrm{A}}+\varepsilon_{\mathrm{B}})/2$ is the average energy of the modes, and $\varepsilon_{\mathrm{z}} =(\varepsilon_{\mathrm{A}}-\varepsilon_{\mathrm{B}})/2$ can be interpreted as the effective Zeeman term in the GL Hamiltonian $H_\mathrm{GL}$. These terms are given by 
\begin{equation}
\varepsilon_{\mathrm{s}} =\ln\left(\frac{T}{\sqrt{T^0_{\mathrm{A}} T^0_{\mathrm{B}}}}\right), \quad \quad \varepsilon_{\mathrm{z}}=\frac{1}{2} \ln\left(\frac{T^0_\mathrm{B}}{T^0_{\mathrm{A}}}\right).
\end{equation}
The average energy $\varepsilon_\mathrm{s}$ smoothly decreases with temperature, but the effective Zeeman term $\varepsilon_z$ is temperature-independent. Its sign and magnitude 
\begin{equation}
\varepsilon_z=\frac{\lambda_\mathrm{B}-\lambda_\mathrm{A}}{2 \lambda_\mathrm{A} \lambda_\mathrm{B}}\approx \frac{1}{\sqrt{\lambda_{\mathrm{A}} \lambda_{\mathrm{B}}}} \frac{\delta}{\epsilon_\mathrm{F}^\mathrm{h}} 
\end{equation}
are given by the sublattice asymmetry $\delta$ and vanish if $\delta=0$.

\begin{figure}[t]
	\begin{center}
		\includegraphics[trim=0cm 0cm 0cm 0.2cm, clip, width=1.0\columnwidth]{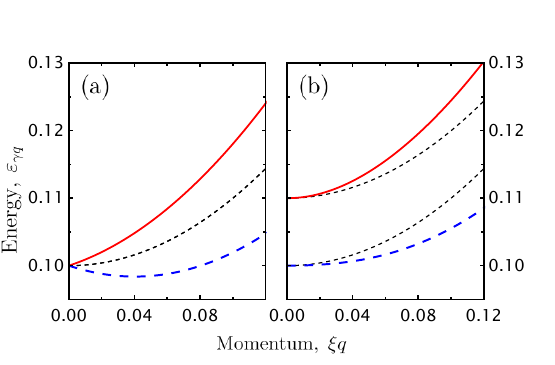}
		\caption{Dispersion relation for fluctuating CPs at $\epsilon_\mathrm{z}=0$ (a) and $\epsilon_\mathrm{z}=0.1$ (b) for $\xi_\star/\xi=0.08$. The dashed black lines correspond to the spectrum without pseudospin-orbit interactions, $\xi_\star/\xi=0$. For a small asymmetry $\epsilon_\mathrm{z}<  \varepsilon_{\mathrm{M}}$ between sublattices, the effective pseudospin-orbit interactions reshape the lower parabolic eigenmode into a Mexican-hat-shaped curve. As a result, the system is unstable toward the Cooper pairing state with finite momentum (Fulde-Ferrell-Larkin-Ovchinnikov state).}
		\label{Fig2}
	\end{center}
\end{figure}

If the effective Zeeman term is smaller $\epsilon_\mathrm{z}<\epsilon_{\mathrm{M}}$ than $\epsilon_{\mathrm{M}}=\xi_\star^2/2\xi^2$, the effective spin-orbit interactions reshape the parabolic lower eigenmode $\varepsilon_{-,\vec{q}}$ into a Mexican-hat curve, as presented in Fig.~\ref{Fig2}-a. A minimum occurs at finite momentum $q_\mathrm{M}$ with an energy gain of $\Delta\varepsilon_\mathrm{M}$ with respect to $\varepsilon_{-,\vec{q}}$ at $\vec{q}=0$, as follows: 
\begin{equation*}
\xi q_\mathrm{M}=\frac{\xi_\star}{\xi}\frac{\sqrt{\varepsilon_\mathrm{M}^2-\varepsilon_z^2}}{2 \varepsilon_\mathrm{M}}, \quad\quad \Delta\varepsilon_{\mathrm{M}}=-\frac{\left(\varepsilon_\mathrm{M}-|\varepsilon_{\mathrm{z}}|\right)^2}{2 \varepsilon_{\mathrm{M}}}.
\end{equation*}
As a result, the electron-hole bilayer is unstable toward Cooper pairing with finite momentum (Fulde-Ferrell state~\cite{FF}) or a non-uniform state (Larkin-Ovchinnikov state~\cite{LO}). The nature of the condensed state can be addressed within the GL theory (with quartic terms in the free energy functional $F_{\mathrm{GL}}$), but such an analysis is beyond the scope of the present paper, which focuses on fluctuating CPs. It should be noted that the FFLO phases are stabilized solely by the pseudospin-orbit interactions and do not require the electron-hole density imbalance~\cite{LOFFSeradjeh,EfimkinLOFF}. Since contribution of effective pseudospin-orbit interactions to the kinetic energy is small compared with the usual quadratic term, this regime $\epsilon_\mathrm{z}<\epsilon_{\mathrm{M}}$ is realized only at vanishingly small energy asymmetry $\delta$ between graphene sublattices.   

If the effective Zeeman term is sufficiently large ( $\epsilon_\mathrm{z}>\epsilon_{\mathrm{M}}$) the effective pseudospin-orbit interactions impact the slope of the dispersion curves $\varepsilon_{\gamma\vec{q}}$ presented in Fig.~\ref{Fig2}-b, but do not reshape them.

The intertwining between $\Delta_{t\vec{r}}^\mathrm{A}$ and $\Delta_{t\vec{r}}^\mathrm{B}$ is manifested in the nontrivial geometry and topology of the $H_\mathrm{GL}$ spectrum. The nontrivial geometry is intricately related to the momentum space texture for the unit vector $\vec{n}_\vec{q}=\vec{h}_\vec{q}/|\vec{h}_\vec{q}|$ defined within the parametrization of the GL Hamiltonian  $H_\mathrm{GL}(\vec{q})=\hat{1}\cdot h^0_\vec{q}  + \hat{{\bm \sigma}}_\mathrm{P}\cdot \vec{h}_\vec{q}$ in terms of Pauli matrices $\hat{{\bm \sigma}}_\mathrm{P}$. The vector $\vec{n}_\vec{q}$ follows the topological momentum space meron texture. Depending on the sign of $\epsilon_\mathrm{z}$, the vector points up or down at the momentum origin $\vec{q}=0$. For a large momentum $\vec{q}$, the vector $\vec{n}_\vec{q}$ lies in the horizontal plane and follows the vortex-like texture.  Being projected to the Riemann sphere, the vector field spans the half that dictates the topological Chern number to be a half-integer $|\mathcal C^\mathrm{C}_\gamma|=1/2$.

In formal mathematics, the nontrivial geometry of the $H_\mathrm{GL}$ spectrum can be characterized by the generalized Berry connection $\hat{\vec{\mathcal A}}_\vec{q}$, which also includes the off-diagonal matrix elements, as follows:
\begin{equation}
\label{BerryConnectionCP}
\vec{\mathcal A}_\vec{q}^{\gamma\gamma'}=\vphantom{0}_{\mathrm{C}}\langle \gamma \vec{q}|i\nabla_\vec{q}|\gamma'\vec{q}\rangle_\mathrm{C},\quad \quad T_{\gamma \vec{q}}^{\alpha \beta}=\vec{\mathcal A}_{\vec{q}\alpha}^{\gamma\bar{\gamma}} \vec{\mathcal A}_{\vec{q}\beta}^{\bar{\gamma}\gamma}.     
\end{equation}
Here, we have also introduced the quantum geometric tensor $\hat{T}_{\gamma \vec{q}}$ for fluctuating CPs with $\alpha$ and $\beta$ as Cartesian indices ($x$ and $y$)~\cite{QFT1,QFT2,QFT3}. The real part of the tensor defines the quantum metric $\hat{G}_{\gamma \vec{q}}=\mathrm{Re}[\hat{T}_{\gamma \vec{q}}]$, which allows one to measure distances between eigenstates $|\gamma\vec{q}\rangle_\mathrm{C}$ in momentum space. The quantum metric for fluctuating CPs is given by 
\begin{equation}
\label{Quantummetric}
\begin{split}
\hat{G}_{\gamma \vec{q}}= 
\frac{1}{8 d_\vec{q}^4} \big[  \xi_\star^4 \vec{q}^2 [\sigma_\mathrm{P}^\mathrm{z} \cos(2\phi_\vec{q}) +  \sigma_\mathrm{P}^\mathrm{x} \sin(2\phi_\vec{q})] \\
-\xi_\star^2 (d_\vec{q}^2+\varepsilon_\mathrm{z}^2) \big].
\end{split}
\end{equation}
The second term is independent of the direction of momentum $\vec{q}$, and the first term has quadrupole symmetry. The imaginary part of the quantum geometric tensor for fluctuating CPs defines their nontrivial Berry curvature $\Omega^{\mathrm{C}}_{\gamma\vec{q}}=-2 \epsilon_{\alpha \beta z} \mathrm{Im}[\hat{T}_{\gamma \vec{q}}^{\alpha \beta}]$, which matches with the another definition $\Omega^\mathrm{C}_{\gamma\vec{q}}= [{\bm \nabla_\vec{q}}\times \vec{\mathcal{A}}^{\gamma\gamma}_{\gamma \vec{q}}]_\mathrm{z}$. The Berry curvature has the opposite sign for two eigenmodes, is valley-dependent, and is given by
\begin{equation}
\Omega^{\mathrm{C}}_{\gamma\vec{q}}=\frac{\zeta \gamma \epsilon_\mathrm{z}}{d_\vec{q}^{\frac{3}{2}}}, \quad \quad 
\mathcal C^\mathrm{C}_\gamma=\int\frac{d\vec{q}}{2\pi} \Omega_{\gamma\vec{q}}^{\mathrm{C}}=\frac{\zeta\gamma \varepsilon_{z}}{2|\varepsilon_z|}.
\end{equation}
Being inherited from the Dirac holes, the band geometry is also characterized by the half-integer Chern number $\mathcal C^\mathrm{C}_\gamma$. Therefore, we interpret the hybrid electron-hole fluctuating CPs to be topologically nontrivial.

%It should be mentioned that the sum of Chern numbers for fluctuating CPs involving holes from different valleys is again zero. 

%\begin{equation}
%\begin{split}
%\label{CPSpinor}
%|+,\vec{q}\rangle_\mathrm{C} =\begin{pmatrix}  - \alpha %\cos\left(\frac{\vartheta_\vec{q}}{2}\right) \\  %\sin\left(\frac{\theta_\vec{q}}{2}\right) e^{i \alpha %\phi_\vec{q}}\end{pmatrix}, \quad |-,\vec{q}\rangle_\mathrm{C} %=\begin{pmatrix} \sin\left(\frac{\vartheta_\vec{q}}{2}\right) %e^{- i \alpha \phi_\vec{q}} \\  %\cos\left(\frac{\vartheta_\vec{q}}{2}\right)\end{pmatrix}. 
%\end{split}
%\end{equation}
%Here,  $\phi_\vec{q}$ is the polar angle for vector $\vec{q}$  and $\cos\left(\vartheta_\vec{q}\right)=\epsilon_\mathrm{z}/\sqrt{\epsilon^2_\mathrm{z}+\xi_\star^2 \vec{q}^2}$. The intertwining of two modes with the valley-dependent phase factor $e^{- i z \phi_\vec{q}}$ results in the nontrivial geometry of fluctuating CPs. The latter can be illustrated in different ways.

The nontrivial geometries of fluctuating CPs closely follow the geometry for Dirac holes in graphene. However, there are two major differences between Dirac holes and fluctuating CPs. First, the pair of massive Dirac bands is separated by the global gap (i.e., present for all momenta), which is not the case for fluctuating CP eigenmodes. Besides the existence of bands with nontrivial Chern numbers, the stability of the edge modes at boundaries and domain walls also requires a global energy gap (in the Discussions section, we discuss approaches to overcome this obstacle). However, the absence of this gap is not an obstacle for anomalous transport phenomena, which rely on the nonzero Berry curvature of eigenmodes and will be considered in the following sections. Second, fluctuating CPs are not bosonic quasiparticles (with Hermitian Hamiltonian dynamics) but overdamped bosonic modes. For this reason, the corresponding anomalous transport phenomena are outside the range of present theories~\cite{AHEReview1,AHE2,AHE3}.

\section{V. Kinetic equation}
The dynamics of fluctuating CPs are governed by the time-dependent GL (TDGL) equation (see Ref.~\cite{LarkinVarlamov} and Appendix C). Components of fluctuating CPs have opposite charges and are spatially separated. For these reasons, coupling with electric potentials in both layers ($\phi^{\mathrm{e}}_{t \vec{r}}$ and $\phi^{\mathrm{h}}_{t \vec{r}})$ can be introduced via the Peierls substitution  $\partial_t\rightarrow \partial_t+i e \phi^{\mathrm{eh}}_{t \vec{r}}$ with  $\phi^{\mathrm{eh}}_{t \vec{r}}=\phi^{\mathrm{e}}_{t \vec{r}}-\phi^{\mathrm{h}}_{t \vec{r}}$. As a result, the TDGL equation can be presented as follows:   
\begin{equation}
\label{TDGL0}
\begin{split}
\tau^* \left(\partial_t+i e \phi^{\mathrm{eh}}_{t \vec{r}}\right)\Delta=-\hat{H}_\mathrm{GL}\Delta+\eta_{t\vec{r}}, \\ \langle \eta_{t\vec{r}} \eta^\dagger_{t'\vec{r}'} \rangle= 2 T \tau' \; \hat{1}\; \delta_{tt'}\delta_{r\vec{r}'}.
\end{split}
\end{equation}
Here, $\tau=\tau'+i \tau''$, where $\tau'=\pi\hbar /8T$ gives the dissipation rate of fluctuating CPs and $\tau''=\hbar/ \lambda (v_\mathrm{e}+v_\mathrm{h}) p_\mathrm{F}$ weights the Hermitian part of their dynamics. 
The external complex field $\eta_{t\vec{r}}=\{\eta_{t\vec{r}}^\mathrm{A},\eta_{t\vec{r}}^\mathrm{B}\}$ is the Langevin noise. Its presence is dictated by the fluctuation-dissipation theorem, and the correlation function $\langle \eta_{t\vec{r}} \eta^\dagger_{t'\vec{r}'} \rangle$, which is free of temporal and spatial correlations (white noise), is proportional to the relaxation rate $\tau'$ but does not depend on $\tau''$. In the considered weak-to-moderate Cooper pairing regime, their ratio is small $\tau''/\tau'\sim T/\lambda (v_\mathrm{e}+v_\mathrm{h}) p_\mathrm{F}$; hence, the dissipation of fluctuating CPs is an essential component in their dynamics. These CPs cannot be interpreted as bosonic quasiparticles, but are instead overdamped bosonic modes.  

The contribution of fluctuating CPs in transport phenomena can be addressed by applying the linear response approach to the TDGL equation. These calculations are presented in Appendix B. In the main part of this paper, we follow another route. As it is presented in Appendix C, the TDGL equation, Eq.~(\ref{TDGL0}), can be transformed into a kinetic equation for the distribution function of fluctuating CPs $n_{\gamma}(\vec{R},\vec{q})$, which is given by:
\begin{equation}
\label{DensityMatrixDiagonal2}
\begin{split}
    \partial_t n_{\gamma}+ e\vec{E}^{\mathrm{eh}}\partial_\vec{q}n_\vec{\gamma}+\frac{\tau''}{|\tau|^2} \Big(\partial_\vec{q} \varepsilon_{\gamma \vec{q}}+ \vec{u}_{\gamma \vec{q}}^\Omega  + \\ \vec{u}_{\gamma \vec{q}}^\mathrm{G} \Big)  \partial_\vec{R}  n_\gamma =-\frac{2\varepsilon_{\gamma\vec{q}} \tau'}{|\tau|^2} (n_\gamma-n_\gamma^0).
    \end{split}
\end{equation}
 Here $n^0_{\gamma \vec{q}}=T/\varepsilon_{\gamma\vec{q}}$ is the equilibrium classical distribution function for fluctuating CPs. We have introduced two distinct anomalous velocities $\vec{u}_{\gamma \vec{q}}^\Omega$ and $\vec{u}_{\gamma \vec{q}}^{\mathrm{G}}$, which are generated by the nonzero Berry curvature $\Omega_{\gamma \vec{q}}$ and quantum metric $\hat{G}_{\gamma \vec{q}}$, respectively. Their explicit expressions are given by 
\begin{equation}
\label{AnomalousVelocities1}
\begin{split}
\vec{u}_{\gamma \vec{q}}^\Omega= \Omega_{\gamma \vec{q}}[ \vec{e}_\mathrm{z}\times e\vec{E}^{\mathrm{eh}}]  \frac{\tau'' (\varepsilon _{\gamma\vec{q}}-\varepsilon_{\bar{\gamma}\vec{q}})^2 |\tau|^2}{|\varepsilon_{\gamma\vec{q}}\tau^*+\varepsilon_{\bar{\gamma}\vec{q}}\tau|^2},  \\   \vec{u}_{\gamma \vec{q}}^{\mathrm{G}} = \hat{G}_{\gamma \vec{q}}\cdot e\vec{E}^{\mathrm{eh}}  \frac{ \tau'(\varepsilon_{\gamma \vec{q}}^2-\varepsilon_{\bar{\gamma}\vec{q}}^2)|\tau|^2}{|\varepsilon_{\gamma\vec{q}}\tau^*+\varepsilon_{\bar{\gamma}\vec{q}}\tau|^2}. 
\end{split}
\end{equation}
The nontrivial Berry curvature is known to generate an anomalous velocity in fermionic and bosonic systems~\cite{BFReview1}. The presence of an additional anomalous velocity $\vec{u}_{\gamma \vec{q}}^{\mathrm{G}}$ is a unique feature of fluctuating CP dynamics and is intricately related to their dissipative nature. If we omit the dissipation and rescale the time ($\tau'=0$ and $t\rightarrow \tau'' t$), the TDGL equation reduces to a Schrodinger-like  equation $i\partial_t \Delta=\hat{H}_\mathrm{GL}\Delta$. At this limit, the additional term $\vec{u}_{\gamma \vec{q}}^{\mathrm{G}} = 0$ vanishes, and the conventional expression $\vec{u}_{\gamma \vec{q}}^\Omega= \Omega_{\gamma \vec{q}}[ \vec{e}_\mathrm{z}\times e\vec{E}^{\mathrm{eh}}]$ for the anomalous velocity is recovered. 

The derived kinetic equation for fluctuating CPs, Eq.~(\ref{DensityMatrixDiagonal2}), is the main result of this section and is a key to the anomalous transport phenomena mediated by fluctuating CPs. In Appendix C, we demonstrate that resulting predictions agree with ones derived by applying the linear response approach to the TDGL equation.

\section{VI. Paraconductivity}
\begin{figure}[t]
	\begin{center}
    	\includegraphics[trim=0cm 5cm 0.5cm 0.8cm, clip, width=0.9\columnwidth]{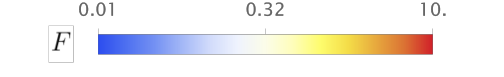}
		\includegraphics[trim=0cm 0cm 0.5cm 0cm, clip, width=0.95\columnwidth]{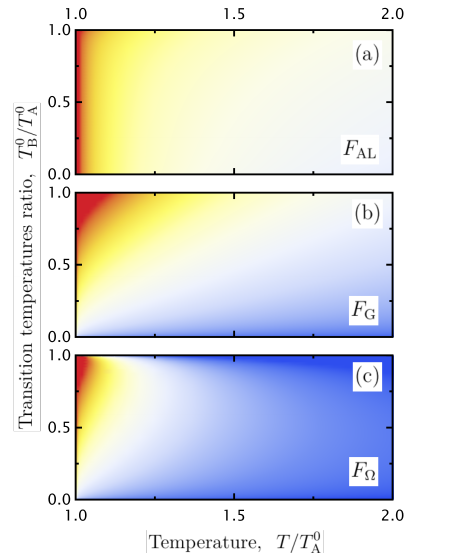}
		\caption{Dependence of the $F$-functions introduced in Eq.~(\ref{ConductivityTemperature2}) on temperature $T/T^0_\mathrm{A}$ and the ratio between transition temperatures $T^0_\mathrm{B}/T^0_\mathrm{A}$ for two competing pairing channels. Both conventional and anomalous contributions to the conductivity tensor exhibit critical behavior and diverge at the transition temperature.}
		\label{Fig3}
	\end{center}
\end{figure}

Due to the spatial separation of electrons and holes forming fluctuating CPs, their motion induces electric currents in both layers. These currents have opposite directions but the same magnitude (as well as the same conductivity and transconductivity, which is defined as the response of an electric current in one layer to an electric field in another layer). The electric field difference $\vec{E}^{\mathrm{eh}}$ shifts the distribution function $n_{\gamma \vec{q}}^0\rightarrow n_{\gamma \vec{q}}^0+n_{\gamma \vec{q}}^1$ and induces electric currents in both layers as
\begin{equation}
\label{CurrentContributions1}
\vec{J}^{\mathrm{e(h)}}=\pm \sum_{\vec{p}\gamma}\partial_\vec{q} \varepsilon_{\gamma \vec{q}} n_{\gamma \vec{q}}^1,  \quad  n_{\gamma\vec{q}}^1= -\frac{|\tau|^2 \vec{E}^{\mathrm{eh}} \partial_\vec{q} n_{\gamma\vec{q}}^0}{2 \tau' \varepsilon_{\gamma \vec{q}}}.
\end{equation}
The electric currents are parallel to $\vec{E}^{\mathrm{eh}}$, and the corresponding contribution to the longitudinal conductivity represents the sum of \emph{conventional} Aslamazov-Larkin paraconductivities for two pairing channels. Besides, the electric field difference $ \vec{E}^{\mathrm{eh}}$ induces an additional contribution via the anomalous velocities as
\begin{equation}
\label{CurrentContributions2}
\vec{J}^{\mathrm{e(h)}}=\pm \sum_{\gamma\vec{q}}\left(u_{\gamma \vec{q}}^{\mathrm{\Omega}}+u_{\gamma \vec{q}}^\mathrm{G}\right) n_{\gamma\vec{q}}^0.
\end{equation}
These two terms can be interpreted as \emph{anomalous} Aslamazov-Larkin paraconductivities. The velocity $u_{\gamma \vec{q}}^{\mathrm{\Omega}}$ is perpendicular to the electric field difference $\vec{E}^{\mathrm{eh}}$ and is responsible for the anomalous contribution to the Hall conductivity. The velocity $u_{\gamma \vec{q}}^{\mathrm{G}}$ has both components, which are perpendicular/parallel to the electric field difference $\vec{E}^{\mathrm{eh}}$. However, the off-diagonal elements of the quantum metric $\hat{G}_{\gamma \vec{q}}$ given by Eq.~(\ref{Quantummetric}) have quadrupole symmetry and vanish over angle integration. As a result, only the second term in Eq.~(\ref{Quantummetric}) survives, and the quantum-metric-induced anomalous velocity $u_{\gamma\vec{q}}^{\mathrm{G}}$ provides an additional contribution only to the longitudinal conductivity. An explicit evaluation of the different contributions to the conductivity is presented in Appendix B. To the leading order in $\xi_\star/\xi$ and $\tau''/\tau'$ (it should be noted that within this approximation the effect of the effective pseudospin-orbit interactions at the dispersion of fluctuating CPs is neglected), the expressions ($\sigma_{\mathrm{L}}\equiv\sigma_{xx}$ and $\sigma_{\mathrm{H}}\equiv\sigma_{xy}$) are given by 
\begin{equation}
\label{ConductivityTemperature1}
\begin{split}
\sigma_{\mathrm{L}}^{\mathrm{AL}}=\frac{e^2}{2\pi \hbar} F_{\mathrm{AL}}, \;\;  \sigma_{\mathrm{L}}^{\mathrm{G}}=\frac{e^2}{2\pi \hbar}\left(\frac{\xi_\star}{\xi}\right)^2 F_\mathrm{G}, \\  \sigma_{\mathrm{H}}^{\mathrm{\Omega}}=\zeta\frac{e^2}{2\pi \hbar}  \left(\frac{\xi_\star}{\xi}\right)^2 \frac{\tau''}{\tau'}\; F_{\mathrm{\Omega}}.
\end{split}
\end{equation}
The temperature behaviors of these terms are governed by dimensionless functions $F\equiv F(\varepsilon_\mathrm{A},\varepsilon_\mathrm{B})$. The latter terms depend only on $\varepsilon_\mathrm{A}=\ln[T/T^0_\mathrm{A}]$ and $\varepsilon_\mathrm{B}=\ln[T/T^0_\mathrm{B}]$ and are given by
\begin{equation}
\label{ConductivityTemperature2}
\begin{split}
F_{\mathrm{AL}}=\frac{\pi}{32}\left(\frac{1}{\varepsilon_{\mathrm{A}}}+\frac{1}{\varepsilon_{\mathrm{B}}} \right), \\
F_\mathrm{G}=\frac{\pi}{8(\varepsilon_\mathrm{A}-\varepsilon_\mathrm{B})^2} \ln \left[\frac{(\varepsilon_\mathrm{A}+\varepsilon_\mathrm{B})^2}{4\varepsilon_\mathrm{A}\varepsilon_\mathrm{B}}\right],\\
F_\mathrm{\Omega}=\frac{\pi}{8(\varepsilon_\mathrm{A}-\varepsilon_\mathrm{B})}\left\{  \frac{\ln \left[\varepsilon_\mathrm{A}/\varepsilon_\mathrm{B}\right]}{\varepsilon_\mathrm{A}-\varepsilon_\mathrm{B}} -\frac{2}{\varepsilon_\mathrm{A}+\varepsilon_\mathrm{B}}\right\}.
\end{split}
\end{equation}
Without loss of generality, we can assume that $T^0_\mathrm{A}>T^0_\mathrm{B}$. It is instructive to introduce the dimensionless temperature $T/T^0_\mathrm{A}$ and the relation of transition temperatures for two channels $T^0_\mathrm{B}/T^0_\mathrm{A}$. The corresponding dependence for the $F$-functions is presented in Fig.~\ref{Fig3}.  

The factor $F_{\mathrm{AL}}$ represents a sum of two independent terms describing the conventional Aslamazov-Larkin effect for each of two competing channels. This factor reaches a maximum when two channels are degenerate ($T^0_{\mathrm{A}}=T^0_{\mathrm{B}}$), and the corresponding result is two times larger than that for the case in which only one channel is present (e.g., $T^0_{\mathrm{B}}/T^0_{\mathrm{A}}\rightarrow 0$). In the vicinity of the transition temperature $T_0$, the factor  $F_{\mathrm{AL}}\sim (T-T_0)^{-1}$ exhibits critical behavior, which, for the case of degeneracy, can be presented as 
\begin{equation}
\label{CriticalAL}
F_{\mathrm{AL}}=\frac{\pi}{16}\frac{1}{\ln\left[T/T_0 \right]}\rightarrow \frac{\pi}{16}\frac{T_0}{T-T_0},  
\end{equation}  
and diverges at the transition temperature.

The term $F_{\mathrm{G}}$ relies on the channel intertwining and also reaches a maximum when the two channels are degenerate. In this case, the expression for $F_{\mathrm{G}}$ has a stronger divergence 
\begin{equation}
\label{CriticalG}
F_\mathrm{G}=\frac{\pi}{32}\frac{1}{\ln^2\left[T/T_0\right]}\rightarrow \frac{\pi}{32} \left(\frac{T_0}{T-T_0}\right)^2    
\end{equation}
than the value $F_\mathrm{AL}$ given by Eq.~(\ref{CriticalAL}). As a result, $F_{\mathrm{G}}$ exceeds $F_{\mathrm{AL}}$ in the broad parameter range around the degeneracy point. However, the corresponding term $\sigma_{\mathrm{L}}^{\mathrm{G}}$ in Eq.~(\ref{ConductivityTemperature1}) has a small prefactor $(\xi_\star/\xi)^2$, which causes the anomalous contribution $\sigma_{L}^{\mathrm{G}}$ to be small compared with the conventional contribution $\sigma_{L}^{\mathrm{AL}}$. 

The factor  $F_{\mathrm{\Omega}}$ also relies on channel intertwining, but vanishes when the two channels are degenerate, $F_{\mathrm{\Omega}}=0$. The Berry curvatures for fluctuating CPs have opposite signs for the two modes, and their contributions vanish when the two modes have the same thermal population. As a result, the factor $F_{\mathrm{\Omega}}$ reaches a maximum for finite but small splitting between transition temperatures for two channels.

\section{VII. Transport phenomena}
The interplay of the conventional Aslamazov-Larkin effect and the effect of Coulomb drag due to momentum transfer between layers has already been discussed in detail. Here, we discuss possible experimental setups that can be used to observe the anomalous Aslamazov-Larkin contributions to the conductivity (and transconductivity) tensor mediated by fluctuating CPs with nontrivial geometries.

First, we recall that the anomalous contribution $\sigma_{\mathrm{L}}^{\mathrm{G}}$, which originates from the quantum geometric tensor for fluctuating CPs, is small compared with $\sigma_{\mathrm{L}}^{\mathrm{AL}}$; thus, this contribution is difficult to observe unless $\xi_\star$ is comparable to $\xi$. For this reason, we focus on the transverse anomalous contribution $\sigma_{\mathrm{H}}^{\mathrm{\Omega}}$ that arises from the nonzero Berry curvature for fluctuating CPs. The sign of the contribution is valley-dependent, and no current appears~\footnote{It should be noted that a valley current is induced and can be detected via nonlocal measurements~\cite{VelleyCurrent1,VelleyCurrent2}} until the symmetry between valleys is broken. In this section, we consider a regime of ultimate asymmetry, in which holes from only one of two valleys participate in fluctuating CPs. We discuss possible approaches to achieve this regime in the Discussions section.

\begin{figure}[t]
	\begin{center}
		\includegraphics[trim=2cm 14cm 17cm 3cm, clip, width=1.0\columnwidth]{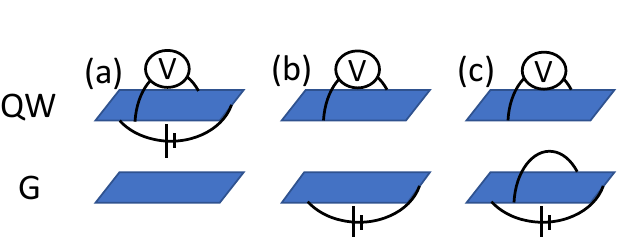}
		\caption{Possible experimental setups with different connections to an  external circuit. As discussed in the main text, setups (a) and (c) directly probe the transverse anomalous Aslamazov-Larkin paraconductivity mediated by fluctuating CPs, but this is not the case for setup (b). }
		\label{Fig4}
	\end{center}
\end{figure}

Second, we note the presence of single-particle contributions to the Hall conductivity of graphene ($\sigma^\mathrm{h}_\mathrm{xy}=\zeta e^2\delta/2\pi \hbar |\epsilon_\mathrm{F}^\mathrm{h}|$ per valley in the absence of valley asymmetry). This contribution can easily exceed the small contribution mediated by fluctuating CPs. For this reason, we consider only experimental setups with a voltage probe in a semiconductor QW layer, where electrons are free of nontrivial geometries. The three possible setups are sketched in Fig.~\ref{Fig4}.

The transport phenomena in an electron-hole bilayer can be described by utilizing a generalized conductivity matrix $\hat{\sigma}$, which connects currents $\hat{J}=\hat{\sigma} \hat{E}$ in two layers $J=\{J_x^\mathrm{e},J_y^\mathrm{e},J_y^\mathrm{h},J_y^\mathrm{h}\}$ with the corresponding electric fields $E=\{E_x^\mathrm{e},E_y^\mathrm{e},E_y^\mathrm{h},E_y^\mathrm{h}\}$. If we recall that fluctuating CPs are excited by an electric field difference $\vec{E}^\mathrm{eh}$ and their motion generates opposite currents in two layers, the conductivity matrix can be presented as
\begin{equation*}
\hat{\sigma}=\begin{pmatrix} \sigma_\mathrm{L}^\mathrm{e} + \sigma_\mathrm{L}^\mathrm{C}  & - \sigma_\mathrm{H}^\mathrm{C} & - \sigma_\mathrm{D} - \sigma_\mathrm{L}^\mathrm{C} & \sigma_\mathrm{H}^\mathrm{C} \\ \sigma_\mathrm{H}^\mathrm{C} & \sigma_\mathrm{L}^\mathrm{e}+\sigma_\mathrm{L}^\mathrm{C} & -\sigma_\mathrm{H}^\mathrm{C} & -\sigma_\mathrm{D} - \sigma_\mathrm{L}^\mathrm{C} \\ -\sigma_\mathrm{D} - \sigma_\mathrm{L}^\mathrm{C} & \sigma_\mathrm{H}^\mathrm{C} & \sigma_\mathrm{L}^\mathrm{h} + \sigma_\mathrm{L}^\mathrm{C} & -\sigma_\mathrm{H}^\mathrm{h} - \sigma_\mathrm{H}^\mathrm{C}  \\ - \sigma_\mathrm{H}^\mathrm{C} & -\sigma_\mathrm{D} - \sigma_\mathrm{L}^\mathrm{C} &   \sigma_\mathrm{H}^\mathrm{h} + \sigma_\mathrm{H}^\mathrm{C} & \sigma_\mathrm{L}^\mathrm{h} + \sigma_\mathrm{L}^\mathrm{C}   
\end{pmatrix}.
\end{equation*}
Here, $\sigma_\mathrm{L}^{\mathrm{e}(\mathrm{h})}$ is the conductivity of the electrons (holes), and $\sigma_\mathrm{D}$ is the contribution to the transconductivity due to momentum transfer between layers.  $\sigma_\mathrm{H}^{\mathrm{h}}$ is the anomalous conductivity of Dirac holes, which is nonzero if the symmetry between valleys is broken. $\sigma_\mathrm{L}^\mathrm{C}=\sigma_\mathrm{L}^\mathrm{AH}+\sigma_\mathrm{L}^\mathrm{G}$ and $\sigma_\mathrm{H}^\mathrm{C}=\sigma_\mathrm{H}^\mathrm{\Omega}$ are contributions of fluctuating CPs in the conductivity tensor. It is instructive to discuss the setups sketched in Fig.~\ref{Fig4} separately.

For the setup shown in Fig.~\ref{Fig5}-a, an electric current is induced in the semiconductor QW. The corresponding Hall resistance of the QW is given by  
\begin{equation}
\label{DragResistivity1}
\rho^{\mathrm{ee}}_\mathrm{H}=\frac{E_
\mathrm{y}^{\mathrm{e}}}{I_{\mathrm{x}}^{\mathrm{e}}}=-\frac{\sigma_\mathrm{H}^\mathrm{C}}{(\sigma_\mathrm{L}^\mathrm{e})^2},
\end{equation} 
which is proportional to the anomalous Aslamazov-Larkin paraconductivity mediated by fluctuating CPs. Importantly, this scheme represents a smoking gun experiment, but does not require independent electrical contacts for the two layers (which are essential for Coulomb drag measurements). Apparently, the anomalous Hall effect in a semiconductor QW is induced by non-perturbative interlayer correlations with chiral Dirac electrons in graphene.

For the setup shown in Fig.~\ref{Fig4}-b, the electric current is induced in graphene. At first glance, this setup appears to be convenient for probing transverse anomalous paraconductivity, but this is actually not the case. The corresponding Hall drag resistance is given by 
\begin{equation}
\label{DragResistivity2}
\bar{\rho}^\mathrm{he}_\mathrm{H}=\frac{E_\mathrm{y}^{\mathrm{e}}}{I_\mathrm{x}^{\mathrm{h}}}=\frac{\sigma_\mathrm{H}^\mathrm{C}}{\sigma_\mathrm{L}^\mathrm{e} \sigma _\mathrm{L}^\mathrm{h}}-\frac{\sigma_\mathrm{D}}{\sigma_\mathrm{L}^\mathrm{e}\sigma_\mathrm{L}^\mathrm{h}} \frac{\sigma_\mathrm{H}^\mathrm{h}}{\sigma_\mathrm{L}^\mathrm{h}},
\end{equation}
which has an additional term with the opposite sign. This term is nonzero even in the absence of fluctuating CPs. This resistance originates from the interplay between the anomalous Hall effect in graphene and the conventional longitudinal Coulomb effect due to momentum transfer between layers. The former results in transverse charge accumulation and momentum flow in graphene, while the latter induces a current in the semiconductor QW. 

The transverse accumulation of charge carriers at graphene sides can be avoided by applying a shortcut, as shown for the setup in Fig.~\ref{Fig4}-c. As a result, the Hall drag resistance 
\begin{equation}
\label{DragResistivity3}
 \rho^\mathrm{he}_\mathrm{H}=\frac{E_\mathrm{y}^{\mathrm{e}}}{I_\mathrm{x}^{\mathrm{h}}}=\frac{\sigma_\mathrm{H}^\mathrm{C}}{\sigma_\mathrm{L}^\mathrm{e} \sigma_{\mathrm{L}}^{\mathrm{h}}}   
\end{equation}
has a single term, which is proportional to the anomalous Hall paraconductivity mediated by fluctuating CPs. It should be noted that this term has the same temperature dependence (up to a prefactor) as the Hall resistivity of QWs, given by Eq.~(\ref{DragResistivity1}). Observations of consistent temperature behavior would be strong evidence of an anomalous Hall response mediated by fluctuating CPs with nontrivial geometries.

\section{VIII. Discussions}

\begin{figure}[t]
	\begin{center}
		\includegraphics[trim=0cm 0cm 0cm 0cm, clip, width=0.88\columnwidth]{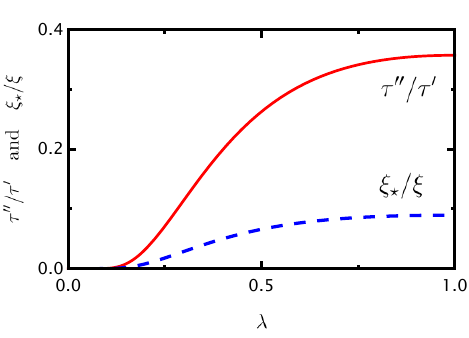}
		\caption{Dependence of length $\xi_\star/\xi$ and time $\tau''/\tau'$  scale ratios on the dimensionless coupling constant $\lambda$. Being small at $\lambda\ll1$, the ratios grow drastically with coupling strength. However, being derived within the weak coupling theory, the values for the scale ratios at $\lambda\sim1$ can only be used for guidance.}
		\label{Fig5}
	\end{center}
\end{figure}

For estimations, we use a set of parameters that are relevant for the hybrid graphene-GaAs bilayer reported in Ref.~\cite{HybridDrag}. We chose different densities for electrons $n_\mathrm{e}\approx 1.2\; 10^{11}\;\hbox{cm}^{-2}$ and holes $n_\mathrm{h}\approx 2.4\; 10^{11}\;\hbox{cm}^{-2}$ because holes have an additional valley degree of freedom. As a result, the Fermi momenta $p_\mathrm{F}$ for charge carriers in both layers are the same, but the Fermi energies ($\epsilon_\mathrm{F}^\mathrm{h}\approx 57~\hbox{meV}$ and $\epsilon_\mathrm{F}^\mathrm{e}\approx 3.8~\hbox{meV}$) and Fermi velocities ($v_\mathrm{e}\approx 1.5 \; 10^{7}\; \hbox{cm}/\hbox{s}$ and $v_\mathrm{h}\approx  \; 10^{8} \; \hbox{cm}/\hbox{s}$)  are drastically different. For the transition temperature $T\approx1\hbox{K}$ and $p_0=0.5 p_\mathrm{F}$, the dimensionless coupling constant $\lambda$ can be estimated as $\lambda\approx0.27$. The resulting ratios of the spatial scale $\xi_\star/\xi=0.022$ and time scale $\tau''/\tau'=0.028$ are quite small. If we approximate the conductivities by the values  $1/\sigma_\mathrm{L}^\mathrm{e}\approx0.3~\mathrm{k\Omega}$ and $1/\sigma_\mathrm{L}^\mathrm{h}\approx0.18~\mathrm{k\Omega}$ measured in the experiment, the order of magnitude for the transverse resistivities can be estimated as $\rho^\mathrm{ee}_\mathrm{H}\sim 51\; \mu \Omega$ and $\rho^\mathrm{he}_\mathrm{H}\sim 30\; \mu \Omega$, which can be probed by state-of-the-art transport experiments.    

The ratio of length $\xi_\star/\xi$ and time $\tau''/\tau'$ scales are expected to grow dramatically with coupling strength $\lambda$. According to microscopic theories~\cite{EHBG4,EHMultiband1,EHMultiband2,EHMultiband3}, electron-hole bilayers experience the crossover from the weak coupling regime to the strong coupling regime as the density of charge carries decreases. In the QW-QW bilayers with the conventional spectrum, the strong coupling regime corresponds to the Bose-Einstein condensate (BEC) of indirect excitons that represent a bound state of electron and hole~\cite{EHCros}. In the systems with the gapless Dirac spectrum the strong coupling regime corresponds to the multiband BCS-like paired state~\cite{EHMultiband1,EHMultiband2,EHMultiband3} where pairing correlations also span to remote bands (valence band in the layer with excess of electrons and conduction band in the layer with excess of holes).  The strong coupling regime for the hybrid graphene-GaAs bilayers has not been addressed microscopically yet. For the sake of estimations, we fix the charge carrier concentrations and treat $\lambda$ as an independent parameter. The corresponding dependence of the ratios is presented in Fig.~\ref{Fig5}. The ratio $\xi_\star/\xi$ is much smaller than $\tau''/\tau'$ due to the small factor $v_\mathrm{e}/(v_\mathrm{e}+v_\mathrm{h})$ in front of $\xi_\star$. Its small magnitude reflects the fact that the electron velocity is much smaller than the velocity of Dirac holes. As a result, the Cooper pair momentum is mostly carried by the electron, which is free of nontrivial geometries. However, it should be stressed that the calculations presented in Fig.~\ref{Fig5} are based on the weak coupling BCS theory with a contact pseudopotential and can be used only for guidance. For $\lambda\approx0.8$ the ratios of scales are $\xi_\star/\xi=0.08$ and $\tau''/\tau'=0.32$. As a result, the order of magnitude for the transverse resistivities can be estimated as $\rho^\mathrm{ee}_\mathrm{H}\sim 8.9\; m \Omega$ and $\rho^\mathrm{he}_\mathrm{H}\sim 5.3\; m\Omega$. 
 
The anomalous Hall current in semiconductor QWs requires the symmetry between two valleys in graphene to be broken. The selective formation of fluctuating CPs involving holes only from one of two valleys can be achieved in the presence of a circularly polarized light-induced population imbalance of Dirac holes~\cite{ValleyPOlarization1,ValleyPOlarization2,ValleyPOlarization3}. Due to the exceptional sensitivity of CP condensation in weak-to-moderate coupling regime to the electron-hole imbalance~\cite{LOFFSeradjeh,EfimkinLOFF}, even a small asymmetry $\delta\varepsilon_\mathrm{F}\sim T_0$ is sufficient to suppress correlations with one of two valleys. 

The presence of two Dirac cones with opposite chiralities can be avoided in hybrid double-layer systems formed by magnetic topological insulator films (e.g., $\hbox{MnB}_2 \hbox{Te}_4$) and semiconductor QWs. Recently discovered magnetic topological insulators~\cite{MagneticTI1,MagneticTI2,MagneticTI3,MagneticTI4} have already been successfully incorporated in different heterostructures, and these hybrid double-layer systems are within present technological capabilities. 

In the present work, we have focused on the Aslamazov-Larkin effect, which describes the direct contribution of fluctuating CPs to the conductivity tensor. The Maki-Thompson contribution also exists~\cite{DragHu,DragMink1,DragMink2}, originating from the Andreev scattering of electrons and holes for fluctuating CPs. However, the interplay between the nontrivial geometry of fluctuating CPs and Andreev scattering is beyond the scope of the present paper.

The absence of a global gap in the spectrum of the GL Hamiltonian does not permit topologically stable edges or domain wall modes. In other systems with intertwined modes and no global gap (e.g., topological exciton-polaritons~\cite{Topolaritons1,Topolaritons2}), this obstacle is usually overcome by applying an external periodic potential with triagonal (or hexagonal) symmetry. Hexagonal periodic textures of sublattice asymmetry $\delta(\vec{r})$ can be easily engineered via a small twist between graphene and its substrate and follow the resulting Moire pattern. However, the physics of fluctuating CPs and the possibility of edge modes in this regime are beyond the scope of the present paper. 

%\textcolor{red}{BKT} Due to the two-dimensional nature of the considered double-layer system, the long-range order of the CP condensate is fragile and limited by phase fluctuations. In the thermodynamic limit, the transition is smooth and becomes the Berezinxkii--Losterlitz Thouless transition describing the proliferation/binding of vorticities. However, phase fluctuations require large scales  $\ln[L/\xi]\ll 1$ to be well developed and are limited in mesoscopic samples $\ln[L/\xi]~1$. In this case, the Gaussian fluctuations survive except in the vicinity of the critical temperature. 

In recent years, the Cooper pairing of spin-orbit coupled fermions has attracted much attention~\cite{FFLOReviewColdAtoms1,FFLOReviewColdAtoms2}. In particular, research has shown that the nonzero quantum metric of fermionic states result in an additional contribution to the superfluid density and are also manifested in other phenomena~\cite{QuantumMetricSuperfluid1,QuantumMetricSuperfluid2,QuantumMetricSuperfluid3,QuantumMetricSuperfluid4,QFT2}. Moreover, the additional contribution to the superfluid density is dominant if the fermion dispersion is sufficiently  flat. While there are some mathematical connections between these results and the anomalous contribution to paraconductivity $\sigma^\mathrm{G}_\mathrm{L}$ derived in this paper, there is an essential difference between these phenomena. In our paper, $\sigma^\mathrm{G}_\mathrm{L}$ is intricately related to the quantum metric characterizing the spectrum of the GL Hamiltonian, but not the spectrum of Bogoliubov quasiparticles in a state with an equilibrium CP condensate.

Fluctuating CPs formed by electrons and holes are the precursors of their equilibrium condensation. The demonstrated presence of two (almost) degenerate competing channels and the instability toward Cooper pairing with a finite center of mass momentum suggest that the physics of equilibrium hybrid electron-hole condensate are very rich and unconventional. We
leave further investigations of this aspect for future work. 

To conclude, we have demonstrated that the spectrum of fluctuating electron-hole CPs in a hybrid graphene-GaAs bilayer is topologically nontrivial. Their nontrivial geometries are manifested in anomalous Aslamazov-Larkin contributions to the longitudinal and transverse conductivities. The contributions exhibit critical behavior and are singular at the transition temperature, and we have discussed possible setups for their experimental observation. 

\section{Acknowledgments}
D. K. E. acknowledges useful discussions with 
D. Culcer and O. Bleu. This research has been supported by the Australian Research Council Centre of Excellence in Future Low-Energy Electronics Technologies.
\bibliography{EHReferences}

\begin{thebibliography}{84}%
\makeatletter
\providecommand \@ifxundefined [1]{%
 \@ifx{#1\undefined}
}%
\providecommand \@ifnum [1]{%
 \ifnum #1\expandafter \@firstoftwo
 \else \expandafter \@secondoftwo
 \fi
}%
\providecommand \@ifx [1]{%
 \ifx #1\expandafter \@firstoftwo
 \else \expandafter \@secondoftwo
 \fi
}%
\providecommand \natexlab [1]{#1}%
\providecommand \enquote  [1]{``#1''}%
\providecommand \bibnamefont  [1]{#1}%
\providecommand \bibfnamefont [1]{#1}%
\providecommand \citenamefont [1]{#1}%
\providecommand \href@noop [0]{\@secondoftwo}%
\providecommand \href [0]{\begingroup \@sanitize@url \@href}%
\providecommand \@href[1]{\@@startlink{#1}\@@href}%
\providecommand \@@href[1]{\endgroup#1\@@endlink}%
\providecommand \@sanitize@url [0]{\catcode `\\12\catcode `\$12\catcode
  `\&12\catcode `\#12\catcode `\^12\catcode `\_12\catcode `\%12\relax}%
\providecommand \@@startlink[1]{}%
\providecommand \@@endlink[0]{}%
\providecommand \url  [0]{\begingroup\@sanitize@url \@url }%
\providecommand \@url [1]{\endgroup\@href {#1}{\urlprefix }}%
\providecommand \urlprefix  [0]{URL }%
\providecommand \Eprint [0]{\href }%
\providecommand \doibase [0]{http://dx.doi.org/}%
\providecommand \selectlanguage [0]{\@gobble}%
\providecommand \bibinfo  [0]{\@secondoftwo}%
\providecommand \bibfield  [0]{\@secondoftwo}%
\providecommand \translation [1]{[#1]}%
\providecommand \BibitemOpen [0]{}%
\providecommand \bibitemStop [0]{}%
\providecommand \bibitemNoStop [0]{.\EOS\space}%
\providecommand \EOS [0]{\spacefactor3000\relax}%
\providecommand \BibitemShut  [1]{\csname bibitem#1\endcsname}%
\let\auto@bib@innerbib\@empty
%</preamble>
\bibitem [{\citenamefont {Xiao}\ \emph {et~al.}(2010)\citenamefont {Xiao},
  \citenamefont {Chang},\ and\ \citenamefont {Niu}}]{BFReview1}%
  \BibitemOpen
  \bibfield  {author} {\bibinfo {author} {\bibfnamefont {D.}~\bibnamefont
  {Xiao}}, \bibinfo {author} {\bibfnamefont {M.-C.}\ \bibnamefont {Chang}}, \
  and\ \bibinfo {author} {\bibfnamefont {Q.}~\bibnamefont {Niu}},\ }\bibfield
  {title} {\bibinfo {title} {\emph {Berry phase effects on electronic
  properties}},\ }\href {\doibase 10.1103/RevModPhys.82.1959} {\bibfield
  {journal} {\bibinfo  {journal} {Rev. Mod. Phys.}\ }\textbf {\bibinfo {volume}
  {82}},\ \bibinfo {pages} {1959} (\bibinfo {year} {2010})}\BibitemShut
  {NoStop}%
\bibitem [{\citenamefont {Resta}(2000)}]{BFReview2}%
  \BibitemOpen
  \bibfield  {author} {\bibinfo {author} {\bibfnamefont {R.}~\bibnamefont
  {Resta}},\ }\bibfield  {title} {\bibinfo {title} {\emph {Manifestations of
  Berry{\textquotesingle}s phase in molecules and condensed matter}},\ }\href
  {\doibase 10.1088/0953-8984/12/9/201} {\bibfield  {journal} {\bibinfo
  {journal} {Journal of Physics: Condensed Matter}\ }\textbf {\bibinfo {volume}
  {12}},\ \bibinfo {pages} {R107} (\bibinfo {year} {2000})}\BibitemShut
  {NoStop}%
\bibitem [{\citenamefont {Nagaosa}\ \emph
  {et~al.}(2010{\natexlab{a}})\citenamefont {Nagaosa}, \citenamefont {Sinova},
  \citenamefont {Onoda}, \citenamefont {MacDonald},\ and\ \citenamefont
  {Ong}}]{AHEReview}%
  \BibitemOpen
  \bibfield  {author} {\bibinfo {author} {\bibfnamefont {N.}~\bibnamefont
  {Nagaosa}}, \bibinfo {author} {\bibfnamefont {J.}~\bibnamefont {Sinova}},
  \bibinfo {author} {\bibfnamefont {S.}~\bibnamefont {Onoda}}, \bibinfo
  {author} {\bibfnamefont {A.~H.}\ \bibnamefont {MacDonald}}, \ and\ \bibinfo
  {author} {\bibfnamefont {N.~P.}\ \bibnamefont {Ong}},\ }\bibfield  {title}
  {\bibinfo {title} {\emph {Anomalous Hall effect}},\ }\href {\doibase
  10.1103/RevModPhys.82.1539} {\bibfield  {journal} {\bibinfo  {journal} {Rev.
  Mod. Phys.}\ }\textbf {\bibinfo {volume} {82}},\ \bibinfo {pages} {1539}
  (\bibinfo {year} {2010}{\natexlab{a}})}\BibitemShut {NoStop}%
\bibitem [{\citenamefont {Schnyder}\ \emph {et~al.}(2008)\citenamefont
  {Schnyder}, \citenamefont {Ryu}, \citenamefont {Furusaki},\ and\
  \citenamefont {Ludwig}}]{TopologicalClassification}%
  \BibitemOpen
  \bibfield  {author} {\bibinfo {author} {\bibfnamefont {A.~P.}\ \bibnamefont
  {Schnyder}}, \bibinfo {author} {\bibfnamefont {S.}~\bibnamefont {Ryu}},
  \bibinfo {author} {\bibfnamefont {A.}~\bibnamefont {Furusaki}}, \ and\
  \bibinfo {author} {\bibfnamefont {A.~W.~W.}\ \bibnamefont {Ludwig}},\
  }\bibfield  {title} {\bibinfo {title} {\emph {Classification of topological
  insulators and superconductors in three spatial dimensions}},\ }\href
  {\doibase 10.1103/PhysRevB.78.195125} {\bibfield  {journal} {\bibinfo
  {journal} {Phys. Rev. B}\ }\textbf {\bibinfo {volume} {78}},\ \bibinfo
  {pages} {195125} (\bibinfo {year} {2008})}\BibitemShut {NoStop}%
\bibitem [{\citenamefont {Hasan}\ and\ \citenamefont
  {Kane}(2010)}]{TopologicalReview1}%
  \BibitemOpen
  \bibfield  {author} {\bibinfo {author} {\bibfnamefont {M.~Z.}\ \bibnamefont
  {Hasan}}\ and\ \bibinfo {author} {\bibfnamefont {C.~L.}\ \bibnamefont
  {Kane}},\ }\bibfield  {title} {\bibinfo {title} {\emph {Colloquium:
  Topological insulators}},\ }\href {\doibase 10.1103/RevModPhys.82.3045}
  {\bibfield  {journal} {\bibinfo  {journal} {Rev. Mod. Phys.}\ }\textbf
  {\bibinfo {volume} {82}},\ \bibinfo {pages} {3045} (\bibinfo {year}
  {2010})}\BibitemShut {NoStop}%
\bibitem [{\citenamefont {Qi}\ and\ \citenamefont
  {Zhang}(2011)}]{TopologicalReview2}%
  \BibitemOpen
  \bibfield  {author} {\bibinfo {author} {\bibfnamefont {X.-L.}\ \bibnamefont
  {Qi}}\ and\ \bibinfo {author} {\bibfnamefont {S.-C.}\ \bibnamefont {Zhang}},\
  }\bibfield  {title} {\bibinfo {title} {\emph {Topological insulators and
  superconductors}},\ }\href {\doibase 10.1103/RevModPhys.83.1057} {\bibfield
  {journal} {\bibinfo  {journal} {Rev. Mod. Phys.}\ }\textbf {\bibinfo {volume}
  {83}},\ \bibinfo {pages} {1057} (\bibinfo {year} {2011})}\BibitemShut
  {NoStop}%
\bibitem [{\citenamefont {Sato}\ and\ \citenamefont
  {Ando}(2017)}]{TopologicalReview3}%
  \BibitemOpen
  \bibfield  {author} {\bibinfo {author} {\bibfnamefont {M.}~\bibnamefont
  {Sato}}\ and\ \bibinfo {author} {\bibfnamefont {Y.}~\bibnamefont {Ando}},\
  }\bibfield  {title} {\bibinfo {title} {\emph {Topological superconductors: a
  review}},\ }\href {\doibase 10.1088/1361-6633/aa6ac7} {\bibfield  {journal}
  {\bibinfo  {journal} {Reports on Progress in Physics}\ }\textbf {\bibinfo
  {volume} {80}},\ \bibinfo {pages} {076501} (\bibinfo {year}
  {2017})}\BibitemShut {NoStop}%
\bibitem [{\citenamefont {Lozovik}\ and\ \citenamefont
  {Yudson}(1975)}]{LozovikYudson1}%
  \BibitemOpen
  \bibfield  {author} {\bibinfo {author} {\bibfnamefont {Y.~E.}\ \bibnamefont
  {Lozovik}}\ and\ \bibinfo {author} {\bibfnamefont {V.}~\bibnamefont
  {Yudson}},\ }\bibfield  {title} {\bibinfo {title} {\emph {Feasibility of
  superfluidity of paired spatially separated electrons and holes; a new
  superconductivity mechanism}},\ }\href@noop {} {\bibfield  {journal}
  {\bibinfo  {journal} {JETP Lett.}\ }\textbf {\bibinfo {volume} {22}},\
  \bibinfo {pages} {274} (\bibinfo {year} {1975})}\BibitemShut {NoStop}%
\bibitem [{\citenamefont {Lozovik}\ and\ \citenamefont
  {Yudson}(1976)}]{LozovikYudson2}%
  \BibitemOpen
  \bibfield  {author} {\bibinfo {author} {\bibfnamefont {Y.~E.}\ \bibnamefont
  {Lozovik}}\ and\ \bibinfo {author} {\bibfnamefont {V.}~\bibnamefont
  {Yudson}},\ }\bibfield  {title} {\bibinfo {title} {\emph {A new mechanism for
  superconductivity: pairing between spatially separated electrons and
  holes}},\ }\href@noop {} {\bibfield  {journal} {\bibinfo  {journal} {Sov.
  Phys. JETP}\ }\textbf {\bibinfo {volume} {44}},\ \bibinfo {pages} {389}
  (\bibinfo {year} {1976})}\BibitemShut {NoStop}%
\bibitem [{\citenamefont {Shevchenko}(1976)}]{Shevchenko}%
  \BibitemOpen
  \bibfield  {author} {\bibinfo {author} {\bibfnamefont {S.~I.}\ \bibnamefont
  {Shevchenko}},\ }\bibfield  {title} {\bibinfo {title} {\emph {Theory of
  superconductivity in the systems with pairing of spatially separated
  electrons and holes}},\ }\href@noop {} {\bibfield  {journal} {\bibinfo
  {journal} {Sov. J. Low Temp. Phys}\ }\textbf {\bibinfo {volume} {2}},\
  \bibinfo {pages} {251} (\bibinfo {year} {1976})}\BibitemShut {NoStop}%
\bibitem [{\citenamefont {Saberi-Pouya}\ \emph {et~al.}(2020)\citenamefont
  {Saberi-Pouya}, \citenamefont {Conti}, \citenamefont {Perali}, \citenamefont
  {Croxall}, \citenamefont {Hamilton}, \citenamefont {Peeters},\ and\
  \citenamefont {Neilson}}]{GaAsRecent1}%
  \BibitemOpen
  \bibfield  {author} {\bibinfo {author} {\bibfnamefont {S.}~\bibnamefont
  {Saberi-Pouya}}, \bibinfo {author} {\bibfnamefont {S.}~\bibnamefont {Conti}},
  \bibinfo {author} {\bibfnamefont {A.}~\bibnamefont {Perali}}, \bibinfo
  {author} {\bibfnamefont {A.~F.}\ \bibnamefont {Croxall}}, \bibinfo {author}
  {\bibfnamefont {A.~R.}\ \bibnamefont {Hamilton}}, \bibinfo {author}
  {\bibfnamefont {F.~m. c.~M.}\ \bibnamefont {Peeters}}, \ and\ \bibinfo
  {author} {\bibfnamefont {D.}~\bibnamefont {Neilson}},\ }\bibfield  {title}
  {\bibinfo {title} {\emph {Experimental conditions for the observation of
  electron-hole superfluidity in GaAs heterostructures}},\ }\href {\doibase
  10.1103/PhysRevB.101.140501} {\bibfield  {journal} {\bibinfo  {journal}
  {Phys. Rev. B}\ }\textbf {\bibinfo {volume} {101}},\ \bibinfo {pages}
  {140501} (\bibinfo {year} {2020})}\BibitemShut {NoStop}%
\bibitem [{\citenamefont {Pikulin}\ and\ \citenamefont
  {Hyart}(2014)}]{GaAsRecent2}%
  \BibitemOpen
  \bibfield  {author} {\bibinfo {author} {\bibfnamefont {D.~I.}\ \bibnamefont
  {Pikulin}}\ and\ \bibinfo {author} {\bibfnamefont {T.}~\bibnamefont
  {Hyart}},\ }\bibfield  {title} {\bibinfo {title} {\emph {Interplay of Exciton
  Condensation and the Quantum Spin Hall Effect in
  $\mathrm{InAs}/\mathrm{GaSb}$ Bilayers}},\ }\href {\doibase
  10.1103/PhysRevLett.112.176403} {\bibfield  {journal} {\bibinfo  {journal}
  {Phys. Rev. Lett.}\ }\textbf {\bibinfo {volume} {112}},\ \bibinfo {pages}
  {176403} (\bibinfo {year} {2014})}\BibitemShut {NoStop}%
\bibitem [{\citenamefont {Fil}\ and\ \citenamefont
  {Shevchenko}(2018)}]{EHreview}%
  \BibitemOpen
  \bibfield  {author} {\bibinfo {author} {\bibfnamefont {D.~V.}\ \bibnamefont
  {Fil}}\ and\ \bibinfo {author} {\bibfnamefont {S.~I.}\ \bibnamefont
  {Shevchenko}},\ }\bibfield  {title} {\bibinfo {title} {\emph {Electron-hole
  Superconductivity (Review)}},\ }\href {\doibase 10.1063/1.5052674} {\bibfield
   {journal} {\bibinfo  {journal} {Low Temperature Physics}\ }\textbf {\bibinfo
  {volume} {44}},\ \bibinfo {pages} {867} (\bibinfo {year} {2018})},\ \Eprint
  {http://arxiv.org/abs/https://doi.org/10.1063/1.5052674}
  {https://doi.org/10.1063/1.5052674} \BibitemShut {NoStop}%
\bibitem [{\citenamefont {Min}\ \emph {et~al.}(2008)\citenamefont {Min},
  \citenamefont {Bistritzer}, \citenamefont {Su},\ and\ \citenamefont
  {MacDonald}}]{EH1}%
  \BibitemOpen
  \bibfield  {author} {\bibinfo {author} {\bibfnamefont {H.}~\bibnamefont
  {Min}}, \bibinfo {author} {\bibfnamefont {R.}~\bibnamefont {Bistritzer}},
  \bibinfo {author} {\bibfnamefont {J.-J.}\ \bibnamefont {Su}}, \ and\ \bibinfo
  {author} {\bibfnamefont {A.~H.}\ \bibnamefont {MacDonald}},\ }\bibfield
  {title} {\bibinfo {title} {\emph {Room-temperature superfluidity in graphene
  bilayers}},\ }\href {\doibase 10.1103/PhysRevB.78.121401} {\bibfield
  {journal} {\bibinfo  {journal} {Phys. Rev. B}\ }\textbf {\bibinfo {volume}
  {78}},\ \bibinfo {pages} {121401} (\bibinfo {year} {2008})}\BibitemShut
  {NoStop}%
\bibitem [{\citenamefont {Lozovik}\ and\ \citenamefont {Sokolik}(2008)}]{EH2}%
  \BibitemOpen
  \bibfield  {author} {\bibinfo {author} {\bibfnamefont {Y.~E.}\ \bibnamefont
  {Lozovik}}\ and\ \bibinfo {author} {\bibfnamefont {A.~A.}\ \bibnamefont
  {Sokolik}},\ }\bibfield  {title} {\bibinfo {title} {\emph {Electron-hole pair
  condensation in a graphene bilayer}},\ }\href {\doibase
  10.1134/S002136400801013X} {\bibfield  {journal} {\bibinfo  {journal} {JETP
  Letters}\ }\textbf {\bibinfo {volume} {87}},\ \bibinfo {pages} {55} (\bibinfo
  {year} {2008})}\BibitemShut {NoStop}%
\bibitem [{\citenamefont {Zhang}\ and\ \citenamefont {Joglekar}(2008)}]{EH3}%
  \BibitemOpen
  \bibfield  {author} {\bibinfo {author} {\bibfnamefont {C.-H.}\ \bibnamefont
  {Zhang}}\ and\ \bibinfo {author} {\bibfnamefont {Y.~N.}\ \bibnamefont
  {Joglekar}},\ }\bibfield  {title} {\bibinfo {title} {\emph {Excitonic
  condensation of massless fermions in graphene bilayers}},\ }\href {\doibase
  10.1103/PhysRevB.77.233405} {\bibfield  {journal} {\bibinfo  {journal} {Phys.
  Rev. B}\ }\textbf {\bibinfo {volume} {77}},\ \bibinfo {pages} {233405}
  (\bibinfo {year} {2008})}\BibitemShut {NoStop}%
\bibitem [{\citenamefont {Kharitonov}\ and\ \citenamefont
  {Efetov}(2008)}]{EH4}%
  \BibitemOpen
  \bibfield  {author} {\bibinfo {author} {\bibfnamefont {M.~Y.}\ \bibnamefont
  {Kharitonov}}\ and\ \bibinfo {author} {\bibfnamefont {K.~B.}\ \bibnamefont
  {Efetov}},\ }\bibfield  {title} {\bibinfo {title} {\emph {Electron screening
  and excitonic condensation in double-layer graphene systems}},\ }\href
  {\doibase 10.1103/PhysRevB.78.241401} {\bibfield  {journal} {\bibinfo
  {journal} {Phys. Rev. B}\ }\textbf {\bibinfo {volume} {78}},\ \bibinfo
  {pages} {241401} (\bibinfo {year} {2008})}\BibitemShut {NoStop}%
\bibitem [{\citenamefont {Zhang}\ and\ \citenamefont {Rossi}(2013)}]{EH5}%
  \BibitemOpen
  \bibfield  {author} {\bibinfo {author} {\bibfnamefont {J.}~\bibnamefont
  {Zhang}}\ and\ \bibinfo {author} {\bibfnamefont {E.}~\bibnamefont {Rossi}},\
  }\bibfield  {title} {\bibinfo {title} {\emph {Chiral Superfluid States in
  Hybrid Graphene Heterostructures}},\ }\href {\doibase
  10.1103/PhysRevLett.111.086804} {\bibfield  {journal} {\bibinfo  {journal}
  {Phys. Rev. Lett.}\ }\textbf {\bibinfo {volume} {111}},\ \bibinfo {pages}
  {086804} (\bibinfo {year} {2013})}\BibitemShut {NoStop}%
\bibitem [{\citenamefont {Lozovik}\ \emph {et~al.}(2012)\citenamefont
  {Lozovik}, \citenamefont {Ogarkov},\ and\ \citenamefont {Sokolik}}]{EH6}%
  \BibitemOpen
  \bibfield  {author} {\bibinfo {author} {\bibfnamefont {Y.~E.}\ \bibnamefont
  {Lozovik}}, \bibinfo {author} {\bibfnamefont {S.~L.}\ \bibnamefont
  {Ogarkov}}, \ and\ \bibinfo {author} {\bibfnamefont {A.~A.}\ \bibnamefont
  {Sokolik}},\ }\bibfield  {title} {\bibinfo {title} {\emph {Condensation of
  electron-hole pairs in a two-layer graphene system: Correlation effects}},\
  }\href {\doibase 10.1103/PhysRevB.86.045429} {\bibfield  {journal} {\bibinfo
  {journal} {Phys. Rev. B}\ }\textbf {\bibinfo {volume} {86}},\ \bibinfo
  {pages} {045429} (\bibinfo {year} {2012})}\BibitemShut {NoStop}%
\bibitem [{\citenamefont {Perali}\ \emph {et~al.}(2013)\citenamefont {Perali},
  \citenamefont {Neilson},\ and\ \citenamefont {Hamilton}}]{EHBG1}%
  \BibitemOpen
  \bibfield  {author} {\bibinfo {author} {\bibfnamefont {A.}~\bibnamefont
  {Perali}}, \bibinfo {author} {\bibfnamefont {D.}~\bibnamefont {Neilson}}, \
  and\ \bibinfo {author} {\bibfnamefont {A.~R.}\ \bibnamefont {Hamilton}},\
  }\bibfield  {title} {\bibinfo {title} {\emph {High-Temperature Superfluidity
  in Double-Bilayer Graphene}},\ }\href {\doibase
  10.1103/PhysRevLett.110.146803} {\bibfield  {journal} {\bibinfo  {journal}
  {Phys. Rev. Lett.}\ }\textbf {\bibinfo {volume} {110}},\ \bibinfo {pages}
  {146803} (\bibinfo {year} {2013})}\BibitemShut {NoStop}%
\bibitem [{\citenamefont {Neilson}\ \emph {et~al.}(2014)\citenamefont
  {Neilson}, \citenamefont {Perali},\ and\ \citenamefont {Hamilton}}]{EHBG2}%
  \BibitemOpen
  \bibfield  {author} {\bibinfo {author} {\bibfnamefont {D.}~\bibnamefont
  {Neilson}}, \bibinfo {author} {\bibfnamefont {A.}~\bibnamefont {Perali}}, \
  and\ \bibinfo {author} {\bibfnamefont {A.~R.}\ \bibnamefont {Hamilton}},\
  }\bibfield  {title} {\bibinfo {title} {\emph {Excitonic superfluidity and
  screening in electron-hole bilayer systems}},\ }\href {\doibase
  10.1103/PhysRevB.89.060502} {\bibfield  {journal} {\bibinfo  {journal} {Phys.
  Rev. B}\ }\textbf {\bibinfo {volume} {89}},\ \bibinfo {pages} {060502}
  (\bibinfo {year} {2014})}\BibitemShut {NoStop}%
\bibitem [{\citenamefont {Su}\ and\ \citenamefont {MacDonald}(2017)}]{EHBG3}%
  \BibitemOpen
  \bibfield  {author} {\bibinfo {author} {\bibfnamefont {J.-J.}\ \bibnamefont
  {Su}}\ and\ \bibinfo {author} {\bibfnamefont {A.~H.}\ \bibnamefont
  {MacDonald}},\ }\bibfield  {title} {\bibinfo {title} {\emph {Spatially
  indirect exciton condensate phases in double bilayer graphene}},\ }\href
  {\doibase 10.1103/PhysRevB.95.045416} {\bibfield  {journal} {\bibinfo
  {journal} {Phys. Rev. B}\ }\textbf {\bibinfo {volume} {95}},\ \bibinfo
  {pages} {045416} (\bibinfo {year} {2017})}\BibitemShut {NoStop}%
\bibitem [{\citenamefont {Conti}\ \emph {et~al.}(2017)\citenamefont {Conti},
  \citenamefont {Perali}, \citenamefont {Peeters},\ and\ \citenamefont
  {Neilson}}]{EHBG4}%
  \BibitemOpen
  \bibfield  {author} {\bibinfo {author} {\bibfnamefont {S.}~\bibnamefont
  {Conti}}, \bibinfo {author} {\bibfnamefont {A.}~\bibnamefont {Perali}},
  \bibinfo {author} {\bibfnamefont {F.~M.}\ \bibnamefont {Peeters}}, \ and\
  \bibinfo {author} {\bibfnamefont {D.}~\bibnamefont {Neilson}},\ }\bibfield
  {title} {\bibinfo {title} {\emph {Multicomponent Electron-Hole Superfluidity
  and the BCS-BEC Crossover in Double Bilayer Graphene}},\ }\href {\doibase
  10.1103/PhysRevLett.119.257002} {\bibfield  {journal} {\bibinfo  {journal}
  {Phys. Rev. Lett.}\ }\textbf {\bibinfo {volume} {119}},\ \bibinfo {pages}
  {257002} (\bibinfo {year} {2017})}\BibitemShut {NoStop}%
\bibitem [{\citenamefont {Saberi-Pouya}\ \emph {et~al.}(2018)\citenamefont
  {Saberi-Pouya}, \citenamefont {Zarenia}, \citenamefont {Perali},
  \citenamefont {Vazifehshenas},\ and\ \citenamefont {Peeters}}]{EHBG5}%
  \BibitemOpen
  \bibfield  {author} {\bibinfo {author} {\bibfnamefont {S.}~\bibnamefont
  {Saberi-Pouya}}, \bibinfo {author} {\bibfnamefont {M.}~\bibnamefont
  {Zarenia}}, \bibinfo {author} {\bibfnamefont {A.}~\bibnamefont {Perali}},
  \bibinfo {author} {\bibfnamefont {T.}~\bibnamefont {Vazifehshenas}}, \ and\
  \bibinfo {author} {\bibfnamefont {F.~M.}\ \bibnamefont {Peeters}},\
  }\bibfield  {title} {\bibinfo {title} {\emph {High-temperature electron-hole
  superfluidity with strong anisotropic gaps in double phosphorene
  monolayers}},\ }\href {\doibase 10.1103/PhysRevB.97.174503} {\bibfield
  {journal} {\bibinfo  {journal} {Phys. Rev. B}\ }\textbf {\bibinfo {volume}
  {97}},\ \bibinfo {pages} {174503} (\bibinfo {year} {2018})}\BibitemShut
  {NoStop}%
\bibitem [{\citenamefont {Debnath}\ \emph {et~al.}(2017)\citenamefont
  {Debnath}, \citenamefont {Barlas}, \citenamefont {Wickramaratne},
  \citenamefont {Neupane},\ and\ \citenamefont {Lake}}]{EHBG6}%
  \BibitemOpen
  \bibfield  {author} {\bibinfo {author} {\bibfnamefont {B.}~\bibnamefont
  {Debnath}}, \bibinfo {author} {\bibfnamefont {Y.}~\bibnamefont {Barlas}},
  \bibinfo {author} {\bibfnamefont {D.}~\bibnamefont {Wickramaratne}}, \bibinfo
  {author} {\bibfnamefont {M.~R.}\ \bibnamefont {Neupane}}, \ and\ \bibinfo
  {author} {\bibfnamefont {R.~K.}\ \bibnamefont {Lake}},\ }\bibfield  {title}
  {\bibinfo {title} {\emph {Exciton condensate in bilayer transition metal
  dichalcogenides: Strong coupling regime}},\ }\href {\doibase
  10.1103/PhysRevB.96.174504} {\bibfield  {journal} {\bibinfo  {journal} {Phys.
  Rev. B}\ }\textbf {\bibinfo {volume} {96}},\ \bibinfo {pages} {174504}
  (\bibinfo {year} {2017})}\BibitemShut {NoStop}%
\bibitem [{\citenamefont {Hu}\ \emph {et~al.}(2020)\citenamefont {Hu},
  \citenamefont {Hyart}, \citenamefont {Pikulin},\ and\ \citenamefont
  {Rossi}}]{EHBG4twisted}%
  \BibitemOpen
  \bibfield  {author} {\bibinfo {author} {\bibfnamefont {X.}~\bibnamefont
  {Hu}}, \bibinfo {author} {\bibfnamefont {T.}~\bibnamefont {Hyart}}, \bibinfo
  {author} {\bibfnamefont {D.~I.}\ \bibnamefont {Pikulin}}, \ and\ \bibinfo
  {author} {\bibfnamefont {E.}~\bibnamefont {Rossi}},\ }\href@noop {} {\bibinfo
  {title} {\emph {Quantum-metric-enabled exciton condensate in double twisted
  bilayer graphene}}} (\bibinfo {year} {2020}),\ \Eprint
  {http://arxiv.org/abs/2008.03241} {arXiv:2008.03241 [cond-mat.mes-hall]}
  \BibitemShut {NoStop}%
\bibitem [{\citenamefont {Seradjeh}\ \emph {et~al.}(2009)\citenamefont
  {Seradjeh}, \citenamefont {Moore},\ and\ \citenamefont {Franz}}]{EHTI1}%
  \BibitemOpen
  \bibfield  {author} {\bibinfo {author} {\bibfnamefont {B.}~\bibnamefont
  {Seradjeh}}, \bibinfo {author} {\bibfnamefont {J.~E.}\ \bibnamefont {Moore}},
  \ and\ \bibinfo {author} {\bibfnamefont {M.}~\bibnamefont {Franz}},\
  }\bibfield  {title} {\bibinfo {title} {\emph {Exciton Condensation and Charge
  Fractionalization in a Topological Insulator Film}},\ }\href {\doibase
  10.1103/PhysRevLett.103.066402} {\bibfield  {journal} {\bibinfo  {journal}
  {Phys. Rev. Lett.}\ }\textbf {\bibinfo {volume} {103}},\ \bibinfo {pages}
  {066402} (\bibinfo {year} {2009})}\BibitemShut {NoStop}%
\bibitem [{\citenamefont {Efimkin}\ \emph {et~al.}(2012)\citenamefont
  {Efimkin}, \citenamefont {Lozovik},\ and\ \citenamefont
  {Sokolik}}]{EfimkinEHTI}%
  \BibitemOpen
  \bibfield  {author} {\bibinfo {author} {\bibfnamefont {D.~K.}\ \bibnamefont
  {Efimkin}}, \bibinfo {author} {\bibfnamefont {Y.~E.}\ \bibnamefont
  {Lozovik}}, \ and\ \bibinfo {author} {\bibfnamefont {A.~A.}\ \bibnamefont
  {Sokolik}},\ }\bibfield  {title} {\bibinfo {title} {\emph {Electron-hole
  pairing in a topological insulator thin film}},\ }\href {\doibase
  10.1103/PhysRevB.86.115436} {\bibfield  {journal} {\bibinfo  {journal} {Phys.
  Rev. B}\ }\textbf {\bibinfo {volume} {86}},\ \bibinfo {pages} {115436}
  (\bibinfo {year} {2012})}\BibitemShut {NoStop}%
\bibitem [{Note1()}]{Note1}%
  \BibitemOpen
  \bibinfo {note} {It should be noted, that these CPs are formed not by
  electrons, but by spatially separated electrons and holes}\BibitemShut
  {NoStop}%
\bibitem [{\citenamefont {Efimkin}\ and\ \citenamefont
  {Lozovik}(2013{\natexlab{a}})}]{TunnelingEfimkin1}%
  \BibitemOpen
  \bibfield  {author} {\bibinfo {author} {\bibfnamefont {D.~K.}\ \bibnamefont
  {Efimkin}}\ and\ \bibinfo {author} {\bibfnamefont {Y.~E.}\ \bibnamefont
  {Lozovik}},\ }\bibfield  {title} {\bibinfo {title} {\emph {Fluctuational
  internal Josephson effect in a topological insulator film}},\ }\href
  {\doibase 10.1103/PhysRevB.88.085414} {\bibfield  {journal} {\bibinfo
  {journal} {Phys. Rev. B}\ }\textbf {\bibinfo {volume} {88}},\ \bibinfo
  {pages} {085414} (\bibinfo {year} {2013}{\natexlab{a}})}\BibitemShut
  {NoStop}%
\bibitem [{\citenamefont {Efimkin}\ \emph {et~al.}(2020)\citenamefont
  {Efimkin}, \citenamefont {Burg}, \citenamefont {Tutuc},\ and\ \citenamefont
  {MacDonald}}]{TunnelingEfimkin2}%
  \BibitemOpen
  \bibfield  {author} {\bibinfo {author} {\bibfnamefont {D.~K.}\ \bibnamefont
  {Efimkin}}, \bibinfo {author} {\bibfnamefont {G.~W.}\ \bibnamefont {Burg}},
  \bibinfo {author} {\bibfnamefont {E.}~\bibnamefont {Tutuc}}, \ and\ \bibinfo
  {author} {\bibfnamefont {A.~H.}\ \bibnamefont {MacDonald}},\ }\bibfield
  {title} {\bibinfo {title} {\emph {Tunneling and fluctuating electron-hole
  Cooper pairs in double bilayer graphene}},\ }\href {\doibase
  10.1103/PhysRevB.101.035413} {\bibfield  {journal} {\bibinfo  {journal}
  {Phys. Rev. B}\ }\textbf {\bibinfo {volume} {101}},\ \bibinfo {pages}
  {035413} (\bibinfo {year} {2020})}\BibitemShut {NoStop}%
\bibitem [{\citenamefont {Hu}(2000)}]{DragHu}%
  \BibitemOpen
  \bibfield  {author} {\bibinfo {author} {\bibfnamefont {B.~Y.-K.}\
  \bibnamefont {Hu}},\ }\bibfield  {title} {\bibinfo {title} {\emph
  {Prospecting for the Superfluid Transition in Electron-Hole Coupled Quantum
  Wells Using Coulomb Drag}},\ }\href {\doibase 10.1103/PhysRevLett.85.820}
  {\bibfield  {journal} {\bibinfo  {journal} {Phys. Rev. Lett.}\ }\textbf
  {\bibinfo {volume} {85}},\ \bibinfo {pages} {820} (\bibinfo {year}
  {2000})}\BibitemShut {NoStop}%
\bibitem [{\citenamefont {Mink}\ \emph {et~al.}(2012)\citenamefont {Mink},
  \citenamefont {Stoof}, \citenamefont {Duine}, \citenamefont {Polini},\ and\
  \citenamefont {Vignale}}]{DragMink1}%
  \BibitemOpen
  \bibfield  {author} {\bibinfo {author} {\bibfnamefont {M.~P.}\ \bibnamefont
  {Mink}}, \bibinfo {author} {\bibfnamefont {H.~T.~C.}\ \bibnamefont {Stoof}},
  \bibinfo {author} {\bibfnamefont {R.~A.}\ \bibnamefont {Duine}}, \bibinfo
  {author} {\bibfnamefont {M.}~\bibnamefont {Polini}}, \ and\ \bibinfo {author}
  {\bibfnamefont {G.}~\bibnamefont {Vignale}},\ }\bibfield  {title} {\bibinfo
  {title} {\emph {Probing the Topological Exciton Condensate via Coulomb
  Drag}},\ }\href {\doibase 10.1103/PhysRevLett.108.186402} {\bibfield
  {journal} {\bibinfo  {journal} {Phys. Rev. Lett.}\ }\textbf {\bibinfo
  {volume} {108}},\ \bibinfo {pages} {186402} (\bibinfo {year}
  {2012})}\BibitemShut {NoStop}%
\bibitem [{\citenamefont {Mink}\ \emph {et~al.}(2013)\citenamefont {Mink},
  \citenamefont {Stoof}, \citenamefont {Duine}, \citenamefont {Polini},\ and\
  \citenamefont {Vignale}}]{DragMink2}%
  \BibitemOpen
  \bibfield  {author} {\bibinfo {author} {\bibfnamefont {M.~P.}\ \bibnamefont
  {Mink}}, \bibinfo {author} {\bibfnamefont {H.~T.~C.}\ \bibnamefont {Stoof}},
  \bibinfo {author} {\bibfnamefont {R.~A.}\ \bibnamefont {Duine}}, \bibinfo
  {author} {\bibfnamefont {M.}~\bibnamefont {Polini}}, \ and\ \bibinfo {author}
  {\bibfnamefont {G.}~\bibnamefont {Vignale}},\ }\bibfield  {title} {\bibinfo
  {title} {\emph {Unified Boltzmann transport theory for the drag resistivity
  close to an interlayer-interaction-driven second-order phase transition}},\
  }\href {\doibase 10.1103/PhysRevB.88.235311} {\bibfield  {journal} {\bibinfo
  {journal} {Phys. Rev. B}\ }\textbf {\bibinfo {volume} {88}},\ \bibinfo
  {pages} {235311} (\bibinfo {year} {2013})}\BibitemShut {NoStop}%
\bibitem [{\citenamefont {Efimkin}\ and\ \citenamefont
  {Lozovik}(2013{\natexlab{b}})}]{DragEfimkin}%
  \BibitemOpen
  \bibfield  {author} {\bibinfo {author} {\bibfnamefont {D.~K.}\ \bibnamefont
  {Efimkin}}\ and\ \bibinfo {author} {\bibfnamefont {Y.~E.}\ \bibnamefont
  {Lozovik}},\ }\bibfield  {title} {\bibinfo {title} {\emph {Drag effect and
  Cooper electron-hole pair fluctuations in a topological insulator film}},\
  }\href {\doibase 10.1103/PhysRevB.88.235420} {\bibfield  {journal} {\bibinfo
  {journal} {Phys. Rev. B}\ }\textbf {\bibinfo {volume} {88}},\ \bibinfo
  {pages} {235420} (\bibinfo {year} {2013}{\natexlab{b}})}\BibitemShut
  {NoStop}%
\bibitem [{\citenamefont {Varlamov}\ \emph {et~al.}(2018)\citenamefont
  {Varlamov}, \citenamefont {Galda},\ and\ \citenamefont {Glatz}}]{Varlamov}%
  \BibitemOpen
  \bibfield  {author} {\bibinfo {author} {\bibfnamefont {A.~A.}\ \bibnamefont
  {Varlamov}}, \bibinfo {author} {\bibfnamefont {A.}~\bibnamefont {Galda}}, \
  and\ \bibinfo {author} {\bibfnamefont {A.}~\bibnamefont {Glatz}},\ }\bibfield
   {title} {\bibinfo {title} {\emph {Fluctuation spectroscopy: From
  Rayleigh-Jeans waves to Abrikosov vortex clusters}},\ }\href {\doibase
  10.1103/RevModPhys.90.015009} {\bibfield  {journal} {\bibinfo  {journal}
  {Rev. Mod. Phys.}\ }\textbf {\bibinfo {volume} {90}},\ \bibinfo {pages}
  {015009} (\bibinfo {year} {2018})}\BibitemShut {NoStop}%
\bibitem [{\citenamefont {Skocpol}\ and\ \citenamefont
  {Tinkham}(1975)}]{SkocpolTinkham}%
  \BibitemOpen
  \bibfield  {author} {\bibinfo {author} {\bibfnamefont {W.~J.}\ \bibnamefont
  {Skocpol}}\ and\ \bibinfo {author} {\bibfnamefont {M.}~\bibnamefont
  {Tinkham}},\ }\bibfield  {title} {\bibinfo {title} {\emph {Fluctuations near
  superconducting phase transitions}},\ }\href {\doibase
  10.1088/0034-4885/38/9/001} {\bibfield  {journal} {\bibinfo  {journal}
  {Reports on Progress in Physics}\ }\textbf {\bibinfo {volume} {38}},\
  \bibinfo {pages} {1049} (\bibinfo {year} {1975})}\BibitemShut {NoStop}%
\bibitem [{\citenamefont {Larkin}\ and\ \citenamefont
  {Valralmov}(2005)}]{LarkinVarlamov}%
  \BibitemOpen
  \bibfield  {author} {\bibinfo {author} {\bibfnamefont {I.}~\bibnamefont
  {Larkin}}\ and\ \bibinfo {author} {\bibfnamefont {A.}~\bibnamefont
  {Valralmov}},\ }\href@noop {} {\emph {\bibinfo {title} {Theory of
  Fluctuations in Superconductors}}}\ (\bibinfo  {publisher} {Clarendon,
  Oxford},\ \bibinfo {year} {2005})\BibitemShut {NoStop}%
\bibitem [{\citenamefont {Aslamazov}\ and\ \citenamefont
  {A.I.}(1975)}]{AslamazovlarkinDiamagnetsim}%
  \BibitemOpen
  \bibfield  {author} {\bibinfo {author} {\bibfnamefont {L.~G.}\ \bibnamefont
  {Aslamazov}}\ and\ \bibinfo {author} {\bibfnamefont {L.}~\bibnamefont
  {A.I.}},\ }\bibfield  {title} {\bibinfo {title} {\emph {Fluctuation-induced
  magnetic susceptibility of superconductors and normal metals}},\ }\href@noop
  {} {\bibfield  {journal} {\bibinfo  {journal} {Sov. Phys. JETP}\ }\textbf
  {\bibinfo {volume} {40}},\ \bibinfo {pages} {321} (\bibinfo {year}
  {1975})}\BibitemShut {NoStop}%
\bibitem [{\citenamefont {Aslamasov}\ and\ \citenamefont
  {Larkin}(1968)}]{AslamazovlarkinConductivity}%
  \BibitemOpen
  \bibfield  {author} {\bibinfo {author} {\bibfnamefont {L.}~\bibnamefont
  {Aslamasov}}\ and\ \bibinfo {author} {\bibfnamefont {A.}~\bibnamefont
  {Larkin}},\ }\bibfield  {title} {\bibinfo {title} {\emph {The influence of
  fluctuation pairing of electrons on the conductivity of normal metal}},\
  }\href {\doibase https://doi.org/10.1016/0375-9601(68)90623-3} {\bibfield
  {journal} {\bibinfo  {journal} {Physics Letters A}\ }\textbf {\bibinfo
  {volume} {26}},\ \bibinfo {pages} {238} (\bibinfo {year} {1968})}\BibitemShut
  {NoStop}%
\bibitem [{\citenamefont {Burg}\ \emph {et~al.}(2018)\citenamefont {Burg},
  \citenamefont {Prasad}, \citenamefont {Kim}, \citenamefont {Taniguchi},
  \citenamefont {Watanabe}, \citenamefont {MacDonald}, \citenamefont
  {Register},\ and\ \citenamefont {Tutuc}}]{TunnelingExp}%
  \BibitemOpen
  \bibfield  {author} {\bibinfo {author} {\bibfnamefont {G.~W.}\ \bibnamefont
  {Burg}}, \bibinfo {author} {\bibfnamefont {N.}~\bibnamefont {Prasad}},
  \bibinfo {author} {\bibfnamefont {K.}~\bibnamefont {Kim}}, \bibinfo {author}
  {\bibfnamefont {T.}~\bibnamefont {Taniguchi}}, \bibinfo {author}
  {\bibfnamefont {K.}~\bibnamefont {Watanabe}}, \bibinfo {author}
  {\bibfnamefont {A.~H.}\ \bibnamefont {MacDonald}}, \bibinfo {author}
  {\bibfnamefont {L.~F.}\ \bibnamefont {Register}}, \ and\ \bibinfo {author}
  {\bibfnamefont {E.}~\bibnamefont {Tutuc}},\ }\bibfield  {title} {\bibinfo
  {title} {\emph {Strongly Enhanced Tunneling at Total Charge Neutrality in
  Double-Bilayer Graphene-${\mathrm{WSe}}_{2}$ Heterostructures}},\ }\href
  {\doibase 10.1103/PhysRevLett.120.177702} {\bibfield  {journal} {\bibinfo
  {journal} {Phys. Rev. Lett.}\ }\textbf {\bibinfo {volume} {120}},\ \bibinfo
  {pages} {177702} (\bibinfo {year} {2018})}\BibitemShut {NoStop}%
\bibitem [{\citenamefont {Wang}\ \emph {et~al.}(2019)\citenamefont {Wang},
  \citenamefont {Rhodes}, \citenamefont {Watanabe}, \citenamefont {Taniguchi},
  \citenamefont {Hone}, \citenamefont {Shan},\ and\ \citenamefont
  {Mak}}]{LumExp}%
  \BibitemOpen
  \bibfield  {author} {\bibinfo {author} {\bibfnamefont {Z.}~\bibnamefont
  {Wang}}, \bibinfo {author} {\bibfnamefont {D.~A.}\ \bibnamefont {Rhodes}},
  \bibinfo {author} {\bibfnamefont {K.}~\bibnamefont {Watanabe}}, \bibinfo
  {author} {\bibfnamefont {T.}~\bibnamefont {Taniguchi}}, \bibinfo {author}
  {\bibfnamefont {J.~C.}\ \bibnamefont {Hone}}, \bibinfo {author}
  {\bibfnamefont {J.}~\bibnamefont {Shan}}, \ and\ \bibinfo {author}
  {\bibfnamefont {K.~F.}\ \bibnamefont {Mak}},\ }\bibfield  {title} {\bibinfo
  {title} {\emph {Evidence of high-temperature exciton condensation in
  two-dimensional atomic double layers}},\ }\href {\doibase
  10.1038/s41586-019-1591-7} {\bibfield  {journal} {\bibinfo  {journal}
  {Nature}\ }\textbf {\bibinfo {volume} {574}},\ \bibinfo {pages} {76}
  (\bibinfo {year} {2019})}\BibitemShut {NoStop}%
\bibitem [{\citenamefont {Morath}\ \emph {et~al.}(2009)\citenamefont {Morath},
  \citenamefont {Seamons}, \citenamefont {Reno},\ and\ \citenamefont
  {Lilly}}]{DragExp1}%
  \BibitemOpen
  \bibfield  {author} {\bibinfo {author} {\bibfnamefont {C.~P.}\ \bibnamefont
  {Morath}}, \bibinfo {author} {\bibfnamefont {J.~A.}\ \bibnamefont {Seamons}},
  \bibinfo {author} {\bibfnamefont {J.~L.}\ \bibnamefont {Reno}}, \ and\
  \bibinfo {author} {\bibfnamefont {M.~P.}\ \bibnamefont {Lilly}},\ }\bibfield
  {title} {\bibinfo {title} {\emph {Density imbalance effect on the Coulomb
  drag upturn in an undoped electron-hole bilayer}},\ }\href {\doibase
  10.1103/PhysRevB.79.041305} {\bibfield  {journal} {\bibinfo  {journal} {Phys.
  Rev. B}\ }\textbf {\bibinfo {volume} {79}},\ \bibinfo {pages} {041305}
  (\bibinfo {year} {2009})}\BibitemShut {NoStop}%
\bibitem [{\citenamefont {Croxall}\ \emph {et~al.}(2008)\citenamefont
  {Croxall}, \citenamefont {Das~Gupta}, \citenamefont {Nicoll}, \citenamefont
  {Thangaraj}, \citenamefont {Beere}, \citenamefont {Farrer}, \citenamefont
  {Ritchie},\ and\ \citenamefont {Pepper}}]{DragExp3}%
  \BibitemOpen
  \bibfield  {author} {\bibinfo {author} {\bibfnamefont {A.~F.}\ \bibnamefont
  {Croxall}}, \bibinfo {author} {\bibfnamefont {K.}~\bibnamefont {Das~Gupta}},
  \bibinfo {author} {\bibfnamefont {C.~A.}\ \bibnamefont {Nicoll}}, \bibinfo
  {author} {\bibfnamefont {M.}~\bibnamefont {Thangaraj}}, \bibinfo {author}
  {\bibfnamefont {H.~E.}\ \bibnamefont {Beere}}, \bibinfo {author}
  {\bibfnamefont {I.}~\bibnamefont {Farrer}}, \bibinfo {author} {\bibfnamefont
  {D.~A.}\ \bibnamefont {Ritchie}}, \ and\ \bibinfo {author} {\bibfnamefont
  {M.}~\bibnamefont {Pepper}},\ }\bibfield  {title} {\bibinfo {title} {\emph
  {Anomalous Coulomb Drag in Electron-Hole Bilayers}},\ }\href {\doibase
  10.1103/PhysRevLett.101.246801} {\bibfield  {journal} {\bibinfo  {journal}
  {Phys. Rev. Lett.}\ }\textbf {\bibinfo {volume} {101}},\ \bibinfo {pages}
  {246801} (\bibinfo {year} {2008})}\BibitemShut {NoStop}%
\bibitem [{\citenamefont {Croxall}\ \emph {et~al.}(2009)\citenamefont
  {Croxall}, \citenamefont {Das~Gupta}, \citenamefont {Nicoll}, \citenamefont
  {Beere}, \citenamefont {Farrer}, \citenamefont {Ritchie},\ and\ \citenamefont
  {Pepper}}]{DragExp4}%
  \BibitemOpen
  \bibfield  {author} {\bibinfo {author} {\bibfnamefont {A.~F.}\ \bibnamefont
  {Croxall}}, \bibinfo {author} {\bibfnamefont {K.}~\bibnamefont {Das~Gupta}},
  \bibinfo {author} {\bibfnamefont {C.~A.}\ \bibnamefont {Nicoll}}, \bibinfo
  {author} {\bibfnamefont {H.~E.}\ \bibnamefont {Beere}}, \bibinfo {author}
  {\bibfnamefont {I.}~\bibnamefont {Farrer}}, \bibinfo {author} {\bibfnamefont
  {D.~A.}\ \bibnamefont {Ritchie}}, \ and\ \bibinfo {author} {\bibfnamefont
  {M.}~\bibnamefont {Pepper}},\ }\bibfield  {title} {\bibinfo {title} {\emph
  {Possible effect of collective modes in zero magnetic field transport in an
  electron-hole bilayer}},\ }\href {\doibase 10.1103/PhysRevB.80.125323}
  {\bibfield  {journal} {\bibinfo  {journal} {Phys. Rev. B}\ }\textbf {\bibinfo
  {volume} {80}},\ \bibinfo {pages} {125323} (\bibinfo {year}
  {2009})}\BibitemShut {NoStop}%
\bibitem [{\citenamefont {Gamucci}\ \emph {et~al.}(2014)\citenamefont
  {Gamucci}, \citenamefont {Spirito}, \citenamefont {Carrega}, \citenamefont
  {Karmakar}, \citenamefont {Lombardo}, \citenamefont {Bruna}, \citenamefont
  {Pfeiffer}, \citenamefont {West}, \citenamefont {Ferrari}, \citenamefont
  {Polini},\ and\ \citenamefont {Pellegrini}}]{HybridDrag}%
  \BibitemOpen
  \bibfield  {author} {\bibinfo {author} {\bibfnamefont {A.}~\bibnamefont
  {Gamucci}}, \bibinfo {author} {\bibfnamefont {D.}~\bibnamefont {Spirito}},
  \bibinfo {author} {\bibfnamefont {M.}~\bibnamefont {Carrega}}, \bibinfo
  {author} {\bibfnamefont {B.}~\bibnamefont {Karmakar}}, \bibinfo {author}
  {\bibfnamefont {A.}~\bibnamefont {Lombardo}}, \bibinfo {author}
  {\bibfnamefont {M.}~\bibnamefont {Bruna}}, \bibinfo {author} {\bibfnamefont
  {L.~N.}\ \bibnamefont {Pfeiffer}}, \bibinfo {author} {\bibfnamefont {K.~W.}\
  \bibnamefont {West}}, \bibinfo {author} {\bibfnamefont {A.~C.}\ \bibnamefont
  {Ferrari}}, \bibinfo {author} {\bibfnamefont {M.}~\bibnamefont {Polini}}, \
  and\ \bibinfo {author} {\bibfnamefont {V.}~\bibnamefont {Pellegrini}},\
  }\bibfield  {title} {\bibinfo {title} {\emph {Anomalous low-temperature
  Coulomb drag in graphene-GaAs heterostructures}},\ }\href {\doibase
  10.1038/ncomms6824} {\bibfield  {journal} {\bibinfo  {journal} {Nature
  Communications}\ }\textbf {\bibinfo {volume} {5}},\ \bibinfo {pages} {5824}
  (\bibinfo {year} {2014})}\BibitemShut {NoStop}%
\bibitem [{\citenamefont {Sodemann}\ \emph {et~al.}(2012)\citenamefont
  {Sodemann}, \citenamefont {Pesin},\ and\ \citenamefont
  {MacDonald}}]{EHMultiband1}%
  \BibitemOpen
  \bibfield  {author} {\bibinfo {author} {\bibfnamefont {I.}~\bibnamefont
  {Sodemann}}, \bibinfo {author} {\bibfnamefont {D.~A.}\ \bibnamefont {Pesin}},
  \ and\ \bibinfo {author} {\bibfnamefont {A.~H.}\ \bibnamefont {MacDonald}},\
  }\bibfield  {title} {\bibinfo {title} {\emph {Interaction-enhanced coherence
  between two-dimensional Dirac layers}},\ }\href {\doibase
  10.1103/PhysRevB.85.195136} {\bibfield  {journal} {\bibinfo  {journal} {Phys.
  Rev. B}\ }\textbf {\bibinfo {volume} {85}},\ \bibinfo {pages} {195136}
  (\bibinfo {year} {2012})}\BibitemShut {NoStop}%
\bibitem [{\citenamefont {Lozovik}\ and\ \citenamefont
  {Sokolik}(2010)}]{EHMultiband2}%
  \BibitemOpen
  \bibfield  {author} {\bibinfo {author} {\bibfnamefont {Y.~E.}\ \bibnamefont
  {Lozovik}}\ and\ \bibinfo {author} {\bibfnamefont {A.~A.}\ \bibnamefont
  {Sokolik}},\ }\bibfield  {title} {\bibinfo {title} {\emph {Ultrarelativistic
  electron-hole pairing in graphene bilayer}},\ }\href {\doibase
  10.1140/epjb/e2009-00415-9} {\bibfield  {journal} {\bibinfo  {journal} {The
  European Physical Journal B}\ }\textbf {\bibinfo {volume} {73}},\ \bibinfo
  {pages} {195} (\bibinfo {year} {2010})}\BibitemShut {NoStop}%
\bibitem [{\citenamefont {Conti}\ \emph {et~al.}(2019)\citenamefont {Conti},
  \citenamefont {Perali}, \citenamefont {Peeters},\ and\ \citenamefont
  {Neilson}}]{EHMultiband3}%
  \BibitemOpen
  \bibfield  {author} {\bibinfo {author} {\bibfnamefont {S.}~\bibnamefont
  {Conti}}, \bibinfo {author} {\bibfnamefont {A.}~\bibnamefont {Perali}},
  \bibinfo {author} {\bibfnamefont {F.~M.}\ \bibnamefont {Peeters}}, \ and\
  \bibinfo {author} {\bibfnamefont {D.}~\bibnamefont {Neilson}},\ }\bibfield
  {title} {\bibinfo {title} {\emph {Multicomponent screening and superfluidity
  in gapped electron-hole double bilayer graphene with realistic bands}},\
  }\href {\doibase 10.1103/PhysRevB.99.144517} {\bibfield  {journal} {\bibinfo
  {journal} {Phys. Rev. B}\ }\textbf {\bibinfo {volume} {99}},\ \bibinfo
  {pages} {144517} (\bibinfo {year} {2019})}\BibitemShut {NoStop}%
\bibitem [{Note2()}]{Note2}%
  \BibitemOpen
  \bibinfo {note} {These arguments are valid in the presence of substrate
  induced gap if is magnitude is small}\BibitemShut {NoStop}%
\bibitem [{Note3()}]{Note3}%
  \BibitemOpen
  \bibinfo {note} {It should be noted that after the formal electron-hole
  transformation the connections to the BCS theory of superconductivity
  established in early works on electron-hole Cooper pair condensation~\cite
  {LozovikYudson1,LozovikYudson2,Shevchenko} become apparent.}\BibitemShut
  {Stop}%
\bibitem [{Note4()}]{Note4}%
  \BibitemOpen
  \bibinfo {note} {The presence of two (almost) degenerate competing channels
  also occurs for other hybrid double-layer systems, including bilayer
  graphene-GaAs and monolayer graphene-bilayer graphene}\BibitemShut {NoStop}%
\bibitem [{Note5()}]{Note5}%
  \BibitemOpen
  \bibinfo {note} {In a graphene/graphene bilayer, the angle factor must be
  modified as $\Lambda _{\protect \mathbf {p}\protect \mathbf {p'}}\rightarrow
  \Lambda _{\protect \mathbf {p}\protect \mathbf {p'}}^2$ for intravalley CPs
  and as $\Lambda _{\protect \mathbf {p}\protect \mathbf {p'}}\rightarrow
  |\Lambda _{\protect \mathbf {p}\protect \mathbf {p'}}|^2$ for intervalley
  CPs. In the absence of sublattice asymmetry, $\delta =0$, the modified angle
  factor is the same, $\Lambda _{\protect \mathbf {p}\protect \mathbf
  {p'}}=\protect \qopname \relax o{cos}^2(\phi _{\protect \mathbf {p},\protect
  \mathbf {p}'}/2)$}\BibitemShut {NoStop}%
\bibitem [{Note6()}]{Note6}%
  \BibitemOpen
  \bibinfo {note} {It is instructive to rescale the fields as $\Delta ^\protect
  \mathrm {A}\rightarrow \Delta ^\protect \mathrm {A}/\protect \sqrt {\nu _0}
  s_\protect \mathrm {F}$ and $\Delta ^\protect \mathrm {B}\rightarrow \Delta
  ^\protect \mathrm {B}/\protect \sqrt {\nu _0} c_\protect \mathrm {F}$. See
  Appendix A for detailed derivations.}\BibitemShut {Stop}%
\bibitem [{\citenamefont {Fulde}\ and\ \citenamefont {Ferrell}(1964)}]{FF}%
  \BibitemOpen
  \bibfield  {author} {\bibinfo {author} {\bibfnamefont {P.}~\bibnamefont
  {Fulde}}\ and\ \bibinfo {author} {\bibfnamefont {R.~A.}\ \bibnamefont
  {Ferrell}},\ }\bibfield  {title} {\bibinfo {title} {\emph {Superconductivity
  in a Strong Spin-Exchange Field}},\ }\href {\doibase
  10.1103/PhysRev.135.A550} {\bibfield  {journal} {\bibinfo  {journal} {Phys.
  Rev.}\ }\textbf {\bibinfo {volume} {135}},\ \bibinfo {pages} {A550} (\bibinfo
  {year} {1964})}\BibitemShut {NoStop}%
\bibitem [{\citenamefont {Larkin}\ and\ \citenamefont {Ovchinikov}(1965)}]{LO}%
  \BibitemOpen
  \bibfield  {author} {\bibinfo {author} {\bibfnamefont {I.}~\bibnamefont
  {Larkin}}\ and\ \bibinfo {author} {\bibfnamefont {Y.}~\bibnamefont
  {Ovchinikov}},\ }\bibfield  {title} {\bibinfo {title} {\emph {Nonuniform
  state of superconductors}},\ }\href@noop {} {\bibfield  {journal} {\bibinfo
  {journal} {Sov. Phys. JETP}\ }\textbf {\bibinfo {volume} {20}},\ \bibinfo
  {pages} {762} (\bibinfo {year} {1965})}\BibitemShut {NoStop}%
\bibitem [{\citenamefont {Seradjeh}(2012)}]{LOFFSeradjeh}%
  \BibitemOpen
  \bibfield  {author} {\bibinfo {author} {\bibfnamefont {B.}~\bibnamefont
  {Seradjeh}},\ }\bibfield  {title} {\bibinfo {title} {\emph {Topological
  exciton condensate of imbalanced electrons and holes}},\ }\href {\doibase
  10.1103/PhysRevB.85.235146} {\bibfield  {journal} {\bibinfo  {journal} {Phys.
  Rev. B}\ }\textbf {\bibinfo {volume} {85}},\ \bibinfo {pages} {235146}
  (\bibinfo {year} {2012})}\BibitemShut {NoStop}%
\bibitem [{\citenamefont {Efimkin}\ and\ \citenamefont
  {Lozovik}(2011)}]{EfimkinLOFF}%
  \BibitemOpen
  \bibfield  {author} {\bibinfo {author} {\bibfnamefont {D.~K.}\ \bibnamefont
  {Efimkin}}\ and\ \bibinfo {author} {\bibfnamefont {Y.}~\bibnamefont
  {Lozovik}},\ }\bibfield  {title} {\bibinfo {title} {\emph {Electron-hole
  pairing with nonzero momentum in a graphene bilayer}},\ }\href {\doibase
  10.1134/S1063776111130048} {\bibfield  {journal} {\bibinfo  {journal} {JETP}\
  }\textbf {\bibinfo {volume} {113}},\ \bibinfo {pages} {880} (\bibinfo {year}
  {2011})}\BibitemShut {NoStop}%
\bibitem [{\citenamefont {Page}(1987)}]{QFT1}%
  \BibitemOpen
  \bibfield  {author} {\bibinfo {author} {\bibfnamefont {D.~N.}\ \bibnamefont
  {Page}},\ }\bibfield  {title} {\bibinfo {title} {\emph {Geometrical
  description of Berry's phase}},\ }\href {\doibase 10.1103/PhysRevA.36.3479}
  {\bibfield  {journal} {\bibinfo  {journal} {Phys. Rev. A}\ }\textbf {\bibinfo
  {volume} {36}},\ \bibinfo {pages} {3479} (\bibinfo {year}
  {1987})}\BibitemShut {NoStop}%
\bibitem [{\citenamefont {Rossi}(2021)}]{QFT2}%
  \BibitemOpen
  \bibfield  {author} {\bibinfo {author} {\bibfnamefont {E.}~\bibnamefont
  {Rossi}},\ }\bibfield  {title} {\bibinfo {title} {\emph {Quantum metric and
  correlated states in two-dimensional systems}},\ }\href {\doibase
  https://doi.org/10.1016/j.cossms.2021.100952} {\bibfield  {journal} {\bibinfo
   {journal} {Current Opinion in Solid State and Materials Science}\ }\textbf
  {\bibinfo {volume} {25}},\ \bibinfo {pages} {100952} (\bibinfo {year}
  {2021})}\BibitemShut {NoStop}%
\bibitem [{\citenamefont {Bleu}\ \emph {et~al.}(2018)\citenamefont {Bleu},
  \citenamefont {Solnyshkov},\ and\ \citenamefont {Malpuech}}]{QFT3}%
  \BibitemOpen
  \bibfield  {author} {\bibinfo {author} {\bibfnamefont {O.}~\bibnamefont
  {Bleu}}, \bibinfo {author} {\bibfnamefont {D.~D.}\ \bibnamefont
  {Solnyshkov}}, \ and\ \bibinfo {author} {\bibfnamefont {G.}~\bibnamefont
  {Malpuech}},\ }\bibfield  {title} {\bibinfo {title} {\emph {Measuring the
  quantum geometric tensor in two-dimensional photonic and exciton-polariton
  systems}},\ }\href {\doibase 10.1103/PhysRevB.97.195422} {\bibfield
  {journal} {\bibinfo  {journal} {Phys. Rev. B}\ }\textbf {\bibinfo {volume}
  {97}},\ \bibinfo {pages} {195422} (\bibinfo {year} {2018})}\BibitemShut
  {NoStop}%
\bibitem [{\citenamefont {Nagaosa}\ \emph
  {et~al.}(2010{\natexlab{b}})\citenamefont {Nagaosa}, \citenamefont {Sinova},
  \citenamefont {Onoda}, \citenamefont {MacDonald},\ and\ \citenamefont
  {Ong}}]{AHEReview1}%
  \BibitemOpen
  \bibfield  {author} {\bibinfo {author} {\bibfnamefont {N.}~\bibnamefont
  {Nagaosa}}, \bibinfo {author} {\bibfnamefont {J.}~\bibnamefont {Sinova}},
  \bibinfo {author} {\bibfnamefont {S.}~\bibnamefont {Onoda}}, \bibinfo
  {author} {\bibfnamefont {A.~H.}\ \bibnamefont {MacDonald}}, \ and\ \bibinfo
  {author} {\bibfnamefont {N.~P.}\ \bibnamefont {Ong}},\ }\bibfield  {title}
  {\bibinfo {title} {\emph {Anomalous Hall effect}},\ }\href {\doibase
  10.1103/RevModPhys.82.1539} {\bibfield  {journal} {\bibinfo  {journal} {Rev.
  Mod. Phys.}\ }\textbf {\bibinfo {volume} {82}},\ \bibinfo {pages} {1539}
  (\bibinfo {year} {2010}{\natexlab{b}})}\BibitemShut {NoStop}%
\bibitem [{\citenamefont {Borunda}\ \emph {et~al.}(2007)\citenamefont
  {Borunda}, \citenamefont {Nunner}, \citenamefont {L\"uck}, \citenamefont
  {Sinitsyn}, \citenamefont {Timm}, \citenamefont {Wunderlich}, \citenamefont
  {Jungwirth}, \citenamefont {MacDonald},\ and\ \citenamefont {Sinova}}]{AHE2}%
  \BibitemOpen
  \bibfield  {author} {\bibinfo {author} {\bibfnamefont {M.}~\bibnamefont
  {Borunda}}, \bibinfo {author} {\bibfnamefont {T.~S.}\ \bibnamefont {Nunner}},
  \bibinfo {author} {\bibfnamefont {T.}~\bibnamefont {L\"uck}}, \bibinfo
  {author} {\bibfnamefont {N.~A.}\ \bibnamefont {Sinitsyn}}, \bibinfo {author}
  {\bibfnamefont {C.}~\bibnamefont {Timm}}, \bibinfo {author} {\bibfnamefont
  {J.}~\bibnamefont {Wunderlich}}, \bibinfo {author} {\bibfnamefont
  {T.}~\bibnamefont {Jungwirth}}, \bibinfo {author} {\bibfnamefont {A.~H.}\
  \bibnamefont {MacDonald}}, \ and\ \bibinfo {author} {\bibfnamefont
  {J.}~\bibnamefont {Sinova}},\ }\bibfield  {title} {\bibinfo {title} {\emph
  {Absence of Skew Scattering in Two-Dimensional Systems: Testing the Origins
  of the Anomalous Hall Effect}},\ }\href {\doibase
  10.1103/PhysRevLett.99.066604} {\bibfield  {journal} {\bibinfo  {journal}
  {Phys. Rev. Lett.}\ }\textbf {\bibinfo {volume} {99}},\ \bibinfo {pages}
  {066604} (\bibinfo {year} {2007})}\BibitemShut {NoStop}%
\bibitem [{\citenamefont {Culcer}\ \emph {et~al.}(2017)\citenamefont {Culcer},
  \citenamefont {Sekine},\ and\ \citenamefont {MacDonald}}]{AHE3}%
  \BibitemOpen
  \bibfield  {author} {\bibinfo {author} {\bibfnamefont {D.}~\bibnamefont
  {Culcer}}, \bibinfo {author} {\bibfnamefont {A.}~\bibnamefont {Sekine}}, \
  and\ \bibinfo {author} {\bibfnamefont {A.~H.}\ \bibnamefont {MacDonald}},\
  }\bibfield  {title} {\bibinfo {title} {\emph {Interband coherence response to
  electric fields in crystals: Berry-phase contributions and disorder
  effects}},\ }\href {\doibase 10.1103/PhysRevB.96.035106} {\bibfield
  {journal} {\bibinfo  {journal} {Phys. Rev. B}\ }\textbf {\bibinfo {volume}
  {96}},\ \bibinfo {pages} {035106} (\bibinfo {year} {2017})}\BibitemShut
  {NoStop}%
\bibitem [{Note7()}]{Note7}%
  \BibitemOpen
  \bibinfo {note} {It should be noted that a valley current is induced and can
  be detected via nonlocal measurements~\cite
  {VelleyCurrent1,VelleyCurrent2}}\BibitemShut {NoStop}%
\bibitem [{\citenamefont {L\'opez~R\'{\i}os}\ \emph {et~al.}(2018)\citenamefont
  {L\'opez~R\'{\i}os}, \citenamefont {Perali}, \citenamefont {Needs},\ and\
  \citenamefont {Neilson}}]{EHCros}%
  \BibitemOpen
  \bibfield  {author} {\bibinfo {author} {\bibfnamefont {P.}~\bibnamefont
  {L\'opez~R\'{\i}os}}, \bibinfo {author} {\bibfnamefont {A.}~\bibnamefont
  {Perali}}, \bibinfo {author} {\bibfnamefont {R.~J.}\ \bibnamefont {Needs}}, \
  and\ \bibinfo {author} {\bibfnamefont {D.}~\bibnamefont {Neilson}},\
  }\bibfield  {title} {\bibinfo {title} {\emph {Evidence from Quantum Monte
  Carlo Simulations of Large-Gap Superfluidity and BCS-BEC Crossover in Double
  Electron-Hole Layers}},\ }\href {\doibase 10.1103/PhysRevLett.120.177701}
  {\bibfield  {journal} {\bibinfo  {journal} {Phys. Rev. Lett.}\ }\textbf
  {\bibinfo {volume} {120}},\ \bibinfo {pages} {177701} (\bibinfo {year}
  {2018})}\BibitemShut {NoStop}%
\bibitem [{\citenamefont {Abergel}\ and\ \citenamefont
  {Chakraborty}(2009)}]{ValleyPOlarization1}%
  \BibitemOpen
  \bibfield  {author} {\bibinfo {author} {\bibfnamefont {D.~S.~L.}\
  \bibnamefont {Abergel}}\ and\ \bibinfo {author} {\bibfnamefont
  {T.}~\bibnamefont {Chakraborty}},\ }\bibfield  {title} {\bibinfo {title}
  {\emph {Generation of valley polarized current in bilayer graphene}},\ }\href
  {\doibase 10.1063/1.3205117} {\bibfield  {journal} {\bibinfo  {journal}
  {Applied Physics Letters}\ }\textbf {\bibinfo {volume} {95}},\ \bibinfo
  {pages} {062107} (\bibinfo {year} {2009})},\ \Eprint
  {http://arxiv.org/abs/https://doi.org/10.1063/1.3205117}
  {https://doi.org/10.1063/1.3205117} \BibitemShut {NoStop}%
\bibitem [{\citenamefont {Yao}\ \emph {et~al.}(2008)\citenamefont {Yao},
  \citenamefont {Xiao},\ and\ \citenamefont {Niu}}]{ValleyPOlarization2}%
  \BibitemOpen
  \bibfield  {author} {\bibinfo {author} {\bibfnamefont {W.}~\bibnamefont
  {Yao}}, \bibinfo {author} {\bibfnamefont {D.}~\bibnamefont {Xiao}}, \ and\
  \bibinfo {author} {\bibfnamefont {Q.}~\bibnamefont {Niu}},\ }\bibfield
  {title} {\bibinfo {title} {\emph {Valley-dependent optoelectronics from
  inversion symmetry breaking}},\ }\href {\doibase 10.1103/PhysRevB.77.235406}
  {\bibfield  {journal} {\bibinfo  {journal} {Phys. Rev. B}\ }\textbf {\bibinfo
  {volume} {77}},\ \bibinfo {pages} {235406} (\bibinfo {year}
  {2008})}\BibitemShut {NoStop}%
\bibitem [{\citenamefont {Farajollahpour}\ and\ \citenamefont
  {Phirouznia}(2017)}]{ValleyPOlarization3}%
  \BibitemOpen
  \bibfield  {author} {\bibinfo {author} {\bibfnamefont {T.}~\bibnamefont
  {Farajollahpour}}\ and\ \bibinfo {author} {\bibfnamefont {A.}~\bibnamefont
  {Phirouznia}},\ }\bibfield  {title} {\bibinfo {title} {\emph {The role of the
  strain induced population imbalance in Valley polarization of graphene: Berry
  curvature perspective}},\ }\href {\doibase 10.1038/s41598-017-18238-5}
  {\bibfield  {journal} {\bibinfo  {journal} {Scientific Reports}\ }\textbf
  {\bibinfo {volume} {7}},\ \bibinfo {pages} {17878} (\bibinfo {year}
  {2017})}\BibitemShut {NoStop}%
\bibitem [{\citenamefont {Liu}\ \emph {et~al.}(2020)\citenamefont {Liu},
  \citenamefont {Wang}, \citenamefont {Li}, \citenamefont {Wu}, \citenamefont
  {Li}, \citenamefont {Li}, \citenamefont {He}, \citenamefont {Xu},
  \citenamefont {Zhang},\ and\ \citenamefont {Wang}}]{MagneticTI1}%
  \BibitemOpen
  \bibfield  {author} {\bibinfo {author} {\bibfnamefont {C.}~\bibnamefont
  {Liu}}, \bibinfo {author} {\bibfnamefont {Y.}~\bibnamefont {Wang}}, \bibinfo
  {author} {\bibfnamefont {H.}~\bibnamefont {Li}}, \bibinfo {author}
  {\bibfnamefont {Y.}~\bibnamefont {Wu}}, \bibinfo {author} {\bibfnamefont
  {Y.}~\bibnamefont {Li}}, \bibinfo {author} {\bibfnamefont {J.}~\bibnamefont
  {Li}}, \bibinfo {author} {\bibfnamefont {K.}~\bibnamefont {He}}, \bibinfo
  {author} {\bibfnamefont {Y.}~\bibnamefont {Xu}}, \bibinfo {author}
  {\bibfnamefont {J.}~\bibnamefont {Zhang}}, \ and\ \bibinfo {author}
  {\bibfnamefont {Y.}~\bibnamefont {Wang}},\ }\bibfield  {title} {\bibinfo
  {title} {\emph {Robust axion insulator and Chern insulator phases in a
  two-dimensional antiferromagnetic topological insulator}},\ }\href {\doibase
  10.1038/s41563-019-0573-3} {\bibfield  {journal} {\bibinfo  {journal} {Nature
  Materials}\ }\textbf {\bibinfo {volume} {19}},\ \bibinfo {pages} {522}
  (\bibinfo {year} {2020})}\BibitemShut {NoStop}%
\bibitem [{\citenamefont {Chen}\ \emph {et~al.}(2010)\citenamefont {Chen},
  \citenamefont {Chu}, \citenamefont {Analytis}, \citenamefont {Liu},
  \citenamefont {Igarashi}, \citenamefont {Kuo}, \citenamefont {Qi},
  \citenamefont {Mo}, \citenamefont {Moore}, \citenamefont {Lu}, \citenamefont
  {Hashimoto}, \citenamefont {Sasagawa}, \citenamefont {Zhang}, \citenamefont
  {Fisher}, \citenamefont {Hussain},\ and\ \citenamefont {Shen}}]{MagneticTI2}%
  \BibitemOpen
  \bibfield  {author} {\bibinfo {author} {\bibfnamefont {Y.~L.}\ \bibnamefont
  {Chen}}, \bibinfo {author} {\bibfnamefont {J.-H.}\ \bibnamefont {Chu}},
  \bibinfo {author} {\bibfnamefont {J.~G.}\ \bibnamefont {Analytis}}, \bibinfo
  {author} {\bibfnamefont {Z.~K.}\ \bibnamefont {Liu}}, \bibinfo {author}
  {\bibfnamefont {K.}~\bibnamefont {Igarashi}}, \bibinfo {author}
  {\bibfnamefont {H.-H.}\ \bibnamefont {Kuo}}, \bibinfo {author} {\bibfnamefont
  {X.~L.}\ \bibnamefont {Qi}}, \bibinfo {author} {\bibfnamefont {S.~K.}\
  \bibnamefont {Mo}}, \bibinfo {author} {\bibfnamefont {R.~G.}\ \bibnamefont
  {Moore}}, \bibinfo {author} {\bibfnamefont {D.~H.}\ \bibnamefont {Lu}},
  \bibinfo {author} {\bibfnamefont {M.}~\bibnamefont {Hashimoto}}, \bibinfo
  {author} {\bibfnamefont {T.}~\bibnamefont {Sasagawa}}, \bibinfo {author}
  {\bibfnamefont {S.~C.}\ \bibnamefont {Zhang}}, \bibinfo {author}
  {\bibfnamefont {I.~R.}\ \bibnamefont {Fisher}}, \bibinfo {author}
  {\bibfnamefont {Z.}~\bibnamefont {Hussain}}, \ and\ \bibinfo {author}
  {\bibfnamefont {Z.~X.}\ \bibnamefont {Shen}},\ }\bibfield  {title} {\bibinfo
  {title} {\emph {Massive Dirac Fermion on the Surface of a Magnetically Doped
  Topological Insulator}},\ }\href {\doibase 10.1126/science.1189924}
  {\bibfield  {journal} {\bibinfo  {journal} {Science}\ }\textbf {\bibinfo
  {volume} {329}},\ \bibinfo {pages} {659} (\bibinfo {year}
  {2010})}\BibitemShut {NoStop}%
\bibitem [{\citenamefont {Deng}\ \emph {et~al.}(2020)\citenamefont {Deng},
  \citenamefont {Yu}, \citenamefont {Shi}, \citenamefont {Guo}, \citenamefont
  {Xu}, \citenamefont {Wang}, \citenamefont {Chen},\ and\ \citenamefont
  {Zhang}}]{MagneticTI3}%
  \BibitemOpen
  \bibfield  {author} {\bibinfo {author} {\bibfnamefont {Y.}~\bibnamefont
  {Deng}}, \bibinfo {author} {\bibfnamefont {Y.}~\bibnamefont {Yu}}, \bibinfo
  {author} {\bibfnamefont {M.~Z.}\ \bibnamefont {Shi}}, \bibinfo {author}
  {\bibfnamefont {Z.}~\bibnamefont {Guo}}, \bibinfo {author} {\bibfnamefont
  {Z.}~\bibnamefont {Xu}}, \bibinfo {author} {\bibfnamefont {J.}~\bibnamefont
  {Wang}}, \bibinfo {author} {\bibfnamefont {X.~H.}\ \bibnamefont {Chen}}, \
  and\ \bibinfo {author} {\bibfnamefont {Y.}~\bibnamefont {Zhang}},\ }\bibfield
   {title} {\bibinfo {title} {\emph {Quantum anomalous Hall effect in intrinsic
  magnetic topological insulator MnBi2Te4}},\ }\href {\doibase
  10.1126/science.aax8156} {\bibfield  {journal} {\bibinfo  {journal}
  {Science}\ }\textbf {\bibinfo {volume} {367}},\ \bibinfo {pages} {895}
  (\bibinfo {year} {2020})}\BibitemShut {NoStop}%
\bibitem [{\citenamefont {Trang}\ \emph {et~al.}(2021)\citenamefont {Trang},
  \citenamefont {Li}, \citenamefont {Yin}, \citenamefont {Hwang}, \citenamefont
  {Akhgar}, \citenamefont {Di~Bernardo}, \citenamefont
  {Grubi{\v{s}}i{\'{c}}-{\v{C}}abo}, \citenamefont {Tadich}, \citenamefont
  {Fuhrer}, \citenamefont {Mo}, \citenamefont {Medhekar},\ and\ \citenamefont
  {Edmonds}}]{MagneticTI4}%
  \BibitemOpen
  \bibfield  {author} {\bibinfo {author} {\bibfnamefont {C.~X.}\ \bibnamefont
  {Trang}}, \bibinfo {author} {\bibfnamefont {Q.}~\bibnamefont {Li}}, \bibinfo
  {author} {\bibfnamefont {Y.}~\bibnamefont {Yin}}, \bibinfo {author}
  {\bibfnamefont {J.}~\bibnamefont {Hwang}}, \bibinfo {author} {\bibfnamefont
  {G.}~\bibnamefont {Akhgar}}, \bibinfo {author} {\bibfnamefont
  {I.}~\bibnamefont {Di~Bernardo}}, \bibinfo {author} {\bibfnamefont
  {A.}~\bibnamefont {Grubi{\v{s}}i{\'{c}}-{\v{C}}abo}}, \bibinfo {author}
  {\bibfnamefont {A.}~\bibnamefont {Tadich}}, \bibinfo {author} {\bibfnamefont
  {M.~S.}\ \bibnamefont {Fuhrer}}, \bibinfo {author} {\bibfnamefont {S.-K.}\
  \bibnamefont {Mo}}, \bibinfo {author} {\bibfnamefont {N.~V.}\ \bibnamefont
  {Medhekar}}, \ and\ \bibinfo {author} {\bibfnamefont {M.~T.}\ \bibnamefont
  {Edmonds}},\ }\bibfield  {title} {\bibinfo {title} {\emph {Crossover from 2D
  Ferromagnetic Insulator to Wide Band Gap Quantum Anomalous Hall Insulator in
  Ultrathin MnBi2Te4}},\ }\href {\doibase 10.1021/acsnano.1c03936} {\bibfield
  {journal} {\bibinfo  {journal} {ACS Nano}\ }\textbf {\bibinfo {volume}
  {15}},\ \bibinfo {pages} {13444} (\bibinfo {year} {2021})}\BibitemShut
  {NoStop}%
\bibitem [{\citenamefont {Karzig}\ \emph {et~al.}(2015)\citenamefont {Karzig},
  \citenamefont {Bardyn}, \citenamefont {Lindner},\ and\ \citenamefont
  {Refael}}]{Topolaritons1}%
  \BibitemOpen
  \bibfield  {author} {\bibinfo {author} {\bibfnamefont {T.}~\bibnamefont
  {Karzig}}, \bibinfo {author} {\bibfnamefont {C.-E.}\ \bibnamefont {Bardyn}},
  \bibinfo {author} {\bibfnamefont {N.~H.}\ \bibnamefont {Lindner}}, \ and\
  \bibinfo {author} {\bibfnamefont {G.}~\bibnamefont {Refael}},\ }\bibfield
  {title} {\bibinfo {title} {\emph {Topological Polaritons}},\ }\href {\doibase
  10.1103/PhysRevX.5.031001} {\bibfield  {journal} {\bibinfo  {journal} {Phys.
  Rev. X}\ }\textbf {\bibinfo {volume} {5}},\ \bibinfo {pages} {031001}
  (\bibinfo {year} {2015})}\BibitemShut {NoStop}%
\bibitem [{\citenamefont {Nalitov}\ \emph {et~al.}(2015)\citenamefont
  {Nalitov}, \citenamefont {Solnyshkov},\ and\ \citenamefont
  {Malpuech}}]{Topolaritons2}%
  \BibitemOpen
  \bibfield  {author} {\bibinfo {author} {\bibfnamefont {A.~V.}\ \bibnamefont
  {Nalitov}}, \bibinfo {author} {\bibfnamefont {D.~D.}\ \bibnamefont
  {Solnyshkov}}, \ and\ \bibinfo {author} {\bibfnamefont {G.}~\bibnamefont
  {Malpuech}},\ }\bibfield  {title} {\bibinfo {title} {\emph {Polariton
  $\mathbb{Z}$ Topological Insulator}},\ }\href {\doibase
  10.1103/PhysRevLett.114.116401} {\bibfield  {journal} {\bibinfo  {journal}
  {Phys. Rev. Lett.}\ }\textbf {\bibinfo {volume} {114}},\ \bibinfo {pages}
  {116401} (\bibinfo {year} {2015})}\BibitemShut {NoStop}%
\bibitem [{\citenamefont {Kinnunen}\ \emph {et~al.}(2018)\citenamefont
  {Kinnunen}, \citenamefont {Baarsma}, \citenamefont {Martikainen},\ and\
  \citenamefont {Torma}}]{FFLOReviewColdAtoms1}%
  \BibitemOpen
  \bibfield  {author} {\bibinfo {author} {\bibfnamefont {J.~J.}\ \bibnamefont
  {Kinnunen}}, \bibinfo {author} {\bibfnamefont {J.~E.}\ \bibnamefont
  {Baarsma}}, \bibinfo {author} {\bibfnamefont {J.-P.}\ \bibnamefont
  {Martikainen}}, \ and\ \bibinfo {author} {\bibfnamefont {P.}~\bibnamefont
  {Torma}},\ }\bibfield  {title} {\bibinfo {title} {\emph {The
  Fulde{\textendash}Ferrell{\textendash}Larkin{\textendash}Ovchinnikov state
  for ultracold fermions in lattice and harmonic potentials: a review}},\
  }\href {\doibase 10.1088/1361-6633/aaa4ad} {\bibfield  {journal} {\bibinfo
  {journal} {Reports on Progress in Physics}\ }\textbf {\bibinfo {volume}
  {81}},\ \bibinfo {pages} {046401} (\bibinfo {year} {2018})}\BibitemShut
  {NoStop}%
\bibitem [{\citenamefont {Gubbels}\ and\ \citenamefont
  {Stoof}(2013)}]{FFLOReviewColdAtoms2}%
  \BibitemOpen
  \bibfield  {author} {\bibinfo {author} {\bibfnamefont {K.}~\bibnamefont
  {Gubbels}}\ and\ \bibinfo {author} {\bibfnamefont {H.}~\bibnamefont
  {Stoof}},\ }\href {\doibase https://doi.org/10.1016/j.physrep.2012.11.004}
  {\bibfield  {journal} {\bibinfo  {journal} {Physics Reports}\ }\textbf
  {\bibinfo {volume} {525}},\ \bibinfo {pages} {255 } (\bibinfo {year}
  {2013})},\ \bibinfo {note} {imbalanced Fermi Gases at Unitarity}\BibitemShut
  {NoStop}%
\bibitem [{\citenamefont {Peotta}\ and\ \citenamefont
  {T{\"o}rm{\"a}}(2015)}]{QuantumMetricSuperfluid1}%
  \BibitemOpen
  \bibfield  {author} {\bibinfo {author} {\bibfnamefont {S.}~\bibnamefont
  {Peotta}}\ and\ \bibinfo {author} {\bibfnamefont {P.}~\bibnamefont
  {T{\"o}rm{\"a}}},\ }\bibfield  {title} {\bibinfo {title} {\emph
  {Superfluidity in topologically nontrivial flat bands}},\ }\href@noop {}
  {\bibfield  {journal} {\bibinfo  {journal} {Nature Communications}\ }\textbf
  {\bibinfo {volume} {6}},\ \bibinfo {pages} {8944} (\bibinfo {year}
  {2015})}\BibitemShut {NoStop}%
\bibitem [{\citenamefont {Liang}\ \emph {et~al.}(2017)\citenamefont {Liang},
  \citenamefont {Vanhala}, \citenamefont {Peotta}, \citenamefont {Siro},
  \citenamefont {Harju},\ and\ \citenamefont
  {T\"orm\"a}}]{QuantumMetricSuperfluid2}%
  \BibitemOpen
  \bibfield  {author} {\bibinfo {author} {\bibfnamefont {L.}~\bibnamefont
  {Liang}}, \bibinfo {author} {\bibfnamefont {T.~I.}\ \bibnamefont {Vanhala}},
  \bibinfo {author} {\bibfnamefont {S.}~\bibnamefont {Peotta}}, \bibinfo
  {author} {\bibfnamefont {T.}~\bibnamefont {Siro}}, \bibinfo {author}
  {\bibfnamefont {A.}~\bibnamefont {Harju}}, \ and\ \bibinfo {author}
  {\bibfnamefont {P.}~\bibnamefont {T\"orm\"a}},\ }\bibfield  {title} {\bibinfo
  {title} {\emph {Band geometry, Berry curvature, and superfluid weight}},\
  }\href {\doibase 10.1103/PhysRevB.95.024515} {\bibfield  {journal} {\bibinfo
  {journal} {Phys. Rev. B}\ }\textbf {\bibinfo {volume} {95}},\ \bibinfo
  {pages} {024515} (\bibinfo {year} {2017})}\BibitemShut {NoStop}%
\bibitem [{\citenamefont {Wang}\ \emph {et~al.}(2020)\citenamefont {Wang},
  \citenamefont {Chaudhary}, \citenamefont {Chen},\ and\ \citenamefont
  {Levin}}]{QuantumMetricSuperfluid3}%
  \BibitemOpen
  \bibfield  {author} {\bibinfo {author} {\bibfnamefont {Z.}~\bibnamefont
  {Wang}}, \bibinfo {author} {\bibfnamefont {G.}~\bibnamefont {Chaudhary}},
  \bibinfo {author} {\bibfnamefont {Q.}~\bibnamefont {Chen}}, \ and\ \bibinfo
  {author} {\bibfnamefont {K.}~\bibnamefont {Levin}},\ }\bibfield  {title}
  {\bibinfo {title} {\emph {Quantum geometric contributions to the BKT
  transition: Beyond mean field theory}},\ }\href {\doibase
  10.1103/PhysRevB.102.184504} {\bibfield  {journal} {\bibinfo  {journal}
  {Phys. Rev. B}\ }\textbf {\bibinfo {volume} {102}},\ \bibinfo {pages}
  {184504} (\bibinfo {year} {2020})}\BibitemShut {NoStop}%
\bibitem [{\citenamefont {Iskin}(2018)}]{QuantumMetricSuperfluid4}%
  \BibitemOpen
  \bibfield  {author} {\bibinfo {author} {\bibfnamefont {M.}~\bibnamefont
  {Iskin}},\ }\bibfield  {title} {\bibinfo {title} {\emph {Quantum-metric
  contribution to the pair mass in spin-orbit-coupled Fermi superfluids}},\
  }\href {\doibase 10.1103/PhysRevA.97.033625} {\bibfield  {journal} {\bibinfo
  {journal} {Phys. Rev. A}\ }\textbf {\bibinfo {volume} {97}},\ \bibinfo
  {pages} {033625} (\bibinfo {year} {2018})}\BibitemShut {NoStop}%
\bibitem [{\citenamefont {Shimazaki}\ \emph {et~al.}(2015)\citenamefont
  {Shimazaki}, \citenamefont {Yamamoto}, \citenamefont {Borzenets},
  \citenamefont {Watanabe}, \citenamefont {Taniguchi},\ and\ \citenamefont
  {Tarucha}}]{VelleyCurrent1}%
  \BibitemOpen
  \bibfield  {author} {\bibinfo {author} {\bibfnamefont {Y.}~\bibnamefont
  {Shimazaki}}, \bibinfo {author} {\bibfnamefont {M.}~\bibnamefont {Yamamoto}},
  \bibinfo {author} {\bibfnamefont {I.~V.}\ \bibnamefont {Borzenets}}, \bibinfo
  {author} {\bibfnamefont {K.}~\bibnamefont {Watanabe}}, \bibinfo {author}
  {\bibfnamefont {T.}~\bibnamefont {Taniguchi}}, \ and\ \bibinfo {author}
  {\bibfnamefont {S.}~\bibnamefont {Tarucha}},\ }\bibfield  {title} {\bibinfo
  {title} {\emph {Generation and detection of pure valley current by
  electrically induced Berry curvature in bilayer graphene}},\ }\href {\doibase
  10.1038/nphys3551} {\bibfield  {journal} {\bibinfo  {journal} {Nature
  Physics}\ }\textbf {\bibinfo {volume} {11}},\ \bibinfo {pages} {1032}
  (\bibinfo {year} {2015})}\BibitemShut {NoStop}%
\bibitem [{\citenamefont {Schaibley}\ \emph {et~al.}(2016)\citenamefont
  {Schaibley}, \citenamefont {Yu}, \citenamefont {Clark}, \citenamefont
  {Rivera}, \citenamefont {Ross}, \citenamefont {Seyler}, \citenamefont {Yao},\
  and\ \citenamefont {Xu}}]{VelleyCurrent2}%
  \BibitemOpen
  \bibfield  {author} {\bibinfo {author} {\bibfnamefont {J.~R.}\ \bibnamefont
  {Schaibley}}, \bibinfo {author} {\bibfnamefont {H.}~\bibnamefont {Yu}},
  \bibinfo {author} {\bibfnamefont {G.}~\bibnamefont {Clark}}, \bibinfo
  {author} {\bibfnamefont {P.}~\bibnamefont {Rivera}}, \bibinfo {author}
  {\bibfnamefont {J.~S.}\ \bibnamefont {Ross}}, \bibinfo {author}
  {\bibfnamefont {K.~L.}\ \bibnamefont {Seyler}}, \bibinfo {author}
  {\bibfnamefont {W.}~\bibnamefont {Yao}}, \ and\ \bibinfo {author}
  {\bibfnamefont {X.}~\bibnamefont {Xu}},\ }\bibfield  {title} {\bibinfo
  {title} {\emph {Valleytronics in 2D materials}},\ }\href {\doibase
  10.1038/natrevmats.2016.55} {\bibfield  {journal} {\bibinfo  {journal}
  {Nature Reviews Materials}\ }\textbf {\bibinfo {volume} {1}},\ \bibinfo
  {pages} {16055} (\bibinfo {year} {2016})}\BibitemShut {NoStop}%
\bibitem [{\citenamefont {Wong}\ and\ \citenamefont
  {Tserkovnyak}(2011)}]{TserkovnyakKineticEq}%
  \BibitemOpen
  \bibfield  {author} {\bibinfo {author} {\bibfnamefont {C.~H.}\ \bibnamefont
  {Wong}}\ and\ \bibinfo {author} {\bibfnamefont {Y.}~\bibnamefont
  {Tserkovnyak}},\ }\bibfield  {title} {\bibinfo {title} {\emph {Quantum
  kinetic equation in phase-space textured multiband systems}},\ }\href
  {\doibase 10.1103/PhysRevB.84.115209} {\bibfield  {journal} {\bibinfo
  {journal} {Phys. Rev. B}\ }\textbf {\bibinfo {volume} {84}},\ \bibinfo
  {pages} {115209} (\bibinfo {year} {2011})}\BibitemShut {NoStop}%
\end{thebibliography}%
\newpage
\begin{widetext}

\begin{appendix}

\section{APPENDIX A. FIELD THEORETICAL APPROACH FOR FLUCTUATING COOPER PAIRS}
\subsection{A1. The model} 
This Appendix presents the quantum field theory approach to describe fluctuating Cooper pairs (CPs) in the hybrid graphene/quantum well bilayer. Following the reported experimental setup~\cite{HybridDrag}, we assume the presence of excess electrons in the quantum well (QW) and their deficit in graphene. The electrons can be described by a quantum field $e_{\tau \vec{r}}$ and their thermodynamic action represents an integral over the imaginary Matsubara time $\tau$ and is given by
\begin{equation}
S_\mathrm{e}=\int_0^\beta\int d\vec{r}\;\;e^\dagger_{\tau \vec{r}} \left(\partial_\tau+\frac{\vec{p}^2-p_\mathrm{F}^2}{2 m_\mathrm{e}} \right)  e_{\tau \vec{r}}.
\end{equation}
Electrons have the conventional quadratic dispersion with the mass $m_\mathrm{e}$ and $p_\mathrm{F}$ is their Fermi momentum. 
In the weak-to-moderate coupling regime the electron-hole correlations span only in the vicinity of the Fermi level. As a result, the spectrum of electrons can be linearized and their Green function can be approximated as
\begin{equation}
G^\mathrm{e}(i\omega_n,\vec{p})=\frac{1}{i \omega_n -\epsilon_\vec{p}^\mathrm{e}}, \quad \quad \quad \quad \epsilon_\vec{p}^\mathrm{e}=v_\mathrm{e} (p-p_\mathrm{F}),  \quad\quad \quad \quad v_\mathrm{e}=p_\mathrm{F}/m_\mathrm{h}.
\end{equation}
Here $\omega_n=(2n+1) \pi T$ is the fermionic Matsubara frequency. The low-energy electronic states in graphene are concentrated near two inequivalent valleys  ($\hbox{K}$ and $\hbox{K}'$), which are labeled by the index $\zeta =\pm 1$. They are described by the spinor field operator $h_{\vec{r}}=(h^{\mathrm{A}}_{\vec{r}},h^{\mathrm{B}}_{\vec{r}})$ and their pseudospin corresponds to the the sublattice (A and B) degree of freedom of the honeycomb lattice. We will assume the deficit of electrons in graphene and refer the valence band states as holes, however it is instructive not to perform the transformation to field operators of holes. As a result, the corresponding contribution to the action is given by 
\begin{equation}
\label{GrapheneAction}
S_\mathrm{h}=\int_0^\beta\int d\vec{r}\;\; \left[h^\dagger_{\tau \vec{r}} (\partial_\tau+v (\zeta p_x \sigma_x+ p_y \sigma_y) + \delta \sigma_z + \epsilon_\mathrm{F}^\mathrm{h})  h_{\tau \vec{r}}\right]. 
\end{equation}
Here $v$ is their velocity and $\epsilon_\mathrm{F}^\mathrm{h}$ is the Fermi energy. The small energy asymmetry between sublattices $\delta\ll \epsilon_\mathrm{F}^\mathrm{h}$ can be induced by the substrate engineering. The empty conduction band in graphene can be just truncated, while the energy and the spinor wave function for the valence band electrons are given by
\begin{equation}
\epsilon_\vec{p}^\mathrm{h}=-\sqrt{(vp)^2+\delta^2}, \quad \quad \quad 	|\vec{p}\rangle_\mathrm{h} =\begin{pmatrix} \sin\left(\theta_\vec{p}/2\right) e^{- i \zeta \phi_\vec{p}/2} \\ - \zeta \cos\left(\theta_\vec{p}/2\right) e^{i \zeta \phi_\vec{p}/2}\end{pmatrix}.
\end{equation}
Here $\phi_\vec{p}$ is the polar angle for vector $\vec{p}$ and $\cos(\theta_\vec{p})=\delta/\epsilon_\vec{p}$. The chirality of the pseudospin-momentum coupling for Dirac electrons is valley dependent but not their dispersion. We further assume that the Fermi momentum of electrons matches with the one $p_\mathrm{F}$ for holes, that is the most favorable regime for their Cooper pairing. As a result, the Green function for holes can be approximated as
\begin{equation}
\hat{G}^\mathrm{h}(i\omega_n,\vec{p})=\frac{|\vec{p}\rangle_{\mathrm{h}\;\mathrm{h}}\langle \vec{p}|}{i \omega_n -\epsilon_\vec{p}^\mathrm{h}}, \quad \quad \quad \quad \epsilon_\vec{p}^\mathrm{h}=v_\mathrm{h} (p-p_\mathrm{F}), \quad \quad \quad \quad v_\mathrm{h}=v^2 p_\mathrm{F}/\epsilon_\mathrm{F}^\mathrm{h} . 
\end{equation}
Here we have linearized  the spectrum of Dirac holes and the matrix $|\vec{p}\rangle_{\mathrm{h}\;\mathrm{h}}\langle \vec{p}|$ is given by 
\begin{equation}
 |\vec{p}\rangle_{\mathrm{h}\;\mathrm{h}}\langle \vec{p}| =  \begin{pmatrix} \sin^2\left(\theta_\vec{p}/2\right) & -\frac{\zeta}{2} \sin\left(\theta_\vec{p}\right) e^{- i \zeta \phi_\vec{p}}  \\ - \frac{\zeta}{2}\sin\left(\theta_\vec{p}\right)  e^{ i \zeta \phi_\vec{p}} &\cos^2\left(\theta_\vec{p}/2\right) \end{pmatrix}=\frac{1}{2 \epsilon_\vec{p}^{\mathrm{h}}} \begin{pmatrix} \epsilon_\vec{p}^{\mathrm{h}}-\delta & - v (\zeta p_x - i p_y ) \\ - v (\zeta p_x - i p_y ) & \epsilon_\vec{p}^{\mathrm{h}}+\delta \end{pmatrix}
\end{equation}
The phase intertwining between sublattices is the signature of the nontrivial band geometries for Dirac holes, which has been discussed in the main text and is passed on to the fluctuating CPs.  Fluctuating CPs are formed due to the attractive Coulomb interactions between electrons and holes. Being effectively screened, the interactions can be approximated by the contact pseudopotential with momentum-independent Fourier transform $V$. We also neglect the scattering between two valleys since they are well separated in momentum space. The resulting contribution to the action is given by 
\begin{equation}
S_{\mathrm{int}}=\int_0^\beta\int d\vec{r} \; V e^\dagger_{\tau \vec{r}} h^\dagger_{\tau \vec{r}}   h_{\tau \vec{r}}e_{\tau \vec{r}}\rightarrow  \int_0^\beta\int d\vec{r} \;\left[ \frac{1}{V} \Delta^\dagger_{\tau \vec{r}} \\\Delta_{\tau \vec{r}} + e^\dagger_{\tau \vec{r}}\Delta_{\tau \vec{r}} h_{\tau \vec{r}} + h^\dagger_{\tau \vec{r}} \Delta^\dagger_{\tau \vec{r}} e_{\tau \vec{r}}   \right].
\end{equation} 
Here we have performed the Hubbard-Stratonovich transformation. It eliminates interactions, but introduces the
bosonic field $\Delta_{\tau \vec{r}}=\{ \Delta^\mathrm{A}_{\tau \vec{r}}, \Delta^\mathrm{B}_{\tau \vec{r}} \}$, which describes the selective Cooper pairing with Dirac holes only at one of two sublattices (A and B). The action of the model $S=S_{\mathrm{e}}+S_{\mathrm{h}}+S_{\mathrm{int}}$ is the starting point for the derivation of fluctuating CPs dynamics, which is presented in the next subsection.    
\subsection{A2. Fluctuating Cooper pairs}
We further assume that the system is above the transition temperature. As a result there is no equilibrium Cooper pair condensate $\langle \Delta \rangle = 0$, but there are its Gaussian fluctuations, which are usually interpreted as fluctuating CPs. Integrating fermions out and expanding the resulting action up to the quadratic order in $\Delta_{\tau \vec{r}}$
results in
\begin{equation}
\label{AppSDelta}
S_\Delta=T \sum_{p_n \vec{q}}\bar{\Delta}_{p_n \vec{q}}\left( \frac{1}{V} - \hat{\Pi}(i p_n,\vec{q})\right) \Delta_{p_n \vec{q}}.\end{equation} 
Here $p_n=2\pi n T$ is the bosonic Matsubara frequency.  $\hat{\Pi}(i p_n,\vec{q})$ is the single step pair propagator in the Cooper ladder sum and is given by 
\begin{equation}
\label{AppPi1}
\hat{\Pi}(ip_n,\vec{q})=T \sum_{\omega_n, \vec{p}}  \hat{G}^\mathrm{e}(i \omega_n, \vec{p} ) G^\mathrm{h}(i \omega_n-i p_n, \vec{p}-\vec{q})=-\sum_\vec{p} |\vec{p}\rangle_{\mathrm{e}\;\mathrm{e}}\langle \vec{p}| \frac{\tanh\left(\frac{\xi^\mathrm{e}_{\vec{p}}}{2T}\right)+\tanh\left(\frac{\xi^\mathrm{h}_{\vec{p}-\vec{q}}}{2T}\right)}{2 (ip_n -\xi_{\vec{p}}^\mathrm{e}-\xi_{\vec{p}-\vec{q}}^\mathrm{h})}.
\end{equation} 
It is instructive to shift the momenta $\vec{p}$ and $\vec{p}-\vec{q}$ as follows 
\begin{equation}
\label{AppMomentumShift1}
\begin{split}
\vec{p}\rightarrow \vec{p}+\alpha_\mathrm{e} \vec{q}, \quad \quad \alpha_\mathrm{e}=\frac{v_h}{v_e+v_h}, \\
\vec{p}-\vec{q}\rightarrow \vec{p}-\alpha_\mathrm{h} \vec{q}, \quad \quad \alpha_\mathrm{h}=\frac{v_e}{v_e+v_h}.
\end{split}
\end{equation}
The momenta $\vec{q}$ is the total momentum for fluctuating CPs and $\vec{p}$ can be interpreted as their relative one. The dispersion relations in Eq.~(\ref{AppPi1}) are modified as
\begin{equation}
\xi^\mathrm{e}_{\vec{p}+\alpha_\mathrm{e} \vec{q}}\approx \xi^\mathrm{e}_{\vec{p}}+\frac{v_* \vec{q} \vec{p}}{2 p}, \quad \quad \quad  \xi^\mathrm{h}_{\vec{p}-\alpha_\mathrm{h} \vec{q}}\approx \xi^\mathrm{h}_{\vec{p}}-\frac{v_* \vec{q} \vec{p}}{2p}, \quad \quad \quad \xi^\mathrm{e}_{\vec{p}+\alpha_\mathrm{e} \vec{q}}
+\xi^\mathrm{h}_{\vec{p}-\alpha_\mathrm{h}\vec{q}}\approx 2 v_0 (p-p_\mathrm{F}).
\end{equation}
Here we have introduced $v_0=(v_e+v_h)/2$ and $v_*=2 v_\mathrm{e} v_\mathrm{h}/(v_\mathrm{e}+v_\mathrm{h})$. After the momentum shift, Eq.(\ref{AppMomentumShift1}), the denominator in Eq.~(\ref{AppPi1}) becomes independent on momentum $\vec{q}$. As a result, the expansion of $\hat{\Pi}(ip_n,\vec{q})$ in $p_n$ and $\vec{q}$ becomes straightforward and results in 
\begin{equation}
\hat{\Pi}(ip_n,\vec{q})=\begin{pmatrix} s_\mathrm{F}^2 \Pi_0 (i p_n,\vec{q}) & \zeta c_\mathrm{F} s_\mathrm{F} \frac{\alpha_e q e^{-i \zeta \phi_\vec{q}}}{4 p_\mathrm{F}} \Pi_0 (0,0)\\  \zeta c_\mathrm{F} s_\mathrm{F} \frac{\alpha_e q e^{-i \zeta \phi_\vec{q}}}{4 p_\mathrm{F}} \Pi_0 (0,0) & c_\mathrm{F}^2 \Pi_0 (i p_n,\vec{q})) \end{pmatrix}. 
\end{equation}
The factors \begin{equation}
c_\mathrm{F}^2=\frac{1}{2}\left(1+\frac{\delta}{\epsilon_\mathrm{F}}\right), \quad \quad  \quad  s_\mathrm{F}^2=\frac{1}{2}\left(1-\frac{\delta}{\epsilon_\mathrm{F}}\right).
\end{equation}
give the probability to find the Fermi level hole at sublattice A or B, respectively.  $\Pi_0 (i p_n,\vec{q})$ can be interpreted as a single step pair propagator in the Cooper ladder sum of a bilayer system without the pseudospin degree of freedom and is given by  
\begin{equation}
\label{Pi0}
\Pi_0 (i p_n,\vec{q})=\nu_0\left[\ln\left(\frac{2 e^C \epsilon_\mathrm{C}}{\pi T}\right)-|p_n| \tau'- i p_n \tau''-\xi^2 \vec{q}^2\right], \quad \; \tau'=\frac{\pi}{8T}, \quad  \tau''=  \frac{\ln\left(\frac{2 e^C \epsilon_\mathrm{c}}{\pi T}\right)}{2 v_0 p_\mathrm{F}}, \quad \xi^2= \frac{7\Xi(3) v_*^2 }{32\pi^2 T^2}.
\end{equation}
Here $\epsilon_0=\sqrt{v_\mathrm{e} v_\mathrm{h}} p_0$ is determined by the momentum cutoff $p_0$. $C=0.577$ is the Euiler constant, $\Xi(x)$ is the Zeta function and $\nu_0=p_\mathrm{F}/2\pi\hbar^2 v_0$ is the density of states defined respect to the average Fermi velocity of electrons and holes. If we rescale the bosonic fields as $\Delta^\mathrm{A}\rightarrow \Delta^\mathrm{A}/\sqrt{\nu_0} s_\mathrm{F}$ and $\Delta^\mathrm{B}\rightarrow \Delta^\mathrm{B}/\sqrt{\nu_0} c_\mathrm{F}$, the action describing fluctuating CPs, Eq.~\ref{AppSDelta} $S_\Delta$, simplifies as
\begin{equation}
S_\Delta=T \sum_{p_n \vec{q}}\bar{\Delta}_{p_n \vec{q}} \hat{L}^{-1}(i p_n,\vec{q}) \Delta_{p_n \vec{q}},
\end{equation}
where $\hat{L}^{-1}(i p_n,\vec{q})$ is the inverse dimensionless CP propagator given by
\begin{equation}
\hat{L}^{-1}(i p_n,\vec{q})=|p_n|\tau' -i p_n \tau'' + \hat{H}_\mathrm{GL}, \quad \quad \quad \hat{H}_\mathrm{GL}=\begin{pmatrix} \varepsilon_\mathrm{A}  + \xi^2\vec{q}^2 & - \xi_\star (\zeta q_x-iq_y) \\ - \xi_\star (\zeta q_x-iq_y)& \varepsilon_\mathrm{B}  + \xi^2\vec{q}^2
\end{pmatrix}.
\end{equation}
The Hermitian matrix $\hat{H}_\mathrm{GL}$ can be referred to as the Ginzburg-Landau Hamiltonian. It describes two dissipative bosonic modes ($\Delta_{\mathrm{A}}$ and $\Delta_{\mathrm{A}}$) that are intertwined by the effective pseudospin-orbit interactions with strength parametrized by the length $\xi_\star=\alpha_\mathrm{e} \ln\left(2 e^C \epsilon_\mathrm{c}/\pi T\right)/4 p_\mathrm{F}$. Each mode has a temperature-dependent energy gap 
\begin{equation}
\epsilon_{\mathrm{A}(\mathrm{B})}=\ln \left[\frac{T}{T_{\mathrm{A}(\mathrm{B})}}\right], \quad\quad T_{\mathrm{A}(\mathrm{B})} = \frac{2e^C e^{- \frac{1}{\lambda_{\mathrm{A}(\mathrm{B})}}}}{\pi}, \quad \quad \lambda_{\mathrm{A}}=s_\mathrm{F}^2 \nu_0 V, \quad\quad \lambda_{\mathrm{B}}=c_\mathrm{F}^2 \nu_0 V.
\end{equation}
The gap vanishes at the transition temperature $\varepsilon_{\mathrm{A}(\mathrm{B})}=\ln [T/T^0_{\mathrm{A}(\mathrm{B})}]$ of the Cooper pair condensation with zero center of mass momentum. It should be mentioned that coefficients $\tau''$ and $\xi_\star$ can be rewritten as  $\tau''=1/2 \lambda v_0 p_\mathrm{F}$ and $\xi_{\star}=1/4\lambda p_\mathrm{F}$ where $\lambda=\mathrm{max}[\lambda_\mathrm{A}, \lambda_\mathrm{B}]$ is the coupling constant of the leading channel.

\subsection{A3. The spectrum of Ginzburg-Landau Hamiltonian}
Eigenvalues for $\hat{H}_{\mathrm{GL}}$ are labeled by $\gamma=\pm$ and are given by 
\begin{equation}
\label{AppCPEnergy}
\varepsilon_{\gamma\vec{q}}=\varepsilon_\mathrm{s}+\xi^2 \vec{q}^2+\gamma d_\vec{q},\quad \quad  d_\vec{q}= \sqrt{\varepsilon^2_\mathrm{z}+\xi_\star^2 \vec{q}^2},
 \quad \quad \epsilon_{\mathrm{s}} =\frac{1}{2}(\epsilon_{\mathrm{A}}+\epsilon_{\mathrm{B}}), \quad \quad  \epsilon_{\mathrm{z}} = \frac{1}{2}(\epsilon_{\mathrm{A}}-\epsilon_{\mathrm{B}}).
\end{equation}
They can be interpreted as energies for fluctuating CPs eigenmodes and their dispersion is discussed in detail in the main part of the paper. The fluctuating CPs with finite center of mass momentum $\vec{q}$ represent a superposition of $\Delta_\mathrm{A}$ and $\Delta_\mathrm{B}$ as 
\begin{equation}
\label{AppCPSpinor}
|+,\vec{q}\rangle_\mathrm{C} =\begin{pmatrix}  \cos\left(\vartheta_\vec{q}/2\right) \\ -  \zeta \sin\left(\theta_\vec{q}/2\right) e^{i \zeta \phi_\vec{q}}\end{pmatrix}, \quad \quad |-,\vec{q}\rangle_\mathrm{C} =\begin{pmatrix} \zeta \sin\left(\vartheta_\vec{q}/2\right) e^{- i \zeta \phi_\vec{q}} \\  \cos\left(\vartheta_\vec{q}/2\right)\end{pmatrix}, \quad \quad \cos\left(\vartheta_\vec{q}\right)=\frac{\epsilon_\mathrm{z}}{\sqrt{\epsilon^2_\mathrm{z}+\xi_\star^2 \vec{q}^2}}.
\end{equation}
They are intertwined with the valley dependent phase factor $e^{- i \zeta \phi_\vec{q}}$, which reflects nontrivial geometries for fluctuating CPs. 

The derived expressions, Eqs.~(\ref{AppCPEnergy}) and (\ref{AppCPSpinor}),  allow to perform the spectral decomposition of CP propagator $L(ip_n,\vec{q})$ and its retarded (advanced) $L^{\mathrm{R}(A)}(\omega,\vec{q})$ counterparts as
\begin{equation}
\label{ApSpectralDecomposition}
\hat{L}(i p_n,\vec{q})=\sum_{\gamma} \frac{|\gamma, \vec{q}\rangle_{\mathrm{C}}\;\vphantom{0}_{\mathrm{C}} \langle \gamma, \vec{q}|}{\tau' |p_n|-i\tau'' p_n + \varepsilon_{\gamma\vec{q}}}, \quad \quad \quad \hat{L}^{\mathrm{R}(\mathrm{A})}(\omega,\vec{q})=\sum_{\gamma} \frac{|\gamma, \vec{q}\rangle_{\mathrm{C}}\;\vphantom {0}_{\mathrm{C}} \langle \gamma, \vec{q}|}{\mp i \omega \tau' -\omega \tau'' + \varepsilon_{\gamma\vec{q}}}.      
\end{equation}
These helpful relations will be used routinely in Appendices B and C.

Due to their dissipative nature, fluctuating CPs cannot be interpreted as bosonic quasiparticles, but are instead overdamped bosonic modes. However, $\hat{H}_\mathrm{GL}$ is the Hermitian matrix and its eigenmodes $|\gamma\vec{q}\rangle_\mathrm{C}$ form a complete basis. $\hat{H}_\mathrm{GL}$ can be diagonalized via the unitary transformation governed by the matrix $\hat{U}_\vec{q}$ given by 
\begin{equation}
\label{Unitary1}
\hat{E}_\mathrm{GL}(\vec{q})=U_\vec{q} \hat{H}_\mathrm{GL}(\vec{q}) U_\vec{q}^\dagger, \quad \quad  \hat{E}_\mathrm{GL}=\begin{pmatrix} \varepsilon_{+,\vec{q}} & 0 \\ 0 & \varepsilon_{-,\vec{q}}
\end{pmatrix}, \quad \quad U_\vec{q}=\begin{pmatrix} \cos\left(\vartheta_\vec{q}/2\right) & \zeta \sin\left(\vartheta_\vec{q}/2\right) e^{- i \zeta \phi_\vec{q}} \\ -  \zeta \sin\left(\theta_\vec{q}/2\right) e^{i \zeta \phi_\vec{q}} & \cos\left(\vartheta_\vec{q}/2\right).
\end{pmatrix}    
\end{equation}
This transformation is used in Appendix C, where a derivation of the kinetic equation for fluctuating CPs is presented.
\section{APPENDIX B. LINEAR RESPONSE THEORY AND PARACONDUCTIVITY}
\subsection{B1. General expression for the conductivity tensor}
This section presents a general expressions for the conductivity tensor, which is derived with the help of the linear response approach. The external electric fields in electron and hole layers that we have disregarded so far can be described by position independent vector potentials $\vec{a}^\mathrm{e}_t$ and $\vec{a}^\mathrm{h}_t$. Components of a fluctuating CP are spatially separated and have the opposite charge. That is why the Peierls substitution introduces the vector potential to the Ginzburg-Landau Hamiltonian as
\begin{equation}
\hat{H}_{\mathrm{GL}}(\vec{q})\rightarrow \hat{H}_{\mathrm{GL}}\left(\vec{q}-\frac{e}{c} \vec{a}^{\mathrm{eh}}_t\right), \quad \quad  \vec{a}^{\mathrm{eh}}_t=\vec{a}^{\mathrm{e}}_t-\vec{a}^{\mathrm{h}}_t.
\end{equation}
Fluctuating CPs are affected by the electric field difference across the bilayer and their motion results in electric currents in both layers. These currents have the same magnitude but the opposite signs. Really, the current operators for two layers have the opposite signs as
\begin{equation}
\label{ApVelcities}
\vec{J}^{\mathrm{e}(\mathrm{h})}(\vec{q})=\pm e \vec{v}(\vec{q}), \quad \quad \quad \vec{v}(\vec{q}) = \partial_\vec{q}H_{\mathrm{GL}}(\vec{q}) 
\end{equation}
Within the Kubo linear response theory, the contribution of fluctuating CPs to the DC conductivity (and transconductivity $\sigma^\mathrm{D}_{\alpha\beta}=-\sigma_{\alpha\beta}$) tensor $\sigma_{\alpha\beta}=e^2 \Im[\chi^\mathrm{R}_{\alpha \beta}(\omega)]/\omega$ is related to the imaginary part of the retarded velocity-velocity correlation function $\chi^\mathrm{R}_{\alpha \beta}(\omega)$. It can be obtained by the analytical continuation from the corresponding Matsubara correlation function $\chi_{\alpha \beta}(i \Omega_n)$,  which is given by 
\begin{equation}
\chi_{\alpha \beta}(i \Omega_n)=T \sum_{p_n \vec{q}} \mathrm{Tr}[\hat{v}_\alpha(\vec{q})\hat{L}(i p_n+i\Omega_n,\vec{q}) \hat{v}_\beta(\vec{q}) \hat{L}(i p_n,\vec{q})].    
\end{equation}
If we use the spectral representation of for the Cooper pair propagator, presented in Appendix A, the summation over Matsubara frequencies and the analytical continuation are straightforward and results in  
\begin{equation}
\label{ApConductivity2}
\sigma_{\alpha \beta}=\sum_{\vec{q}\gamma \gamma'}\left(\mathrm{Re}[Q_{\alpha \beta}^{\gamma\gamma'}(\vec{q})] J_\mathrm{Re}^{\gamma \gamma'}(\vec{q})+  
\mathrm{Im}[Q_{xy}^{\gamma\gamma'}(\vec{q})] J_\mathrm{Im}^{\gamma \gamma'}(\vec{q})\right).
\end{equation}
Here $Q_{\alpha\beta}^{\gamma\gamma'}$ represents the product of matrix elements for velocity operators and is given by
\begin{equation}
\label{ApQexpressions2}
Q_{\alpha \beta}^{\gamma\gamma'}(\vec{q})=\langle \gamma \vec{q}|v_\alpha(\vec{q})| \gamma' \vec{q} \rangle_\mathrm{C} \;\vphantom{0}_{\mathrm{C}} \langle \gamma' \vec{q}| v_\beta(\vec{q}) |\gamma \vec{q}\rangle_\mathrm{C}.
\end{equation}
The factors $J_\mathrm{Re}^{\gamma \gamma'}(p)$ and $J_\mathrm{Im}^{\gamma \gamma'}(p)$ are given by
\begin{equation}
\label{ApJIntegrals1}
\begin{split}
J_\mathrm{Re}^{\gamma \gamma'}=\int \frac{d \omega}{ \pi} \mathrm{Im}[L_\gamma^\mathrm{R}] \mathrm{Im}[L_{\gamma'}^\mathrm{R}] \left(-\frac{\partial n_\mathrm{B}(\omega)}{\partial \omega}\right), \quad \quad
J_\mathrm{Im}^{\gamma \gamma'}=\int \frac{d \omega}{ \pi}\left[\mathrm{Im}[L_\gamma^\mathrm{R}] \frac{\partial \mathrm{Re}[L_{\gamma'}^\mathrm{R}] }{\partial \omega}-\mathrm{Im}[L_{\gamma'}^\mathrm{R}] \frac{\partial \mathrm{Re}[L_{\gamma}^\mathrm{R}] }{\partial \omega}\right]. 
\end{split}
\end{equation}
Here we use the compact notation $L^\mathrm{R}_\gamma\equiv L^\mathrm{R}_{\gamma}(\omega,\vec{q})$ for all Green functions inside the integrals. It should be noted that these relations are very general and do not rely on any specific frequency or momentum dependence of the Green function $L^\mathrm{R}_\gamma(\omega, \vec{q})$. For the Green functions in Eq.~(\ref{ApSpectralDecomposition}), which describe fluctuating CPs, we get
\begin{equation}
\mathrm{Re}[L_\gamma^\mathrm{R}(\omega, \vec{q})]=\frac{\varepsilon_{\gamma\vec{q}} (p)+\omega \tau''}{(\omega\tau')^2+(\varepsilon_{\gamma\vec{q}} (p)+\omega \tau'')^2}, \quad \quad   \mathrm{Im}[L_\gamma^\mathrm{R}(\omega, \vec{p})]=\frac{\omega \tau'}{(\omega\tau')^2+(\varepsilon_{\gamma\vec{q}} (p)+\omega \tau'')^2}.\end{equation}
The integration in Eq.~(\ref{ApJIntegrals1}) can be performed and results in 
\begin{equation}
\label{ApJIntegrals2}
J_\mathrm{Re}^{\gamma \gamma'}=\left(\frac{T}{\varepsilon_{\gamma\vec{q}}}+\frac{T}{\varepsilon_{\gamma'\vec{q}}}\right) \frac{\tau'|\tau|^2}{|\varepsilon_{\gamma\vec{q}}\tau+\varepsilon_{\gamma'\vec{q}}\tau^*| ^2}, \quad \quad  J_\mathrm{Im}^{\gamma \gamma'}=\left(\frac{T}{\varepsilon_{\gamma\vec{q}}}-\frac{T}{E _{\gamma'}}\right) \frac{\tau''|\tau|^2}{|\varepsilon_{\gamma\vec{q}}\tau+\varepsilon_{\gamma'\vec{q}}\tau^*| ^2}.  
\end{equation}
It is instructive to consider intrachannel ($\gamma=\gamma'$) and intermode ($\gamma\neq\gamma'$) contributions to the conductivity tensor separately. The factor $J_\mathrm{Im}^{\gamma \gamma'}$ vanishes at $\gamma'=\gamma$ and the intramode contribution can be written as 
\begin{equation}
\label{ApSigmaIntra}
\sigma_{\alpha \alpha}^{\mathrm{AL}}=\sum_{\gamma\vec{q}} \frac{\mathrm{Re}[Q_{\alpha \alpha}^{\gamma \gamma}(\vec{q})]}{2 \varepsilon^3_{\gamma\vec{q}}} \frac{T|\tau|^2}{\tau'}.
\end{equation}
Here we have taken into account that $Q_{\alpha \bar{\alpha}}^{\gamma \gamma}=0$ and intramode transitions contribute only to the longitudinal conductivity. As it is discussed in the main part of paper, $\sigma_{\alpha \alpha}^{\mathrm{AL}}$ represents a sum of two independent terms describing the conventional Aslamazov-Larkin effect for each of two competing channels. It is instructive to separate the intermode contribution into two terms as follows
\begin{equation}
\label{ApSigmaInterOmega}
\sigma_{\alpha \bar{\alpha}}^{\mathrm{\Omega}}\; =\; \sum_{\gamma\vec{q}} \frac{\mathrm{Im}[Q_{\alpha \bar{\alpha}}^{\gamma \bar{\gamma}}]\tau''|\tau|^2}{|\varepsilon_{\gamma\vec{q}}\tau+\varepsilon_{\bar{\gamma}\vec{q}}\tau^*| ^2}\left(\frac{T}{\varepsilon_{\gamma\vec{q}}}-\frac{T}{\varepsilon _{\bar{\gamma}\vec{q}}}\right) \; = \; \sum_{\gamma\vec{q}} \frac{2 \mathrm{Im}[Q_{\alpha \bar{\alpha}}^{\gamma \bar{\gamma}}]\tau''|\tau|^2}{|\varepsilon_{\gamma\vec{q}}\tau+\varepsilon_{\bar{\gamma}\vec{q}}\tau^*| ^2}\cdot \frac{T}{\varepsilon_{\gamma\vec{q}}},
\end{equation}
\begin{equation}
\label{ApSigmaInterG}
\sigma_{\alpha \beta}^{\mathrm{G}}\;=\;\sum_{\gamma\vec{q}} \frac{ \mathrm{Re}[Q_{\alpha \beta}^{\gamma \bar{\gamma}}]\tau'|\tau|^2}{|\varepsilon_{\gamma\vec{q}}\tau+\varepsilon_{\bar{\gamma}\vec{q}}\tau^*| ^2}\left(\frac{T}{\varepsilon_{\gamma\vec{q}}}+\frac{T}{\varepsilon_{\bar{\gamma} \vec{q}}}\right)\;=\; \sum_{\gamma\vec{q}} \frac{2  \mathrm{Re}[Q_{\alpha \beta}^{\gamma \bar{\gamma}}]\tau'|\tau|^2}{|\varepsilon_{\gamma\vec{q}}\tau+\varepsilon_{\bar{\gamma}\vec{q}}\tau^*| ^2}\cdot \frac{T}{\varepsilon_{\gamma\vec{q}}}.
\end{equation}
Here we have taken into account that $\mathrm{Im}[Q_{\alpha \alpha}^{\gamma \bar{\gamma}}]=0$.   These terms represent anomalous Aslamazov-Larkin contributions to the conductivity. In Appendix C we demonstrate that they are intricately related with the nonzero Berry curvature ($\Omega_{\gamma \vec{q}}$) and quantum metric ($\hat{G}_{\gamma \vec{q}}$) for GL Hamiltonian $H_\mathrm{GL}$.

\subsection{B2. Temperature dependence of the conductivity tensor}
This subsection presents derivation of conventional and anomalous Aslamazov-Larkin contributions to conductivity tensor. The explicit expression for the velocity operator $\vec{v}(\vec{q})$ introduced in Eq.~(\ref{ApVelcities}) are   
\begin{equation}
v_x(\vec{q})=\begin{pmatrix} 2 \xi \vec{q}_x & -\xi_\star \\  -\xi_\star & 2 l \vec{q}_x
\end{pmatrix}, \quad \quad \quad v_y(\vec{q})=\begin{pmatrix} 2 \xi \vec{q}_y & i \xi_\star \\  -i \xi_\star & 2 l \vec{q}_y
\end{pmatrix}.
\end{equation}
As a result, the product of matrix elements for velocity operators $Q_{\alpha\beta}^{\gamma\gamma'}(\vec{q})$ averaged over the direction of momentum $\vec{q}$ is given by
\begin{equation}
\label{ApQExpanded}
\begin{split}
\langle Q_{xx}^{\gamma\gamma}\rangle_{\phi_\vec{q}}= 2 \xi^4 \vec{q}^2+ \frac{\xi_\star^4 \vec{q}^2}{2d_\vec{q}}\approx 2 \xi^4 \vec{q}^2
, \quad \quad \quad \quad \quad \quad \quad \quad \; \; \quad \langle Q_{xy}^{\gamma\gamma}\rangle_{\phi_\vec{q}}=0, \\
\langle Q_{xx}^{\gamma\bar{\gamma}}\rangle_{\phi_\vec{q}}=  \frac{\xi_\star^2}{2} \left(1+ \frac{\varepsilon_{z}^2 }{d_\vec{q}^2} \right)\approx \xi_\star^2 , \quad \quad \quad
\langle Q_{xy}^{\gamma\bar{\gamma}}\rangle_{\phi_\vec{q}}= \zeta\xi_\star^2\frac{\varepsilon_{z} }{d_\vec{q}}\approx \zeta \xi_\star^2   \mathrm{sign}[\varepsilon_\mathrm{z}].
\end{split}
\end{equation}
Here we have also performed the expansion in the ratio of spatial scales, $\xi_\star/\xi\ll1$. Besides, in the leading order in $\xi_\star/\xi$ the effect of the effective pseudospin-orbit interactions at the dispersion of fluctuating CPs can be neglected. As a result, the dispersion relations are approximated as $\varepsilon_{\mathrm{A}(\mathrm{B})}+\xi^2 q^2$. If we also expand in the ratio of time scales $\tau''/\tau'\ll 1$, the expressions for the conductivity tensor Eqs.~(\ref{ApSigmaIntra}), (\ref{ApSigmaInterOmega}), and (\ref{ApSigmaInterG}) can be presented as \begin{equation}
\label{ApConductivityTemperature1}
\sigma_{xx}^{\mathrm{AL}}=\frac{e^2}{2\pi \hbar} \frac{8 T\tau'}{\pi} F_{\mathrm{AL}}, \quad\quad \quad   \sigma_{yx}^{\mathrm{\Omega}}=\zeta\frac{e^2}{2\pi \hbar}  \frac{8 T\tau'}{\pi} \left(\frac{\xi_\star}{l}\right)^2 \frac{\tau''}{\tau'}\; F_{\mathrm{\Omega}}, \quad \quad  \sigma_{xx}^{\mathrm{G}}=\frac{e^2}{2\pi \hbar} \frac{8 T\tau'}{\pi} \left(\frac{\xi_\star}{l}\right)^2 F_\mathrm{G}.
\end{equation}
For the microscopically evaluated fluctuating CPs  relaxation time $\tau'$ given by Eq.~(\ref{Pi0}), the second factor simplifies as $8T\tau'/\pi=1$. The temperature behavior of these terms is governed by dimensionless functions $F\equiv F(\varepsilon_\mathrm{A},\varepsilon_\mathrm{B})$. These functions depend only on $\varepsilon_\mathrm{A}=\ln[T/T^0_\mathrm{A}]$ and $\varepsilon_\mathrm{B}=\ln[T/T^0_\mathrm{B}]$ and are given by
\begin{equation}
\label{ApConductivityTemperature2}
\begin{split}
F_\mathrm{L}^{\mathrm{AL}}=
\frac{\pi}{8}\int_0^\infty \frac{d (\xi^2 q^2)}{2} \left[\frac{\xi^2 q^2}{(\varepsilon_\mathrm{A} + \xi^2 q^2)^3}+\frac{\xi^2 q^2}{(\varepsilon_\mathrm{B}+\xi^2 q^2)^3}\right]=\frac{\pi}{32}\left(\frac{1}{\varepsilon_{\mathrm{A}}}+\frac{1}{\varepsilon_{\mathrm{B}}} \right), \\
F_\mathrm{G}=\frac{\pi}{8} \int_0^\infty \frac{d (\xi^2 q^2)}{2} \frac{1}{(\varepsilon_\mathrm{s}+\xi^2 q^2)(\varepsilon_\mathrm{A}+\xi^2 q^2)(\varepsilon_\mathrm{B}+\xi^2 q^2)}=\frac{\pi}{8(\varepsilon_\mathrm{A}-\varepsilon_\mathrm{B})^2} \ln \left[\frac{(\varepsilon_\mathrm{A}+\varepsilon_\mathrm{A})^2}{4\varepsilon_\mathrm{A}\varepsilon_\mathrm{B}}\right],\\
F_\mathrm{\Omega}=\frac{\pi}{8} \int_0^\infty \frac{d (\xi^2 q^2)}{2} \frac{\varepsilon_z}{(\varepsilon_\mathrm{s}+\xi^2 q^2)^2(\varepsilon_\mathrm{A}+\xi^2 q^2)(\varepsilon_\mathrm{B}+\xi^2 q^2)}=\frac{\pi}{8(\varepsilon_\mathrm{A}-\varepsilon_\mathrm{B})}\left\{  \frac{\ln \left[\varepsilon_\mathrm{A}/\varepsilon_\mathrm{B}\right]}{\varepsilon_\mathrm{A}-\varepsilon_\mathrm{B}} -\frac{2}{\varepsilon_\mathrm{A}+\varepsilon_\mathrm{B}}\right\}.
\end{split}
\end{equation}
As a result, we recover the expressions Eq.~(\ref{ConductivityTemperature1}), which are presented in the main text of the paper. 
\subsection{APPENDIX C. KINETIC EQUATION FOR FLUCTUATING COOPER PAIRS}

\subsection{C1. Time-dependent Ginzburg-Landau equation}
This Appendix is devoted to the time-dependent Ginzburg-Landau (TDGL) equation. It governs the dissipative dynamics of the bosonic field $\Delta\equiv\{ \Delta^\mathrm{A}_{\tau \vec{r}}, \Delta^\mathrm{B}_{\tau \vec{r}} \}$ for fluctuating CPs and is given by 
\begin{equation}
\label{TDGL1}
\tau^* \left(\frac{\partial \Delta_{t\vec{r}}}{\partial t}+i e \phi^{\mathrm{eh}}_{t \vec{r}}\Delta\right)=-\hat{H}_\mathrm{GL}\left( \vec{q}\right)\Delta+\eta_{t\vec{r}}, \quad \quad \quad \langle \eta_{t\vec{r}} \eta^\dagger_{t'\vec{r}'} \rangle= \hat{N} \delta_{tt'}\delta_{r\vec{r}'}, \quad \quad \hat{N} = 2 T \tau' \hat{1}. 
\end{equation}
Here we take into account that components of fluctuating CPs have opposite charges and are spatially separated. For these reasons, coupling with electric potentials in both layers ($\phi^{\mathrm{e}}_{t \vec{r}}$ and $\phi^{\mathrm{h}}_{t \vec{r}})$ can be introduced via the Peierls substitution  $\partial_t\rightarrow \partial_t+i e \phi^{\mathrm{eh}}_{t \vec{r}}$ with  $\phi^{\mathrm{eh}}_{t \vec{r}}=\phi^{\mathrm{e}}_{t \vec{r}}-\phi^{\mathrm{h}}_{t \vec{r}}$. The external complex field $\eta_{t\vec{r}}=\{\eta_{t\vec{r}}^\mathrm{A},\eta_{t\vec{r}}^\mathrm{B}\}$ is the Langevin noise. Its presence is dictated by the fluctuation-dissipation theorem, and the correlation function $\langle \eta_{t\vec{r}} \eta^\dagger_{t'\vec{r}'} \rangle$, which is free of temporal and spatial correlations (white noise), is proportional to the relaxation rate $\tau'$ but does not depend on $\tau''$. In the considered weak-to-moderate Cooper pairing regime, their ratio is small $\tau''/\tau'\ll 1$; hence, the dissipation of fluctuating CPs is an essential component in their dynamics. These CPs cannot be interpreted as bosonic quasiparticles, but are instead overdamped bosonic modes.

The TDGL equation can be derived microscopically (e.g. with the help of the Keldysh diagrammatic technique), or introduced in a phenomenological way as we do here. If we ignore the coupling to electromagnetic field (that can be reintroduced by the Peierls substitution), the dynamics of fluctuating CPs in TDGL is governed by the inverse retarded propagator  $\hat{L}^\mathrm{R}(\omega,\vec{q})$. In equilibrium the bosonic field describing fluctuating CPs is not zero, but is shaped by the noise as $\Delta_{\omega \vec{q}}=\hat{L}^{\mathrm{R}}(\omega,\vec{q}) \eta_{\omega \vec{q}}$  (and $\Delta^\dagger_{\omega \vec{q}}=\eta_{\omega \vec{q}}^\dagger\hat{L}^{\mathrm{A}}(\omega,\vec{q})$ with $\hat{L}^{\mathrm{A}}(\omega,\vec{q})$ is the advanced CP propagator). If we assume that the Langevin noise does not have any temporal and spatial correlations (white noise), the correlation $F^{\mathrm{dyn}}_{\Delta}(\vec{q})=\langle \Delta_{t \vec{q}} \Delta^\dagger_{t\vec{q}}\rangle$ can be presented as
\begin{equation}
\label{FCorellation1} 
 F^{\mathrm{dyn}}_{\Delta}(\vec{q})=\int \frac{d \omega}{ 2\pi} \hat{L}^\mathrm{R}(\omega,\vec{q}) \hat{N} \hat{L}^\mathrm{A}(\omega,\vec{q}),    
 \quad \quad \quad 
 F^{\mathrm{td}}_{\Delta}(\vec{q})=\int_{-\infty}^{\infty}\frac{d \omega}{2\pi i} \frac{T}{\nu_0\omega}\left[\hat{L}^\mathrm{R}(\omega, \vec{q})-\hat{L}^\mathrm{A}(\omega,\vec{q})\right],
\end{equation}
and is determined by the pseudospin structure of the noise correlation function $\hat{N}$. According to the general philosophy of the Langevin approach, $\hat{N}$ is chosen in a such way that the correlation function $F^{\mathrm{dyn}}_{\Delta}(\vec{q})$ matches with the one  $F^{\mathrm{td}}_{\Delta}(\vec{q})=\langle \Delta_{\tau \vec{q}} \Delta^\dagger_{\tau \vec{q}}\rangle$, which is evaluated via the thermodynamic averaging and is also presented in Eq.~(\ref{FCorellation1}). If we use the spectral representation for CP propagators $\hat{L}^{\mathrm{R}(\mathrm{A})}(\omega,\vec{q})$ given by Eq.~(\ref{ApSpectralDecomposition}), the integration over frequency $\omega$ is straightforward and results in 
\begin{equation}
F^{\mathrm{td}}_{\Delta}(\vec{q})=\sum_{\gamma} |\gamma, \vec{q}\rangle_{\mathrm{C}}\;\vphantom {0}_{\mathrm{C}} \langle \gamma, \vec{q}| \frac{T}{\varepsilon_{\gamma\vec{q}}}, \quad \quad F^{\mathrm{dyn}}_{\Delta}(\vec{q})=\sum_{\gamma, \gamma'} \frac{|\gamma, \vec{q}\rangle_{\mathrm{C}}\;\vphantom {0}_{\mathrm{C}} \langle \gamma, \vec{q}| \hat{N} |\gamma', \vec{q}\rangle_{\mathrm{C}}\;\vphantom {0}_{\mathrm{C}} \langle \gamma', \vec{q}|}{\tau' (\varepsilon_{\gamma\vec{q}}+\varepsilon_{\gamma' \vec{q}})-i \tau'' (\varepsilon_{\gamma\vec{q}}-\varepsilon_{\gamma' \vec{q}})}. 
\end{equation}
For the noise correlation function given by the right-hand side expression in Eq.~(\ref{TDGL1}), $F^{\mathrm{td}}_{\Delta}(\vec{q})$ and $F^{\mathrm{dyn}}_{\Delta}(\vec{q})$ match each other. As a result, the description of fluctuating CPs within the quantum field theory approach and within the TDGL framework are consistent with each other. 

\section{C2. Transformation of the TDGL equation into the kinetic equation}
This subsection presents derivation of the kinetic equation for fluctuating CPs. First, it is instructive to introduce a density matrix for fluctuating CPs as $\rho_{\vec{r}\vec{r}'t}=\Delta_{\vec{r} t} \Delta_{\vec{r}' t}^\dagger$. The density matrix satisfies the equation
\begin{equation}
\label{ApDensityMatrixDynamics1}
\frac{\partial\rho_{\vec{r}\vec{r}t}}{\partial t}=-i e (\phi^{\mathrm{eh}}_{\vec{r} t}-\phi^{\mathrm{eh}}_{\vec{r}'t})\rho_{\vec{r}\vec{r}'t}-\frac{\hat{H}_\mathrm{GL} \rho_{\vec{r}\vec{r}'t}}{\tau} - \frac{\rho_{\vec{r}\vec{r}' t} \hat{H}_\mathrm{GL}}{\tau^*} + \frac{ \eta_{\vec{r}t} \Delta^\dagger_{\vec{r}'t}}{\tau} + \frac{\Delta_{\vec{r}t}  \eta^\dagger_{\vec{r}'t} }{\tau^*},
\end{equation}
which can be interpreted as a generalization of the Liouville–von Neumann equation.

Second, we introduce a distribution function $\hat{n}_{\vec{R}\vec{q}t}$ for fluctuating CPs. This function is obtained from the density matrix by the Wigner transformation and the unitary rotation to the eigenmode basis as follows: 
\begin{equation*}
\hat{n}_{\vec{R}\vec{q}t}=U_\vec{q}^\dagger \hat{n}'_{\vec{R}\vec{q}t} U_\vec{q}.
\quad \hat{n}'_{\vec{R}\vec{q}t}=\int d\vec{r} e^{-i \vec{q} \vec{r}} \hat{\rho}_{\vec{\vec{R}+\frac{r}{2}},\vec{\vec{R}-\frac{r}{2}},t}=\int \frac{d\vec{p}}{(2\pi)^2} e^{i \vec{p} \vec{R}} \hat{\rho}_{\vec{\vec{q}+\frac{p}{2}},\vec{\vec{q}-\frac{p}{2}},t}.
\end{equation*}
Here, $U_\vec{q}$ is the unitary matrix that diagonalizes the GL Hamiltonian $\hat{H}_\mathrm{GL}$ as $U_\vec{q}^\dagger \hat{H}_{\mathrm{GL}}U_\vec{q}=\mathrm{diag}\{\varepsilon_{+,\vec{q}},\varepsilon_{-,\vec{q}}\}$ and is given by Eq.~(\ref{Unitary1}). Diagonal elements of the rotated distribution matrix ($n_{++}\equiv n_{+}$ and $n_{--}=n_{-}$) can be interpreted as distribution functions for eigenmodes in phase space ($\vec{R}$,$\vec{q}$). The off-diagonal elements ($n_{+-}$ and $n_{-+}$) are responsible for intermode coherence. For noise-induced equilibrium fluctuating CPs, $\Delta_{\omega \vec{q}}^0=\hat{L}^{\mathrm{R}}(\omega,\vec{q}) \eta_{\omega \vec{q}}$, the distribution function $\hat{n}^0_{\vec{q}}$ is position- and time-independent, as given below:
\begin{equation}
\label{AppEquilibriumDistribution}
\hat{n}^0_{\vec{q}}=\begin{pmatrix} n^0_{+,\vec{q}} & 0 \\   0 & n^0_{-,\vec{q}}
\end{pmatrix}.
\end{equation}
At thermal equilibrium, there is no intermode coherence, and the diagonal terms are given by the classical distribution function $n^0_{\gamma \vec{q}}=T/\varepsilon_{\gamma\vec{q}}$, which confirms the interpretation outlined above for $\hat{n}_{\vec{R}\vec{q}t}$.

Third, we will follow Ref.~\cite{TserkovnyakKineticEq} and perform a gradient expansion of the equation for the matrix distribution function $\hat{n}_{\vec{R}\vec{q}t}$. For a start, we apply the Wigner transformation to Eq.~(\ref{ApDensityMatrixDynamics1}) and get the following equation for $\hat{n}'_{\vec{R}\vec{q}t}$ 
\begin{equation}
\label{ApDensityMatrixDynamics2}
   \partial_t\hat{n}'_{\vec{R}\vec{q}t} = T_1+T_2+T_3.
\end{equation}
The right-hand side is divided into three terms, which need to be dealt with separately. The term $T_1$ appears due to the coupling of fluctuating CPs with external electric field and within leading order of the gradient expression simplifies as 
\begin{equation}
T^1=   -i e \int d\vec{r} e^{-i \vec{q} \vec{r}} (\phi^{\mathrm{eh}}_{\vec{R}+\vec{r}/2, t}-\phi^{\mathrm{eh}}_{\vec{R}+\vec{r}/2,t}) \hat{\rho}_{\vec{R}+\frac{r}{2},\vec{R}-\frac{r}{2},t}\approx-e \vec{E}^{\mathrm{eh}}_{\vec{R} t} \partial_\vec{q} \hat{n}'_{\vec{R}\vec{q}t}.
\end{equation}
Here $E^{\mathrm{eh}}_{\vec{R}t}=- \nabla\phi^{\mathrm{eh}}_{\vec{R}t}$ is an electric field difference between layers. The term $T_2$ describes the dynamics of fluctuating CPs as
\begin{equation}
T_2=-\frac{\hat{H}_\mathrm{GL} \otimes \hat{n}'_{\vec{R}\vec{q}t}}{\tau} - \frac{\hat{n}'_{\vec{R}\vec{q}t} \otimes \hat{H}_\mathrm{GL}}{\tau^*}\approx  -\frac{\hat{H}_\mathrm{GL}(\vec{q}) \hat{n}}{\tau} - \frac{\hat{n} \hat{H}_\mathrm{GL}(\vec{q})}{\tau^*}+\frac{i \hbar }{2}\left(\frac{\partial_\vec{q}\hat{H}_\mathrm{GL} \partial_{\vec{R}}\hat{n}'_{\vec{R}\vec{q}t}}{\tau}- \frac{\partial_{\vec{R}}\hat{n}'_{\vec{R}\vec{q}t} \partial_\vec{q}\hat{H}_\mathrm{GL}}{\tau^*}\right). 
\end{equation}
Here $\otimes$ is the Moyal product, which in the lowest order of the gradient expansion simplifies as
\begin{equation}
 \otimes=\mathrm{exp}\left(\frac{i \hbar}{2} \left[\overleftarrow{\partial}_\vec{R} \overrightarrow{\partial}_\vec{q}-\overleftarrow{\partial}_\vec{q} \overrightarrow{\partial}_\vec{R}\right]\right)\approx 1+\frac{i \hbar}{2} \left[\overleftarrow{\partial}_\vec{R} \overrightarrow{\partial}_\vec{q}-\overleftarrow{\partial}_\vec{q} \overrightarrow{\partial}_\vec{R}\right].   
\end{equation}
The term $T_3$ describes the effect of the Langevin noise and is of the first order in it. That is why we can approximate the order parameter by its value in equilibrium $\Delta_{\omega \vec{q}}^0=\hat{L}^{\mathrm{R}}(\omega,\vec{q}) \eta_{\omega \vec{q}}$ that allows to simplify $T_3$ as follows
\begin{equation}
T_3=\int \frac{d \omega}{ 2\pi}\left(\frac{ \hat{N}\hat{L}^\mathrm{A}(\omega,\vec{q})}{\tau}  + \frac{\hat{L}^\mathrm{A}(\omega,\vec{q}) \hat{N}}{\tau}\right)=\frac{2T \tau'}{|\tau|^2} \hat{1}.
\end{equation}
Here we have used the spectral decomposition for the Cooper pair propagator,  Eq.~(\ref{ApSpectralDecomposition}). If we combine all three terms $T_1$, $T_2$ and $T_3$ together we get a closed-form equation for the distribution function $\hat{n}$ as
\begin{equation}
\label{ApDensityMatrixDynamics3}
\partial_t \hat{n}'_{\vec{R}\vec{q}t} =-e \vec{E}^{\mathrm{eh}}_{\vec{R} t} \partial_\vec{q} \hat{n}'_{\vec{R}\vec{q}t} -\frac{\hat{H}_\mathrm{GL}(\vec{q}) \hat{n}'_{\vec{R}\vec{q}t}}{\tau} - \frac{\hat{n}'_{\vec{R}\vec{q}t} \hat{H}_\mathrm{GL}(\vec{q})}{\tau^*}+\frac{i \hbar }{2}\left(\frac{\partial_\vec{q}\hat{H}_\mathrm{GL} \partial_{\vec{R}}\hat{n}'_{\vec{R}\vec{q}t}}{\tau}- \frac{\partial_{\vec{R}}\hat{n}'_{\vec{R}\vec{q}t} \partial_\vec{q}\hat{H}_\mathrm{GL}}{\tau^*}\right)+ \frac{2T \tau'}{|\tau|^2 \nu_0} \hat{1}. 
\end{equation}
Now we are ready to perform the rotation $\hat{n}=U_\vec{q}^\dagger \hat{n}'_{\vec{R}\vec{q}t} U_\vec{q}$ to the eigenmode basis. Here, $U_\vec{q}$ is the unitary matrix that diagonalizes the GL Hamiltonian $\hat{H}_\mathrm{GL}$ as $U_\vec{q}^\dagger \hat{H}_{\mathrm{GL}}U_\vec{q}=\mathrm{diag}\{\varepsilon_{+,\vec{q}},\varepsilon_{-,\vec{q}}\}$. The rotation of Eq.~(\ref{ApDensityMatrixDynamics3}) results in
\begin{equation}
\label{ApDensityMatrixDynamics4}
\frac{\partial\hat{n}}{\partial t}=-e \vec{E}^{\mathrm{eh}}_{\vec{R} t} D_\vec{q} \hat{n} -\frac{\hat{E}_\mathrm{GL}(\vec{q}) \hat{n}}{\tau} - \frac{\hat{n} \hat{E}_\mathrm{GL}(\vec{q})}{\tau^*}+\frac{i \hbar }{2}\left(\frac{D_\vec{q}\hat{E}_\mathrm{GL} \partial_{\vec{R}}\hat{n}}{\tau}- \frac{\partial_{\vec{R}}\hat{n} D_\vec{q}\hat{E}_\mathrm{GL}}{\tau^*}\right)+ \frac{2T \tau'}{|\tau|^2 \nu_0} \hat{1}. 
\end{equation}
Here $D_\vec{q} n=\partial_\vec{q}-i [\vec{\mathcal A}_\vec{q}, n]$ is the covariant derivative. $\vec{\mathcal A}_\vec{q}=i U_\vec{q}^\dagger \partial_\vec{q} U_\vec{q}$ is the generalized Berry connection, which has been introduced in the main part of the paper as Eq.~(\ref{BerryConnectionCP}). The equation for the diagonal elements of the distribution $\hat{n}_{\gamma}=$ can be presented as 
\begin{equation}
\label{ApPaperKineticEq1}
\begin{split}
    \partial_t n_{\gamma}+ e\vec{E}^{\mathrm{eh}} \left[\partial_\vec{q}n_\vec{\gamma}-i \left(\vec{ \mathcal A}^{\gamma \bar{\gamma}}_\vec{q} n_{\bar{\gamma} \gamma}-\vec{\mathcal A}^{\bar{\gamma}\gamma}_\vec{q} n_{\gamma \bar{\gamma}}\right)\right] + \frac{\tau''}{|\tau|^2} \partial_\vec{q} \varepsilon_{\gamma \vec{q}} \partial_\vec{R} n_\gamma=-\frac{2\varepsilon_{\gamma \vec{q}} \tau'}{|\tau|^2} (n_\gamma-n_\gamma^0).\end{split}
    \end{equation}
The electric field difference, $\vec{E}^{\mathrm{eh}}_{t \vec{R}}=-\partial_\vec{R}\phi^{\mathrm{eh}}_{t \vec{r}})$, not only shifts the distribution functions $n_\gamma$, but also induces intermode coherence. The coupling with the off-diagonal components of the distribution function is governed by the generalized Berry connection matrix $\vec{\mathcal A}_\vec{q}=i U_\vec{q}^\dagger \partial_\vec{q} U_\vec{q}$. The dissipative nature of fluctuating CPs dynamics is manifested in the presence of a relaxation term on the right-hand side of the kinetic equation.  The equation for the off-diagonal elements is kept to the zeroth order of the gradient expansion (all gradient terms including $\partial_t\hat{n}_{\gamma \bar{\gamma}}$ are neglected). As a result, the intermode coherence $n_{\gamma \bar{\gamma}}$ adiabatically follows the disturbed distribution functions $n_\gamma$ as
\begin{equation}
\label{ApPaperInterbandCoherence}
n_{\gamma \bar{\gamma}}=i \vec{A}^{\gamma \bar{\gamma}}_\vec{q}\partial_{\vec{R}} n_+ \frac{(\varepsilon_{\gamma\vec{q}}-\varepsilon_{\bar{\gamma}\vec{q}}) \tau''}{\varepsilon_{\gamma\vec{q}} \tau^*+\varepsilon_{\bar{\gamma}\vec{q}}\tau},\quad \;n_{\bar{\gamma}\gamma}=n_{\gamma \bar{\gamma}}^*. 
\end{equation}
Here we also have followed Ref.~\cite{TserkovnyakKineticEq} and performed the approximation $\partial_\vec{R}n_+\approx \partial_\vec{R} n_-$, which decouples the kinetic equations for $n_+$ and $n_+$. Besides, in this approximation predictions based on the resulting kinetic equations match with the ones based on the TDGL equation for fluctuating CPs and the linear response approach, as we discuss in the next subsection. If we combine expressions Eqs.~(\ref{ApPaperKineticEq1}) and  (\ref{ApPaperInterbandCoherence}), we obtain the kinetic equation, which is presented in the main part of the paper as Eq.~(\ref{DensityMatrixDiagonal2}).

\section{C3. The conductivity tensor}
In this section we demonstrate that the expressions for the conductivity tensor derived with the help of the kinetic equation for fluctuating CPs match with results obtained in Appendix B with the help of the linear response theory. As we discuss in the main text, electric field difference $e \vec{E}^{\mathrm{eh}}$ induces the three contributions to the current in both layers of the different physical origin 
\begin{equation}
\label{AppCurrentContributions1}
\vec{J}^{\mathrm{e(h)}}=\pm \sum_{\vec{p}\gamma}\left[ \partial_\vec{q} \varepsilon_{\gamma \vec{q}} n_{\gamma \vec{q}}^1 + u_{\gamma \vec{q}}^{\mathrm{\Omega}} n_{\gamma\vec{q}}^0+u_{\gamma \vec{q}}^\mathrm{G}n_{\gamma\vec{q}}^0\right],  \quad \quad \quad n_{\gamma\vec{q}}^0=\frac{T}{\varepsilon_{\gamma\vec{q}}}, \quad \quad \quad n_{\gamma\vec{q}}^1= -\frac{|\tau|^2 \vec{E}^{\mathrm{eh}} \partial_\vec{q} n_{\gamma\vec{q}}^0}{2 \tau' \varepsilon_{\gamma \vec{q}}}.
\end{equation}
The first term originates from the shift of the distribution function for fluctuating CPs and represents the conventional Aslamazov-Larkin effect. The corresponding contribution to the conductivity tensor is given by
\begin{equation}
\label{ApSigmaIntraKinetic}
\sigma_{\alpha \beta}^{\mathrm{AL}}=\sum_{\gamma\vec{q}} \frac{\partial_{\vec{q}_\alpha}\varepsilon_{\gamma \vec{q}} \partial_{\vec{q}_\beta}\varepsilon_{\gamma \vec{q}} }{2 \varepsilon^3_{\gamma\vec{q}}} \frac{T|\tau|^2}{\tau'}, \quad \quad \partial_{\vec{q}_\alpha}\varepsilon_{\gamma \vec{q}} \partial_{\vec{q}_\beta}\varepsilon_{\gamma \vec{q}} = \mathrm{Re}[Q_{\alpha \beta}^{\gamma \gamma}(\vec{q})]
\end{equation}
Here we recall that $\langle \gamma \vec{q}|v_\alpha(\vec{q})| \gamma \vec{q} \rangle_\mathrm{C}=\partial_{\vec{q}_\alpha} \varepsilon_{\gamma \vec{q}}$. As a result, this expression matches with the Eq.~(\ref{ApSigmaIntra}) derived in Appendix B. 

The second term originates from the anomalous velocity $u_{\gamma \vec{q}}^{\mathrm{\Omega}}$, which is intricately related with the nonzero Berry curvature of fluctuating CPs. It results only in the Hall conductivity that is given by
\begin{equation}
\label{ApSigmaInterOmegaKinetic}
\sigma_{\alpha\beta}^{\mathrm{\Omega}}=\epsilon_{\alpha\beta z}\sum_{\gamma \vec{q}}\Omega_{\gamma \vec{q}} \frac{(\varepsilon _{\gamma\vec{q}}-\varepsilon_{\bar{\gamma}\vec{q}})^2 |\tau'' \tau|^2}{|\varepsilon_{\gamma\vec{q}}\tau^*+\varepsilon_{\bar{\gamma}\vec{q}}\tau|^2}\; \frac{T}{\varepsilon_{\gamma\vec{q}}}, \quad \quad \Omega_{\gamma \vec{q}}=\frac{2 \mathrm{Im}[Q_{xy}^{\gamma \bar{\gamma}}]}{(\varepsilon_{\gamma\vec{q}}- \varepsilon_{\bar{\gamma}\vec{q}})^2}.
\end{equation}
Here we have rewritten the Berry curvature $\Omega_{\gamma \vec{q}}$ in terms of the product of matrix elements for velocity operators. As a result, this expression matches with Eq.~(\ref{ApSigmaInterOmega}) derived in Appendix B.

The third term originates from the anomalous velocity $u_{\gamma \vec{q}}^{\mathrm{G}}$, which is intricately related with the nonzero quantum metric of fluctuating CPs. It results both in longitudinal and transverse elements of the conductivity tensor that are given by 
\begin{equation}
\label{ApSigmaInterGKinetic}
\sigma_{\alpha\beta}^{\mathrm{G}}=\sum_{\gamma \vec{q}} \hat{G}_{\gamma \vec{q}}\cdot  \frac{(\varepsilon_{\gamma\vec{q}}^2-\varepsilon_{\bar{\gamma}\vec{q}}^2)\tau' |\tau|^2}{|\varepsilon_{\gamma\vec{q}}\tau^*+\varepsilon_{\bar{\gamma}\vec{q}}\tau|^2} \;  \frac{T}{\varepsilon_{\gamma\vec{q}}}, \quad \quad \hat{G}_{\gamma \vec{q}}=-\frac{ \mathrm{Re}[\hat{Q}^{\gamma \bar{\gamma}}]}{(\varepsilon_{\gamma\vec{q}}- \varepsilon_{\bar{\gamma}\vec{q}})^2}.  
\end{equation}
Here we have rewritten the quantum metric $G_{\gamma\vec{q}}$ in terms of the product of velocity operators. If we use the following helpful identity
\begin{equation}
\frac{\varepsilon_{+,\vec{q}}+\varepsilon_{-, \vec{q}}}{\varepsilon_{+,\vec{q}}-\varepsilon_{-, \vec{q}}} \left(\frac{T}{\varepsilon_{+,\vec{q}}}-\frac{T}{\varepsilon_{-,\vec{q}}}\right) = - \left(\frac{T}{\varepsilon_{+,\vec{q}}}+\frac{T}{\varepsilon_{-,\vec{q}}}\right)   
\end{equation}
the expression Eq.~(\ref{ApSigmaInterGKinetic}) matches with the Eq.~(\ref{ApSigmaInterG}) derived in Appendix B. As a result, we conclude that the results obtained with the help of linear response approach and the kinetic equation are consistent with each other. 
\end{appendix}

\end{widetext}

\end{document}